\documentclass[longauth]{aa}  

\usepackage{graphicx}
\usepackage{placeins}
 \usepackage{lscape}
\usepackage{txfonts}
\usepackage{amsmath}
\usepackage{natbib}
\usepackage[switch, modulo]{lineno}
\usepackage{caption}
\usepackage{subcaption}

\newcommand*\patchAmsMathEnvironmentForLineno[1]{%
  \expandafter\let\csname old#1\expandafter\endcsname\csname #1\endcsname
  \expandafter\let\csname oldend#1\expandafter\endcsname\csname end#1\endcsname
  \renewenvironment{#1}%
     {\linenomath\csname old#1\endcsname}%
     {\csname oldend#1\endcsname\endlinenomath}}% 
\newcommand*\patchBothAmsMathEnvironmentsForLineno[1]{%
  \patchAmsMathEnvironmentForLineno{#1}%
  \patchAmsMathEnvironmentForLineno{#1*}}%
\AtBeginDocument{%
\patchBothAmsMathEnvironmentsForLineno{equation}%
\patchBothAmsMathEnvironmentsForLineno{align}%
\patchBothAmsMathEnvironmentsForLineno{flalign}%
\patchBothAmsMathEnvironmentsForLineno{alignat}%
\patchBothAmsMathEnvironmentsForLineno{gather}%
\patchBothAmsMathEnvironmentsForLineno{multline}%
}

\newcommand{\Fermic}{\textit{Fermi}}
\newcommand{\Fermi}{\Fermic\ }
\newcommand{\FermiLATc}{\Fermic-LAT}
\newcommand{\FermiLAT}{\FermiLATc\ }

\newcommand{\Swiftc}{\textit{Swift}}
\newcommand{\Swift}{\Swiftc\ }

\newcommand{\RXTEc}{\textit{RXTE}}
\newcommand{\RXTE}{\RXTEc\ }

\newcommand{\lapp}{\ensuremath{\stackrel{\scriptstyle <}{{}_{\sim}}}}

\usepackage{color}
\usepackage{ulem}

\begin{document}

  \title{Unprecedented study of the broadband
emission of Mrk~421 
    during flaring activity in March 2010
}
   \titlerunning{Mrk~421 in March 2010}
   %\subtitle{Mrk~421 March 2010}

% authors 30.05.2014  Format AA
%

\author{
J.~Aleksi\'c\inst{1} \and
S.~Ansoldi\inst{2} \and
L.~A.~Antonelli\inst{3} \and
P.~Antoranz\inst{4} \and
A.~Babic\inst{5} \and
P.~Bangale\inst{6} \and
U.~Barres de Almeida\inst{6,}\inst{25} \and
J.~A.~Barrio\inst{7} \and
J.~Becerra Gonz\'alez\inst{8,}\inst{26} \and
W.~Bednarek\inst{9}
% ...
\and  \\
(The MAGIC Collaboration) \\
S.~Archambault\inst{33} \and
A.~Archer\inst{34} \and
M.~Beilicke\inst{34} \and
W.~Benbow\inst{35} \and
K.~Berger\inst{36} \and
R.~Bird\inst{37} \and
J.~Biteau\inst{38} \and
J.~H.~Buckley\inst{34} \and
V.~Bugaev\inst{34} \and
M.~Cerruti\inst{35} \and
X.~Chen\inst{39,10} \and
L.~Ciupik\inst{40} \and
E.~Collins-Hughes\inst{37} \and
W.~Cui\inst{41} \and
J.~D.~Eisch\inst{42} \and
A.~Falcone\inst{43} \and
Q.~Feng\inst{41} \and
J.~P.~Finley\inst{41} \and
P.~Fortin\inst{35} \and
L.~Fortson\inst{44} \and
A.~Furniss\inst{38} \and
N.~Galante\inst{35} \and
G.~H.~Gillanders\inst{45} \and
S.~Griffin\inst{33} \and
G.~Gyuk\inst{40} \and
N.~H{\aa}kansson\inst{39} \and
J.~Holder\inst{36} \and
C.~A.~Johnson\inst{38} \and
P.~Kaaret\inst{46} \and
P.~Kar\inst{47} \and
M.~Kertzman\inst{48} \and
D.~Kieda\inst{47} \and
M.~J.~Lang\inst{45} \and
S.~McArthur\inst{49} \and
A.~McCann\inst{50} \and
K.~Meagher\inst{51} \and
J.~Millis\inst{52} \and
P.~Moriarty\inst{45,53} \and
R.~A.~Ong\inst{54} \and
A.~N.~Otte\inst{51} \and
J.~S.~Perkins\inst{26} \and
A.~Pichel\inst{55} \and
M.~Pohl\inst{39,10} \and
A.~Popkow\inst{54} \and
H.~Prokoph\inst{10} \and
E.~Pueschel\inst{37} \and
K.~Ragan\inst{33} \and
L.~C.~Reyes\inst{56} \and
P.~T.~Reynolds\inst{57} \and
G.~T.~Richards\inst{51} \and
E.~Roache\inst{35} \and
A.~C.~Rovero\inst{55} \and
G.~H.~Sembroski\inst{41} \and
K.~Shahinyan\inst{44} \and
D.~Staszak\inst{33} \and
I.~Telezhinsky\inst{39,10} \and
J.~V.~Tucci\inst{41} \and
J.~Tyler\inst{33} \and
A.~Varlotta\inst{41} \and
S.~P.~Wakely\inst{49} \and
R.~Welsing\inst{10} \and
A.~Wilhelm\inst{39,10} \and
D.~A.~Williams\inst{38} \and \\
(The VERITAS Collaboration) \\
S.~Buson\inst{15} \and
J.~Finke\inst{58} \and 
M.~Villata\inst{59} \and
C.~Raiteri\inst{59} \and
H.~D.~Aller\inst{60} \and
M.~F.~Aller\inst{60} \and
A.~Cesarini\inst{61} \and
W.~P.~Chen\inst{62} \and
M.~A.~Gurwell\inst{63} \and
S.~G.~Jorstad\inst{64,65} \and
G.~N.~Kimeridze\inst{67} \and
E.~Koptelova\inst{62,66} \and
O.~M.~Kurtanidze\inst{67,68} \and
S.~O.~Kurtanidze\inst{67} \and
A.~L\"ahteenm\"aki\inst{69,70} \and
V.~M.~Larionov\inst{71,72,73} \and
E.~G.~Larionova\inst{71} \and
H.~C.~Lin\inst{62} \and
B.~McBreen\inst{37} \and 
J.~W.~Moody\inst{74} \and 
D.~A.~Morozova\inst{71} \and
A.~P.~Marscher\inst{64} \and
W.~Max-Moerbeck\inst{75} \and
M.~G.~Nikolashvili\inst{67} \and
M.~Perri\inst{3,76} \and
A.~C.~S.~Readhead\inst{75} \and
J.~L.~Richards\inst{41} \and
J.~A.~Ros\inst{77} \and
A.~C.~Sadun\inst{78} \and
T.~Sakamoto\inst{79} \and
L.~A.~Sigua\inst{67} \and
P.~S.~Smith\inst{80} \and
M.~Tornikoski\inst{69} \and
I.~S.~Troitsky\inst{71} \and
A.~E.~Wehrle\inst{81} \and
B.~Jordan\inst{82}
}
\institute { IFAE, Campus UAB, E-08193 Bellaterra, Spain
\and Universit\`a di Udine, and INFN Trieste, I-33100 Udine, Italy
\and INAF National Institute for Astrophysics, I-00136 Rome, Italy
\and Universit\`a  di Siena, and INFN Pisa, I-53100 Siena, Italy
\and Croatian MAGIC Consortium, Rudjer Boskovic Institute, University of Rijeka and University of Split, HR-10000 Zagreb, Croatia
\and Max-Planck-Institut f\"ur Physik, D-80805 M\"unchen, Germany
\and Universidad Complutense, E-28040 Madrid, Spain
\and Inst. de Astrof\'isica de Canarias, E-38200 La Laguna, Tenerife, Spain
\and University of \L\'od\'z, PL-90236 Lodz, Poland
\and Deutsches Elektronen-Synchrotron (DESY), D-15738 Zeuthen, Germany
\and ETH Zurich, CH-8093 Zurich, Switzerland
\and Universit\"at W\"urzburg, D-97074 W\"urzburg, Germany
\and Centro de Investigaciones Energ\'eticas, Medioambientales y Tecnol\'ogicas, E-28040 Madrid, Spain
\and Institute of Space Sciences, E-08193 Barcelona, Spain
\and Universit\`a di Padova and INFN, I-35131 Padova, Italy
\and Technische Universit\"at Dortmund, D-44221 Dortmund, Germany
\and Unitat de F\'isica de les Radiacions, Departament de F\'isica, and CERES-IEEC, Universitat Aut\`onoma de Barcelona, E-08193 Bellaterra, Spain
\and Universitat de Barcelona, ICC, IEEC-UB, E-08028 Barcelona, Spain
\and Japanese MAGIC Consortium, Division of Physics and Astronomy, Kyoto University, Japan
\and Finnish MAGIC Consortium, Tuorla Observatory, University of Turku and Department of Physics, University of Oulu, Finland
\and Inst. for Nucl. Research and Nucl. Energy, BG-1784 Sofia, Bulgaria
\and Universit\`a di Pisa, and INFN Pisa, I-56126 Pisa, Italy
\and ICREA and Institute of Space Sciences, E-08193 Barcelona, Spain
\and Universit\`a dell'Insubria and INFN Milano Bicocca, Como, I-22100 Como, Italy
\and now at Centro Brasileiro de Pesquisas F\'isicas (CBPF/MCTI), R. Dr. Xavier Sigaud, 150 - Urca, Rio de Janeiro - RJ, 22290-180, Brazil
\and NASA Goddard Space Flight Center, Greenbelt, MD 20771, USA and Department of Physics and Department of Astronomy, University of Maryland, College Park, MD 20742, USA
\and now at Ecole polytechnique f\'ed\'erale de Lausanne (EPFL), Lausanne, Switzerland
\and now at Institut f\"ur Astro- und Teilchenphysik, Leopold-Franzens- Universit\"at Innsbruck, A-6020 Innsbruck, Austria
\and now at Finnish Centre for Astronomy with ESO (FINCA), Turku, Finland
\and now at Astrophysics Science Division, Bhabha Atomic Research Centre, Mumbai 400085, India
\and also at INAF-Trieste
\and now at School of Chemistry \& Physics, University of Adelaide, Adelaide 5005, Australia
\and Physics Department, McGill University, Montreal, QC H3A 2T8, Canada
\and Department of Physics, Washington University, St. Louis, MO 63130, USA
\and Fred Lawrence Whipple Observatory, Harvard-Smithsonian Center for Astrophysics, Amado, AZ 85645, USA
\and Department of Physics and Astronomy and the Bartol Research Institute, University of Delaware, Newark, DE 19716, USA
\and School of Physics, University College Dublin, Belfield, Dublin 4, Ireland
\and Santa Cruz Institute for Particle Physics and Department of Physics, University of California, Santa Cruz, CA 95064, USA
\and Institute of Physics and Astronomy, University of Potsdam, 14476 Potsdam-Golm, Germany
\and Astronomy Department, Adler Planetarium and Astronomy Museum, Chicago, IL 60605, USA
\and Department of Physics and Astronomy, Purdue University, West Lafayette, IN 47907, USA
\and Department of Physics and Astronomy, Iowa State University, Ames, IA 50011, USA
\newpage
\and Department of Astronomy and Astrophysics, 525 Davey Lab, Pennsylvania State University, University Park, PA 16802, USA
\and School of Physics and Astronomy, University of Minnesota, Minneapolis, MN 55455, USA
\and School of Physics, National University of Ireland Galway, University Road, Galway, Ireland
\and Department of Physics and Astronomy, University of Iowa, Van Allen Hall, Iowa City, IA 52242, USA
\and Department of Physics and Astronomy, University of Utah, Salt Lake City, UT 84112, USA
\and Department of Physics and Astronomy, DePauw University, Greencastle, IN 46135-0037, USA
\and Enrico Fermi Institute, University of Chicago, Chicago, IL 60637, USA
\and Kavli Institute for Cosmological Physics, University of Chicago, Chicago, IL 60637, USA
\and School of Physics and Center for Relativistic Astrophysics, Georgia Institute of Technology, 837 State Street NW, Atlanta, GA 30332-0430, USA
\and Department of Physics, Anderson University, 1100 East 5th Street, Anderson, IN 46012, USA
\and Department of Life and Physical Sciences, Galway-Mayo Institute of Technology, Dublin Road, Galway, Ireland
\and Department of Physics and Astronomy, University of California, Los Angeles, CA 90095, USA
\and Instituto de Astronomia y Fisica del Espacio, Casilla de Correo 67 - Sucursal 28, (C1428ZAA) Ciudad Autónoma de Buenos Aires, Argentina
\and Physics Department, California Polytechnic State University, San Luis Obispo, CA 94307, USA
\and Department of Applied Physics and Instrumentation, Cork Institute of Technology, Bishopstown, Cork, Ireland
\and Space Science Division, Naval Research Laboratory, Washington, DC 20375-5352, USA
\and INAF, Osservatorio Astronomico di Torino, I-10025 Pino Torinese (TO), Italy
\and Department of Astronomy, University of Michigan, Ann Arbor, MI 48109-1042, USA
\and Department of Physics, University of Trento, I38050, Povo, Trento, Italy
\and Graduate Institute of Astronomy, National Central University, Jhongli 32054, Taiwan
\and Harvard-Smithsonian Center for Astrophysics, Cambridge, MA 02138, USA
\and Institute for Astrophysical Research, Boston University, 725 Commonwealth Avenue, Boston, MA 02215, USA
\and Astronomical Institute, St. Petersburg State University, Universitetskij Pr. 28, Petrodvorets, 198504 St. Petersburg, Russia
\and Institute of Astronomy, National Tsing Hua University, 101 Guanfu Rd., Hsinchu 30013, Taiwan
\and Abastumani Observatory, Mt. Kanobili, 0301 Abastumani, Georgia
\and Landessternwarte, Zentrum f\"ur Astronomie der Universit\"at Heidelberg, K\"onigstuhl 12, 69117 Heidelberg, Germany
\and Aalto University Mets\"ahovi Radio Observatory, Mets\"ahovintie 114, 02540 Kylm\"al\"a, Finland
\and Aalto University Department of Radio Science and Engineering,  P.O. BOX 13000, FI-00076 AALTO, Finland
\and Astron.\ Inst., St.-Petersburg State University, Russia
\and Pulkovo Observatory, St.-Petersburg, Russia
\and Isaac Newton Institute of Chile, St.-Petersburg Branch, Russia
\and Department of Physics and Astronomy, Brigham Young University, Provo, Utah
\and Cahill Center for Astronomy and Astrophysics, California Institute of Technology, 1200 E California Blvd, Pasadena, CA 91125, USA
\and ASI-Science Data Center, Via del Politecnico, I-00133 Rome, Italy
\and Agrupaci\'o Astron\`omica de Sabadell, Spain
\and Department of Physics, University of Colorado Denver, Denver, Colorado
\and Department of Physics and Mathematics, College of Science and Engineering, Aoyama Gakuin University, 5-10-1 Fuchinobe, 1105 Chuoku, Sagamihara-shi Kanagawa 252-5258, Japan
\and Steward Observatory, University of Arizona, Tucson, AZ 85721, USA
\and Space Science Institute, Boulder, CO 80301, USA
\and School of Cosmic Physics, Dublin Institute for Advanced Studies, Belfield, Dublin 2, Ireland
\newpage
\and {*} Corresponding authors: David Paneque (dpaneque@mpp.mpg.de), Shangyu Sun (sysun@mpp.mpg.de), Hajime Takami (takami@post.kek.jp)
}

   %\date{Received September 15, 1996; accepted March 16, 1997}

% \abstract{}{}{}{}{} 
% 5 {} token are mandatory

% DAvid Paneque (2014/05/30)
% Additional authors to be included for the MW data. List below is
% still growing. 

% Svetlana G. Jorstad\altaffilmark{1,2}, Alan P. Marscher\altaffilmark{1}
% \altaffiltext{1}{Institute for Astrophysical Research, Boston University, 725 Commonwealth Avenue, Boston, MA 02215}
% \altaffiltext{2}{Astronomical Institute, St. Petersburg State University, Universitetskij Pr. 28, Petrodvorets,
%198504 St. Petersburg, Russia}

% Russians (St Petersburg Crimean)
%V.~M.~Larionov \inst{1,2,3} 
%E.~G.~Larionova \inst{1} 
%D.~A.~Morozova \inst{1} 
%I.~S.~Troitsky \inst{1} 
%   \institute{
 % 1
 %  \and   Astron.\ Inst., St.-Petersburg State Univ., Russia
 % 2
 %  \and   Pulkovo Observatory, St.-Petersburg, Russia
 % 3
 %\and Isaac Newton Institute of Chile, St.-Petersburg Branch
 % 4
 %  \and   Physics  Department, University of Crete, Heraklion, Greece

% GASP
% MAssimo
%  J.~A.~Ros                Agrupaci\'o Astron\`omica de Sabadell, Spain  

% New mexico skies
% http://www.nmskies.com/index.html
% Professor J. Ward Moody, Department of Physics and Astronomy, Brigham Young University, Provo, Utah
% Professor Alberto C. Sadun, Department of Physics, University of Colorado Denver, Denver, Colorado

% Swift
% MAtteo PErri
% Andrea Cesarini

% SMA
%   Mark A. Gurwell, Harvard-Smithsonian Center for Astrophysics, Cambridge, MA 02138 USA
%   Ann E. Wehrle, Space Science Institute, 4750 Walnut Street, Suite 205, Boulder, CO 80301

\abstract
{Because of its proximity, Mrk~421 is one of the best sources on which to study the nature of BL Lac objects. Its proximity allows us to characterize its broadband spectral energy distribution (SED).}
{The goal is to better understand the mechanisms responsible for the broadband emission and the temporal evolution of Mrk~421. These mechanisms may also apply to more distant blazars that cannot be studied with the same level of detail.}
{A flare occurring in March 2010 was observed for 13 consecutive
  days (from MJD 55265 to MJD 55277) with unprecedented wavelength
  coverage from radio to very high energy (VHE; $E > 100$ GeV)
  $\gamma$-rays with MAGIC, VERITAS, Whipple, \FermiLATc,
  \textit{MAXI}, \RXTEc, \Swiftc, GASP-WEBT, and several optical and
  radio telescopes. We modeled the day-scale SEDs with one-zone and two-zone synchrotron self-Compton (SSC) models, investigated the physical parameters, and evaluated whether the observed broadband SED variability can be associated with variations in the relativistic particle population.}
{The activity of Mrk~421 initially was high and then slowly decreased
  during the 13-day period. The flux variability was remarkable at the
  X-ray and VHE bands, but it was minor or not significant at the
  other bands. The variability in optical polarization was also
  minor.  These observations revealed an almost linear correlation
  between the X-ray flux at the 2--10 keV band and the VHE
  $\gamma$-ray flux above 200 GeV,  consistent with the $\gamma$-rays
  being produced  by inverse-Compton scattering in the Klein-Nishina
  regime in the framework of SSC models.
The one-zone SSC model can describe 
%% HT141119 %%
%the temporal evolution of the broadband SEDs during the 13 consecutive days,  
the SED of each day for the 13 consecutive days reasonably well, 
which once more shows the success of this standard theoretical scenario to
describe the SEDs of VHE BL Lacs such as Mrk~421. This flaring activity
is also very well described by a two-zone SSC model, where one zone is
responsible for the quiescent emission, while the other smaller zone,
which is spatially separated from the first, contributes to the
daily variable emission occurring at X-rays and VHE $\gamma$-rays. The
second blob is assumed to have a smaller volume and a narrow electron
energy distribution with $3 \times 10^4 < \gamma < 6 \times 10^5$,
where $\gamma$ is the Lorentz factor of the electrons.  Such a two-zone scenario would naturally lead to the correlated variability at the X-ray and VHE bands without variability at the optical/UV band, as well as to shorter timescales for the variability at the X-ray and VHE bands with respect to the variability at the other bands.}
{Both the one-zone and the two-zone SSC models can describe the
  daily SEDs via the variation of only four or five model
  parameters, under the hypothesis that the variability is associated
  mostly with the underlying particle population. This shows that the
  particle acceleration and cooling mechanism that produces the radiating particles might be the main mechanism responsible for the broadband SED variations during the flaring episodes in blazars. The two-zone SSC model provides a better agreement with the observed SED at the narrow peaks of the low- and high-energy bumps during the highest activity, although the reported one-zone SSC model could be further improved by varying the parameters related to the emitting region itself ($\delta$, $B$ and $R$), in addition to the parameters related to the particle population.}

%\keywords{non-thermal radiation mechanisms -- active galaxies -- BL Lacertae objects -- $\gamma$-ray observations}
\keywords{radiation mechanisms: non-thermal -- galaxies: active -- BL Lacertae objects: individual -- gamma rays: galaxies \\
\vspace{0.3cm}
\\ {\bf Online-data.} Multi-wavelength light curves (data in Fig. 1) and broadband spectral energy distributions (the data in Figs. 6, and B1-B4) are available at the CDS via anonymous ftp to cdsarc.u-strasbg.fr (130.79.128.5) or via \url{http://cdsarc.u-strasbg.fr/viz-bin/qcat?J/A+A/578/A22}
}

\maketitle
%\linenumbers
%%%%%%%%%%%%%%%%%%%%%%%%%%%%%%%%%%%%%%%%%%%%%%%%%%%%%%%%%%%%
%%%%%%%%%%%%%%%%%%%%%%%%%%%%%%%%%%%%%%%%%%%%%%%%%%%%%%%%%%%%
\section{Introduction}
%%%%%%%%%%%%%%%%%%%%%%%%%%%%%%%%%%%%%%%%%%%%%%%%%%%%%%%%%%%%
%%%%%%%%%%%%%%%%%%%%%%%%%%%%%%%%%%%%%%%%%%%%%%%%%%%%%%%%%%%%

Markarian 421 (Mrk~421; RA=11$^{\rm h}$4$\arcmin$27.31$\arcsec$,
Dec=38$^\circ$12$\arcmin$31.8$\arcsec$, J2000) is a BL Lac object that is
believed to have a pair of relativistic jets flowing in opposite
directions closely aligned to our line of sight. It is also one of the closest \citep[$z = 0.031$;][]{1991trcb.book.....D} and brightest BL Lac objects in the extragalactic X-ray and very high energy (VHE; $E > 100$ GeV) sky. This object was the first BL Lac object detected by the Energetic Gamma Ray Experiment Telescope \citep[EGRET;][]{lin92} at energies above 100 MeV, and was also the first extragalactic source detected by imaging atmospheric Cherenkov telescopes \citep[IACTs;][]{1992Natur.358..477P}. Mrk~421 is one of the best-studied BL Lac objects at VHE because it can be detected by modern IACTs within several minutes, and its broadband spectral energy distribution (SED) can be well measured by operating instruments covering energies from radio to VHE. Nearly all the IACTs have measured its VHE $\gamma$-ray spectrum \citep{Krennrich2002, Aharonian2002,2002ApJ...579L...9O,Aharonian2003,Aharonian2005,Mrk421MAGIC}.

The SED of a blazar is dominated by the emission of the jet components
magnified by relativistic beaming. The observed spectrum and
polarization indicate that the low-energy bump is synchrotron
radiation of electrons in turbulent magnetic fields in the
jet. Mrk~421 has a peak frequency of the low-energy bump above
$10^{15}$ Hz, and therefore it is categorized as a
high-synchrotron-peaked (HSP) BL Lac object based on the
classification criterion presented in \citet{latsed}. The peak
frequency of the high-energy bump for HSP blazars detected at VHE is
usually below 100 GeV\footnote{See the TeV catalog at \url{http://tevcat.uchicago.edu/}}. This bump may be interpreted as the inverse-Compton scattering of the same population of electrons off synchrotron photons \citep[synchrotron self-Compton, SSC;][]{1992ApJ...397L...5M,dermer, bloom}. Alternatively, hadronic models can also explain this bump \citep[e.g.,][]{1993A&A...269...67M,Mucke}. Although both leptonic and hadronic models can reproduce the time-averaged broadband SED of Mrk~421 \citep[e.g.][]{AbdoMrk421}, it is difficult to produce short-time variability ($< 1$ hour) with hadronic models, which has been observed in Mrk~421 \citep[e.g.][]{1996Natur.383..319G}. Thus, leptonic models are favored, at least in active states. A recent study on Mrk~421 also supports leptonic models during low blazar activity \citep{2015arXiv150202650M}. In leptonic scenarios, one-zone SSC models with an electron distribution described by one or two power-law functions can typically describe the observed SEDs \citep[e.g.,][]{Katar2003,Blazejowski2005,Mrk421Whipple2006,fossati08,Horan2009}.

Because Mrk~421 is bright and highly variable, long-term multiwavelength (MW) monitoring campaigns have been organized to intensely study its SED and its temporal evolution from radio to VHE $\gamma$-rays. Since 2009, an exceptionally long and dense monitoring of the broadband emission of Mrk~421 has been performed. The results of the 2009 MW campaign, which 
%% HT141119 %%
%relate 
is related 
to Mrk~421 during nonflaring (typical) activity, were reported in \cite{AbdoMrk421}. The SED was successfully modeled by both a leptonic and a hadronic model, but the authors commented that the hadronic model required extreme conditions for particle acceleration and confinement. Moreover, the densely sampled SED revealed that the leptonic one-zone SSC model required two breaks in the electron energy distribution (EED) to satisfactorily describe the smooth bumps in the quiescent state SED.

%\footnote{The flux above 200 GeV of the Crab nebula used in this paper is $2.2\times 10^{-10}$ cm$^{-2}$s$^{-1}$.This value is obtained by integrating the fit function published in \cite{Ale12} from 200 GeV to 10 TeV.}

Mrk~421 showed high activity during the entire multi-instrument
campaign in 2010. During the period from March 10 (Modified Julian Day
[MJD] 55265) to March 22 (MJD 55277), the VHE activity decreased from
a high flux $\sim 2$ Crab units (c.u.)\footnote{The VHE flux of the
  Crab nebula between 200 GeV and 10 TeV used in this paper is
  $2.2\times 10^{-10}$ cm$^{-2}$s$^{-1}$ \citep{Ale12}.} down to the
typical value $\sim 0.5$ c.u \citep{2014APh....54....1A}, hence
offering the possibility of studying the evolution of the SED during
the decay of a flaring event. The extensive MW data collected allow measuring the broadband SED over the largest available
fraction of wavelengths with simultaneous observations (mostly within
2-3 hours) during 13 consecutive days. The present study is
unprecedented for any blazar. The SED and indicated physical
parameters in the emission region at different epochs and their
temporal evolution have been studied
\citep[e.g.,][]{2011ApJ...733...14M,AcciariMrk4212011,2012A&A...542A.100A},
but based on sparse sampling. The observational data for 13
consecutive days provide a first opportunity to directly study the temporal evolution of the SED.

In Sect.~\ref{Observation}, we report the observations and data analysis performed with the various instruments. In Sect.~\ref{LightCurves} we present the observational results on multi-band variability. In Sect.~\ref{SED_Model}, all the broadband SEDs during the flaring activity are characterized within two SSC models and physical parameters in emission regions are derived. In Sect.~\ref{Discussion} we discuss the interpretation of the experimental results, and then we summarize this study in Sect.~\ref{Conclusion}. 
Throughout this paper, the $\Lambda$CDM cosmology with $H_0 = 71$ km s$^{-1}$, 
$\Omega_{\rm M} = 0.27$, and $\Omega_{\Lambda} = 0.73$ is adopted.

%%%%%%%%%%%%%%%%%%%%%%%%%%%%%%%%%%%%%%%%%%%%%%%%%%%%%%%%%%%%
%%%%%%%%%%%%%%%%%%%%%%%%%%%%%%%%%%%%%%%%%%%%%%%%%%%%%%%%%%%%
\section{Observation and data analysis} \label{Observation}
%%%%%%%%%%%%%%%%%%%%%%%%%%%%%%%%%%%%%%%%%%%%%%%%%%%%%%%%%%%%
%%%%%%%%%%%%%%%%%%%%%%%%%%%%%%%%%%%%%%%%%%%%%%%%%%%%%%%%%%%%

All instruments that observed Mrk~421 during this campaign are listed in Table\,\ref{TableWithInstruments}. 
The details of observations by each instrument are described below.

%%%%%%%%%%%%%%%%%%%%%%%%%%%%%%%%%%%%%%%%%%%%%%%%%%%%%%%%%%%%
\subsection{MAGIC}
%%%%%%%%%%%%%%%%%%%%%%%%%%%%%%%%%%%%%%%%%%%%%%%%%%%%%%%%%%%%

The 
%% HT141119 %%
%MAGIC 
Major Atmospheric Gamma-ray Imaging Cherenkov (MAGIC)
telescope system consists of two 17-meter telescopes that
are located on the island of La Palma, 2200 m above sea level. Stereo
observation can provide a sensitivity of $\sim 0.008$ c.u. above $\sim$300 GeV in 50 hours of observation and allows detecting VHE photons between 50 GeV and 50 TeV. A detailed description of the performance of the MAGIC stereo system can be found in \cite{Ale12}.

During this flare, MAGIC made 11 observations, all in stereoscopic
mode. The exposure time ranged from $\sim 10$ to $\sim 80$ minutes,
with the zenith angle ranging from 5 to 30 degrees. In total we collected 4.7 hours of good-quality data. The MAGIC data presented in this paper were taken in dark conditions and were not affected by bright moonlight. All these observations were conducted in the false-source-tracking (wobble) mode \citep{fomin1994}: alternatively tracking two positions in the sky that are symmetric with respect to the true source position and $0.4^{\circ}$ away from it. The MAGIC data on MJD 55272 and 55275 suffered from bad weather and occasional technical problems and were therefore removed from the analysis.

The MAGIC data were analyzed using the MAGIC Standard Analysis Software
\citep[MARS; ][]{2010ascl.soft11004M}. In the analysis routine,
signals are first calibrated and then an image-cleaning algorithm that involves the time structure of the shower images, and removes the
contribution from the night sky background is applied. Afterward ,
the shower images are parameterized with an extended set of Hillas
parameters \citep{Hillas:1985ta}, and another parameter, hadronness, to reject background showers resulting from charged cosmic
rays. The hadronness is determined through a random forest
classification \citep{Breiman2001}, which is trained based on shower-image parameters and time information.

Then, all these parameters from the two telescopes are combined
to reconstruct the arrival directions and energies of the $\gamma$-ray
candidate events. The number of signal (excess) events is the number
of events around the source position after subtracting the number
of background events, which is estimated using the number of events in
a source-free region. Flux and a preliminary spectrum are calculated
based on this number. Finally, this preliminary spectrum is unfolded
to correct for the effect of the limited energy resolution of the
detector, as reported in \cite{Albert2007c}, which leads to the final
(true) observed VHE spectrum of the source. 

The systematic uncertainties in the spectral measurements with MAGIC
stereo observations are 11\% in the normalization factor (at $>$300
GeV) and  0.15--0.20 
in the photon index. The error on the flux does not include uncertainty on the energy scale. The energy scale of the MAGIC telescopes is determined with a precision of about 17\% at low energies ($E<$100 GeV) and 15\% at medium energies ($E> 300$ GeV). Further details are reported in \citet{Ale12}.

%%%%%%%%%%%%%%%%%%%%%%%%%%%%%%%%%%%%%%%%%%%%%%%%%%%%%%%%%%%%
\subsection{VERITAS}
%%%%%%%%%%%%%%%%%%%%%%%%%%%%%%%%%%%%%%%%%%%%%%%%%%%%%%%%%%%%

%% HT141119 %%
%VERITAS 
The Very Energetic Radiation Imaging Telescope Array System (VERITAS)
is an array of four imaging atmospheric Cherenkov telescopes 12~m in diameter that are located in southern Arizona \citep{Weekes:2002pi} and are designed to detect emission from astrophysical objects in the energy range from $\sim$100~GeV to greater than 30~TeV. VERITAS has an energy resolution of $\sim 15\%$ and an angular resolution (68\% containment) of $\sim 0.1^\circ$ per event at 1~TeV. A source with a flux of 0.01 c.u. is detected in $\sim 25$ hours of observations, while a \mbox{0.05 c.u.} source is detected in less than 2~hours. The field of view of the VERITAS cameras is $3.5^\circ$. For more details on the VERITAS instrument and its imaging atmospheric Cherenkov technique, see \citet{Perkins2009}.

VERITAS monitored Mrk 421 in March 2010 with a 10-minute run each day on MJD 55260, 55265, 55267-55274. Observations were taken near culmination at zenith angles in the range $18^\circ$ -- $23^\circ$ to benefit from the lowest possible energy threshold. All data were taken in wobble mode where the telescopes are pointed away from the source by $0.5^\circ$ north, south, east, and west to allow for simultaneous background estimation using events from the same field of view.

Before event selection and background subtraction, all shower images are calibrated and cleaned as described in \citet{Cogan:2006} and \citet{Daniel:2007kx}. Following the calibration and cleaning of the data, the events are parameterized using a moment analysis \citep{Hillas:1985ta}. From this moment analysis, scaled parameters are calculated and used to select the $\gamma$-ray-like events \citep{Aharonian:1997rm,Krawczynski:2006ts}. The event-selection cuts are optimized {\it a priori} for a Crab-like source (power-law spectrum photon index $\Gamma = 2.5$ and Crab nebula flux level).

%%%%%%%%%%%%%%%%%%%%%%%%%%%%%%%%%%%%%%%%%%%%%%%%%%%%%%%%%%%%
\subsection{Whipple 10 m}
%%%%%%%%%%%%%%%%%%%%%%%%%%%%%%%%%%%%%%%%%%%%%%%%%%%%%%%%%%%%

The Whipple 10 m $\gamma$-ray telescope was situated at the Fred
Lawrence Whipple Observatory in southern Arizona. It operated in the
300 GeV to 20 TeV energy range, with a peak response energy (for a
Crab-like spectrum) of approximately 400 GeV. The telescope had a
10 meter optical reflector with a camera consisting of 379
photomultiplier tubes, covering a field of view of
$2.6^{\circ}$\citep{2007APh....28..182K}. The Whipple 10-m was
  decommissioned in July 2011.

The Whipple 10 m telescope made ten observations performed in the
ON/OFF and TRK (tracking) modes, in which the telescope tracked the
source, which was centered in the field of view, for 28 minutes (ON
and TRK runs). The duration of the observations ranged from about one to six hours, with half of the observations more than four hours long. The
corresponding OFF run was collected at an offset of 30 minutes from
the source's right ascension for a period of 28 minutes. The two runs
were taken at the same declination over the same range of telescope
azimuth and elevation angles. This removed systematic errors that
depend on slow changes in the atmosphere. In the TRK mode, only ON
runs were taken without corresponding OFF observations, and the
background was estimated from events whose major axis points away
from the center of the camera (the source position). The data set amounts to 36 hours and
was analyzed using the University College Dublin analysis package as
described in \cite{AcciariPhDThesis}. The photon fluxes, initially
derived in Crab units for energies above 400 GeV, were converted into
photon fluxes above 200 GeV using a Crab nebula flux of $2.2 \times
10^{-10}$ cm$^{-2}$ s$^{-1}$ \citep{Ale12}. Because the spectrum of
Mrk~421 is variable (and sometimes slightly harder or softer than that of the Crab nebula), this conversion could overestimate or underestimate the photon fluxes, but
only at the level of $\sim$10\%, which is not critical for the results
reported in this paper.

\begin{table*}
\caption{List of participating instruments in the campaign on Mrk~421 during 2010 March.}\label{TableWithInstruments}
\centering
\begin{tabular}{ll} 
\hline\hline       
Instrument/Observatory        & Energy range covered \\
\hline
MAGIC                 & 0.08--5.0\,TeV                   \\
VERITAS               & 0.2--5.0\,TeV                    \\
Whipple 10-m             &0.4--2.0\,TeV                  \\           
\FermiLAT                & 0.1--400\,GeV                 \\
\Swiftc-BAT                & 14--195\,keV                \\
\RXTEc-PCA                & 3--32\,keV                   \\
\Swiftc-XRT                & 0.3--10\,keV                \\
\RXTEc-ASM                & 2--10\,keV                   \\
\textit{MAXI}         & 2--10\,keV       \\
\Swiftc-UVOT                & UVW1, UVM2, UVW2                \\
Abastumani$^{\dagger}$         & R       band \\
Lulin$^{\dagger}$              & R       band  \\
Roque de los Muchachos (KVA)$^{\dagger}$          & R       band        \\
St. Petersburg$^{\dagger}$             & R band polarization        \\
Sabadell$^{\dagger}$              & R      band         \\
Goddard Robotic Telescope (GRT)                &  R band                 \\
The Remote Observatory for Variable Object Research (ROVOR)        &   B, R, V  bands          \\
New Mexico Skies (NMS)              &   R, V bands           \\
Bradford Robotic Telescope (BRT)     & B, R, V bands    \\
Perkins  &  R band polarization    \\
Steward     &  R band polarization      \\
Crimean     &  R band polarization      \\
Submillimeter Array (SMA)            &  225\,GHz   \\
Mets\"ahovi Radio Observatory$^{\dagger}$                   & 37\,GHz                \\
University of Michigan Radio Astronomy Observatory (UMRAO)$^{\dagger}$                & 8.0, 14.5\,GHz \\
Owens Valley Radio Observatory (OVRO)                & 15\,GHz                                                   \\
\hline
\end{tabular}
\tablefoot{The energy range shown in Column 2 is the actual energy range covered during the Mrk\,421 observations, and not necessarily the nominal energy range of the instrument, which might only be achievable for bright sources and in excellent observing conditions. \\
$^{\dagger}$ through GASP-WEBT program}
\end{table*}

%\noindent 

%%%%%%%%%%%%%%%%%%%%%%%%%%%%%%%%%%%%%%%%%%%%%%%%%%%%%%%%%%%%
\subsection{\FermiLAT}
%%%%%%%%%%%%%%%%%%%%%%%%%%%%%%%%%%%%%%%%%%%%%%%%%%%%%%%%%%%%

The {\it Fermi} Large Area Telescope (LAT) is a $\gamma$-ray telescope
operating from $20$\,MeV to more than 300\,GeV. The LAT is an array of 4 $\times$ 4 identical towers, each one consisting of a tracker (where the photons are pair-converted) and a calorimeter (where the energies of the pair-converted photons are measured).  LAT has a large peak effective area (0.8 m$^2$ for 1 GeV photons), an energy resolution typically better than 10\%, and a field of view of about 2.4 sr with an angular resolution (68\% containment angle) better than 1$^{\circ}$ for energies above 1 GeV. Further details on the description of LAT are given in \citet{Atwood2009} and \citet{Ackermann2012}. The analyses of the \FermiLAT data were performed here with the ScienceTools software package version v9r32p5. We used the reprocessed \FermiLAT events\footnote{See \url{http://fermi.gsfc.nasa.gov/ssc/data/analysis/}\\\url{documentation/Pass7REP_usage.html}} belonging to the P7REP$\_$SOURCE$\_$V15 class that are located in a circular region of interest (ROI) of 
10$^{\circ}$ radius around Mrk~421, after applying a cut of $< 52^{\circ}$ in the rocking angle, and $< 100^{\circ}$ on the zenith angle to reduce contamination from the $\gamma$-rays produced in the upper atmosphere and observed along Earth’s limb. The background model used to extract the $\gamma$-ray signal includes a Galactic diffuse-emission component and an isotropic component.  The model we adopted for the Galactic component is given by the file gll$\_$iem$\_$v05.fit, and the isotropic component, which is the sum of the extragalactic diffuse emission and the residual charged particle background, is parameterized by the file iso$\_$source$\_$v05.txt \footnote{ \url{http://fermi.gsfc.nasa.gov/ssc/data/access/lat/}
\\\url{BackgroundModels.html}}. The normalizations of the two components in the background model were allowed to vary freely during the spectral-point fitting. The spectral parameters were estimated using the unbinned maximum-likelihood technique \citep{1996ApJ...461..396M} in the energy range 300 MeV 
to 300 GeV. We used the P7REP$\_$SOURCE$\_$V15 instrument response function\footnote{See \url{http://fermi.gsfc.nasa.gov/ssc/data/analysis/}} and took into account all the sources from the second \FermiLAT catalog  \citep[2FGL,][]{Nolan2012} that are located within $15^{\circ}$ of Mrk~421. When performing the fit, the spectral parameters of sources within $10^{\circ}$ of Mrk~421 were allowed to vary, while those between 10$^{\circ}$ and 15$^{\circ}$ were fixed to their values from the 2FGL. When performing the likelihood fit in differential energy bins (spectral bins in the SED), the photon indices of the sources were frozen to the best-fit values obtained from the full spectral analysis.

The sensitivity of \FermiLAT is not good enough to detect Mrk~421
within a few hours, and hence we integrated over two days to
have significant detections and to be able to produce $\gamma$-ray
spectra. Despite the two-day integration window, the number of
collected photons above 300 MeV is only about 8 to 15 for each of the two-day intervals. Most of these photons have energies below a few GeV, since photons above 10 GeV are rarely detected from Mrk\,421 in a two-day interval. Upper limits at 95\% confidence level were calculated for the differential energy bins whenever the maximum-likelihood test statistic (TS)\footnote{The maximum-likelihood test statistic TS \citep{1996ApJ...461..396M} is defined as ${\rm TS} = 2 \Delta \log({\rm likelihood})$ between models with and without a point source at the position of Mrk~421.} was below 4, or when the detected signal had fewer than two events. The systematic uncertainty in the flux is dominated by the systematic uncertainty in the effective area, which is 
estimated as $10\%$ below $0.1$\,GeV, $5\%$ in the energy range between 0.3 GeV and 10 GeV and $10\%$ above $10$\,GeV\citep{Ackermann2012}. The systematic uncertainties are substantially smaller than the statistical uncertainties of the data points in the light curve and spectra.

%%%%%%%%%%%%%%%%%%%%%%%%%%%%%%%%%%%%%%%%%%%%%%%%%%%%%%%%%%%%
\subsection{X-ray observations}
%%%%%%%%%%%%%%%%%%%%%%%%%%%%%%%%%%%%%%%%%%%%%%%%%%%%%%%%%%%%

All 11 \Swiftc-XRT \citep{Burrows2005} observations were carried out
using the windowed timing (WT) readout mode. The data set was first
processed with the XRTDAS software package (v.2.9.3) developed at the
ASI Science Data Center (ASDC) and distributed by HEASARC within the
HEASoft package
(v. 6.15.1)\footnote{\url{http://heasarc.gsfc.nasa.gov/docs/software/lheasoft/}\\\url{download.html}}. Event
files were calibrated and cleaned with standard filtering criteria
with the {\it xrtpipeline} task using the calibration files available
in the \Swiftc-XRT CALDB version {\it 20140120}. Events for the spectral analysis were selected within a 20-pixel ($\sim$46 arcsec) radius, which encloses about 90\% of the PSF, centered on the source position. The background was extracted from a nearby circular region of 40 pixel radius. The ancillary response files were generated with the {\it xrtmkarf} task applying corrections for PSF losses and CCD defects using the cumulative exposure map. Before the spectral fitting, the 0.3-10 keV 
source energy spectra were binned to ensure a minimum of 20 counts per bin. The spectra were corrected for absorption with a neutral hydrogen column density fixed to the Galactic 21 cm value in the direction of Mrk~421, namely $1.9 \times 10^{20}$\,cm$^{-2}$ \citep{Kalberla2005}.

The {\it Rossi} X-ray Timing Explorer \citep[\RXTEc;][]{RXTERef}
satellite performed daily pointing observations of Mrk\,421 during the
time interval from MJD 55265 to MJD 55277. The data analysis was
performed using \texttt{FTOOLS} v6.9 and following the procedures and
filtering criteria recommended by the NASA \RXTE Guest Observer
Facility\footnote{\url{http://heasarc.gsfc.nasa.gov/docs/xte/xhp_proc_}\\\url{analysis.html}}. The
observations were filtered following the conservative procedures for
faint sources. Only the first xenon layer of PCU2 was used. We used the
package \texttt{pcabackest} to model the background and the package
\texttt{saextrct} to produce spectra for the source and background
files and the script\footnote{The CALDB files are located at
  \url{http://heasarc.gsfc.nasa.}\\\url{gov/FTP/caldb}}  \texttt{pcarsp} to
produce the response matrix. The PCA average spectra above 3 keV were
fitted using the XSPEC package using a power-law function with an
exponential cutoff (cutoffpl) with the same neutral hydrogen column density as was used in the {\it Swift}-XRT data analysis.
However, since the PCA bandpass starts at 3 keV, the results do not depend strongly on the column density adopted.

We also used data from the all-sky X-ray instruments available in 2010, namely \RXTEc/ASM, \textit{MAXI}, and \Swiftc/BAT. The data from \RXTEc/ASM were obtained from the ASM web page\footnote{See \url{http://xte.mit.edu/ASM_lc.html}} and were filtered according to the  prescription provided in the ASM web page. The daily fluxes from \Swiftc/BAT were gathered from the BAT web page\footnote{See \url{http://swift.gsfc.nasa.gov/docs/swift/results/}\\\url{transients/}} and the daily fluxes from \textit{MAXI} were retrieved from a dedicated \textit{MAXI} web page\footnote{See \url{http://maxi.riken.jp/top/index.php?cid=1&jname=}\\\url{J1104+382}}.

%%%%%%%%%%%%%%%%%%%%%%%%%%%%%%%%%%%%%%%%%%%%%%%%%%%%%%%%%%%%
\subsection{Optical and UV observations} 
\label{OptAndUV}
%%%%%%%%%%%%%%%%%%%%%%%%%%%%%%%%%%%%%%%%%%%%%%%%%%%%%%%%%%%%

The optical fluxes reported in this paper were obtained within the 
GLAST-AGILE Support Program (GASP) within the Whole Earth Blazar
  Telescope (WEBT) \citep[e.g.][]{Villata2008,2009A&A...504L...9V}, with various optical telescopes around the globe. Additionally, many observations were performed with the Perkins, Rovor, New Mexico Skies, and the Bradford telescopes. Optical polarization measurements are also included from the Steward, Crimean, and St Petersburg observatories. All the instruments use the calibration stars reported in \citet{Villata1998} for calibration, and the Galactic extinction was corrected with the reddening corrections given in \citet{schlegel98}. The flux from the host galaxy (which is significant only below $\nu \sim 10^{15}$~Hz) was estimated using the flux values across the R band from \cite{Nilsson2007} and the colors reported in \cite{Fukugita1995}, and then subtracted from the measured flux.

The \Swiftc~Ultraviolet/Optical Telescope \citep[UVOT;][]{Roming2005}
obtained data cycling through each of the three ultraviolet pass
bands, UVW1, UVM2, and UVW2 with central wavelengths of 260 nm, 220
nm, and 193 nm, respectively. The photometry was computed using a
$5$\,arcsec source region around Mrk\,421 using a custom UVOT pipeline
that performs the calibrations presented in
\citet{Poole2008}. Moreover, the custom pipeline also allows for
separate, observation-by-observation, corrections for astrometric
misalignments \citep{AcciariMrk4212011}.  The flux measurements
obtained were corrected for Galactic extinction with E$_{B-V}$=0.015
magnitude \citep[]{schlegel98} at each spectral band
\citep[]{Fitzpatrick99}. The contribution of the host galaxy to the UV fluxes is negligible and hence not considered.

%%%%%%%%%%%%%%%%%%%%%%%%%%%%%%%%%%%%%%%%%%%%%%%%%%%%%%%%%%%%
\subsection{Radio observations}
%%%%%%%%%%%%%%%%%%%%%%%%%%%%%%%%%%%%%%%%%%%%%%%%%%%%%%%%%%%%

The radio data reported in this manuscript were taken with the 14 m Mets\"ahovi Radio Observatory at 37 GHz, the 40 m Owens Valley Radio Observatory (OVRO) telescope at 15 GHz, and the 26 m University of Michigan Radio Astronomy Observatory (UMRAO) at 14.5 GHz. Details of the observing strategy and data reduction are given by \citet[Mets\"ahovi]{Terasranta1998}, \citet[OVRO]{Richards2011}, and \citet[UMRAO]{Aller1985}. The 225\,GHz (1.3 mm) light curve was obtained at the Submillimeter Array (SMA) near the summit of Mauna Kea (Hawaii).  During the period covered in this work, Mrk\,421 was observed as part of a dedicated program to follow sources on the \FermiLAT Monitored Source List (PI:A. Wehrle). Observations of available LAT sources were made periodically for several minutes, and the measured source signal strength was calibrated against known standards, typically solar system objects (Titan, Uranus, Neptune, or Callisto).

Mrk~421 is a point-like and unresolved source for these three single-dish radio instruments and for SMA, which means that the measured fluxes are the flux densities integrated over the full source extension, and hence should be considered as upper limits in the SED model fits reported in this paper. However, it is worth noting that the radio flux of Mrk~421 resolved with the VLBA for a region of 1--2$\times 10^{17}$cm (hence comparable to the size of the blazar emission) is a very large part of the radio flux measured with the single-dish radio instruments \citep[see][]{AbdoMrk421}, and thus it is reasonable to assume that the blazar emission contributes substantially to the radio flux measured by single-dish radio telescopes such as Mets\"ahovi, OVRO and UMRAO. Moreover, there are several works reporting a correlation between radio and GeV emission in blazars as a population \citep[see e.g.][]{2011ApJ...741...30A}, which implies that at least a fraction of the radio emission is connected to the $\
gamma$-ray (blazar) emission.  The 225\,GHz observations from SMA connect the bottom (radio) to the peak (optical/X-rays) of the synchrotron (low-energy) bump of the SED, and hence it is also expected to be strongly dominated by the blazar emission of the source. Therefore,  it seems reasonable to adjust the theoretical model in such a way that the predicted energy flux for the millimeter band is close to the SMA measurement, and the predicted energy flux for the radio band is not too far below the measurements performed by the single-dish instruments.

%%%%%%%%%%%%%%%%%%%%%%%%%%%%%%%%%%%%%%%%%%%%%%%%%%%%%%%%%%%%
%%%%%%%%%%%%%%%%%%%%%%%%%%%%%%%%%%%%%%%%%%%%%%%%%%%%%%%%%%%%
\section{Multiband variability} \label{LightCurves}
%%%%%%%%%%%%%%%%%%%%%%%%%%%%%%%%%%%%%%%%%%%%%%%%%%%%%%%%%%%%
%%%%%%%%%%%%%%%%%%%%%%%%%%%%%%%%%%%%%%%%%%%%%%%%%%%%%%%%%%%%

\begin{figure*}
\centering
\includegraphics[width=17cm]{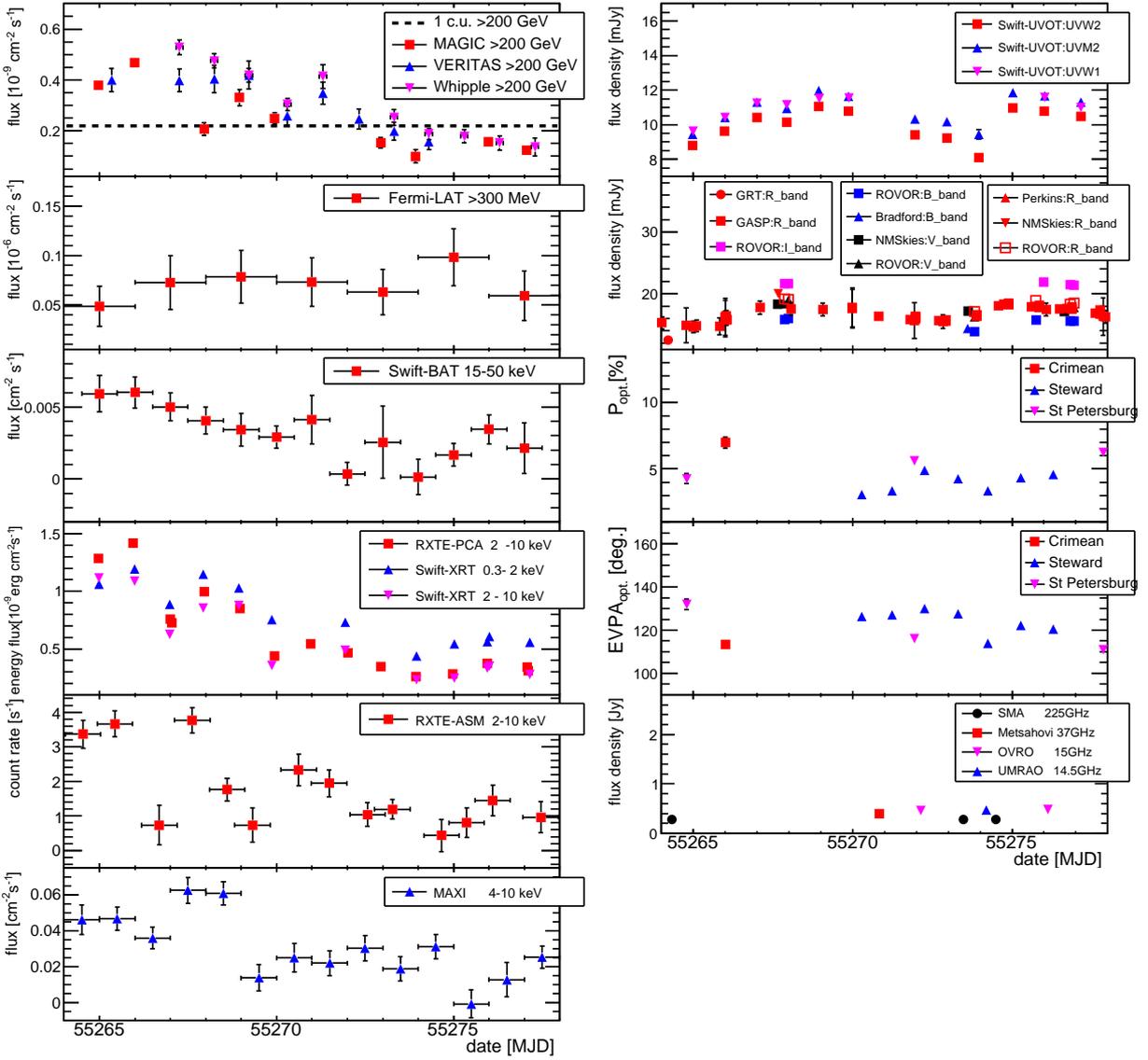}
\caption{Light curves of Mrk~421 between MJD 55264 and 55278, from VHE to radio (including optical polarization). The Whipple data were converted into fluxes above 200 GeV, and the host galaxy contribution was subtracted from the reported optical fluxes. $P_{\rm opt}$ and EVPA$_{\rm opt}$ stand for the polarization degree and the electric vector polarization angle. For details, see text in Sect.~\ref{LightCurves}.}
 \label{fig22}
\end{figure*}

In this section, we present the experimental results derived from the MW campaign observations described in Sect.~\ref{Observation}. Figure~\ref{fig22} shows the multiband light curves during the decline observed between 2010 March 10 (MJD~55265) and 2010 March 22 (MJD~55277).  In the top left panel, the VHE band includes nine observations from MAGIC, nine from VERITAS, and ten from Whipple.

%The Whipple photon fluxes, which were originally computed in relation to the Crab nebula flux above 400 GeV, were converted into photon fluxes above 200 GeV assuming that Mrk~421 and Crab nebula have the same spectral index around this energy range, and using that the Crab nebula flux above 200 GeV is $2.2\times 10^{-10}$cm$^{-2}$s$^{-1}$. 

%The VHE light curve shows a rough decrease in the activity vs. time. 
The flux above 200 GeV decreases roughly steadily with time. Before
MJD~55272 the fluxes are $\sim 1$ -- $2$ c.u., while on subsequent
days they are below 1 c.u., showing that only the decay (perhaps
including the peak) of the flare was observed with the VHE
$\gamma$-ray instruments in 2010 March. It is worth noting that the
VHE flux measured with MAGIC for MJD~55268 is roughly 50\% lower
than that measured with VERITAS  for that day:  $2.1 \pm 0.3$ vs. $4.0
\pm 0.6$ in units of $10^{-10}$ cm$^{-2}$ s$^{-1}$. Taking into
account the measured errors, these fluxes are different by three to four standard deviations. This might result (at least partially) from systematics related to the instruments or observations during that night, but it might also be due to intra-night variability over the MAGIC and VERITAS observation windows, which are about seven hours apart.

The photon flux above 300 MeV (measured by \FermiLAT in two-day long
time intervals) does not show any significant variability. A fit with
a constant line gives a flux level of $(6.8 \pm 0.9) \times 10^{-8}
\text{cm}^{-2} \text{s}^{-1}$, with $\chi^2/{\rm ndf} = 2.5/6$, which is similar to the mean flux of $\sim 7.2\times 10^{-8}
\text{cm}^{-2} \text{s}^{-1}$ observed during the first 1.5 years of
\Fermi\ operation, from 2008 August to 2010 March \citep{AbdoMrk421}.

The variability at the X-ray band as measured with \RXTEc,
\Swiftc~and \textit{MAXI}  is high, with light curves that resemble
those at VHE. The \Swiftc-XRT energy flux at the band 0.3-10 keV
 decreases from $\sim 2.2 \times 10^{-9} \text{erg} ~ \text{cm}^{-2}
 \text{s}^{-1}$ down to $\sim 0.8 \times 10^{-9} \text{erg} ~ \text{cm}^{-2}
 \text{s}^{-1}$. The low X-ray fluxes measured during this 13-day period are
 comparable to the mean 0.3-10 keV X-ray flux of $\sim 0.9 \times 10^{-9} \text{erg} ~\text{cm}^{-2}
 \text{s}^{-1}$, measured  during the first seven years of \Swift
 operation, from 2005 to 2012 \citep{Stroh2013}.

At UV and optical frequencies, the variability is also rather small,
in contrast to the VHE and X-ray bands. The emission at the UV and optical bands is variable. For instance, a constant fit yields  $\chi^2/{\rm ndf}$ of 174/11 and 144/60 for the UVOT-UVM2 and GASP/R band, respectively. 
Hence Mrk~421 showed some activity at these bands, although it is substantially weaker than that shown at VHE and X-rays.
The optical flux at the R band measured during this 13-day period
  is $\sim 16$ mJy ($\sim 24$ mJy if the host galaxy is included), which is
  comparable to the typical flux of $\sim 25$ mJy measured during the
  first eight years of the Tuorla blazar monitoring program, from 2003 to
  2011\footnote{\url{http://users.utu.fi/kani/1m/Mkn_421_jy.html}}.

Optical polarization measurements are also reported in
Fig.~\ref{fig22}.  The errors
on these observations are smaller than 0.1\% and 3$^{\circ}$
 for the polarization degree and the electric vector
polarization angle and are therefore too small to be visible in the
plot. The collected data do not show any flare in the polarization
degree or high rotation in the electric vector polarization angle as
is observed during the flaring activities in other blazars
\citep[e.g.][]{2008Natur.452..966M}. There are some small
variations in the polarization degree and angle, however, but such random fluctuations are common and expected
due to continuous noise processes and not by singular events \citep[see][]{2014ApJ...780...87M}.

In the radio bands, there were only seven observations during this
period, which were performed at frequencies from 14~GHz to
225~GHz. All of them reported a flux of about 0.4--0.5 Jy. We did not find
significant variability in any of these single-dish radio
observations, which are $\lapp$1 hour long. The radio
  fluxes measured during this 13-day period are comparable to the
  typical 15 GHz radio flux of $\sim$0.45 mJy 
  measured during the first three years of the OVRO monitoring
  program, from 2008 to 2011 \citep{Richards2013}.

\begin{figure}
%\resizebox{\hsize}{!}{\includegraphics{March_freq_variability_Instrument_whipple_MAGICd7275c70_betterMAXI.eps}}
\resizebox{\hsize}{!}{\includegraphics{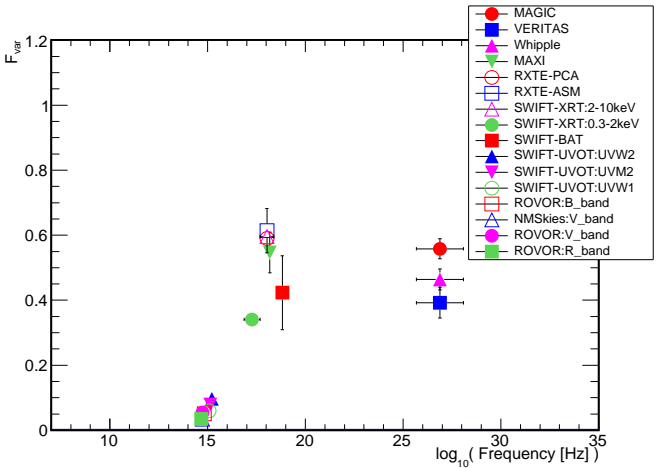}}
\caption{Fractional variability $F_{\rm var}$ as a function of frequency.}
\label{fig24}
\end{figure}
\begin{figure}
%\epsscale{0.75}
\resizebox{\hsize}{!}{\includegraphics{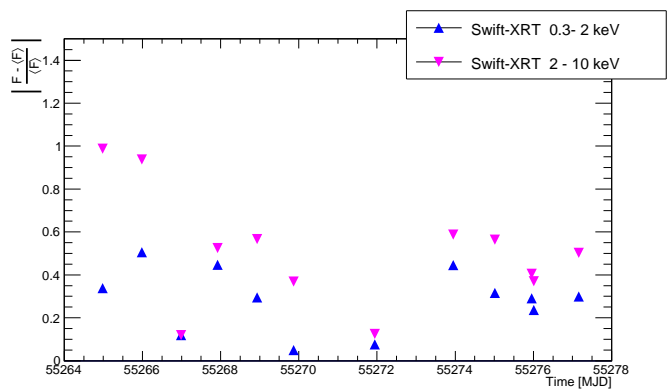}}
\caption{Temporal evolution of the absolute value of the normalized deviation of the \Swiftc-XRT flux, $F_{\rm dev}$. See text for further details.}
\label{fig24b}
\end{figure}
\begin{figure}
\resizebox{\hsize}{!} {\quad \quad \quad \includegraphics{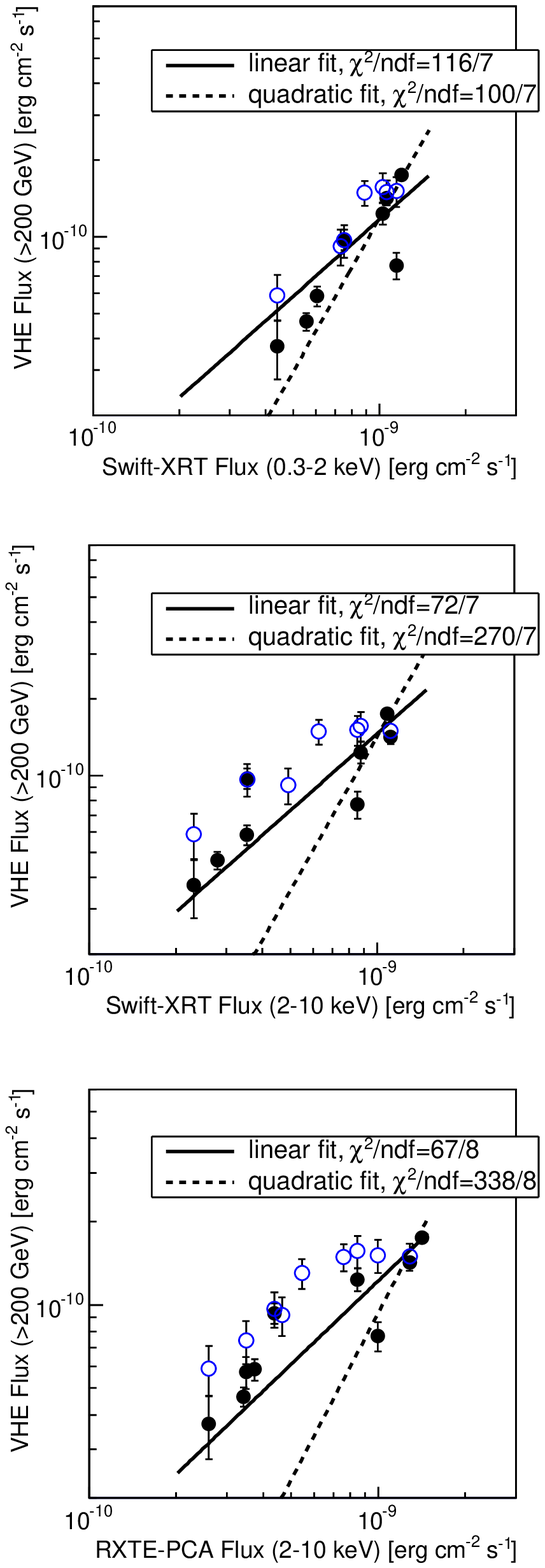} \quad \quad \quad}
\caption{Correlation between VHE $\gamma$-ray flux (MAGIC, black solid
  circles, and VERITAS, blue empty circles) and X-ray fluxes.
  \textbf{Top:} X-ray flux at the 0.3--2 keV band measured with
  \Swiftc-XRT. \textbf{Middle:} X-ray flux at the 2--10 keV band measured with \Swiftc-XRT. \textbf{Bottom:} X-ray flux at the 2--10 keV band measured with \RXTEc-PCA. The lines show the fits with linear and quadratic functions. Only MAGIC data points were used for the fits to ensure VHE-X-ray simultaneity (see Appendix~\ref{sec:simultaneity}).}
\label{fig24m}
\end{figure}

To quantify the overall variability during these 13 consecutive days, we followed the method provided in \cite{Vaughan2003}. The fractional variability $F_{\rm var}$ at each energy band is computed as 
\begin{equation}
F_{\rm var} = \sqrt{\frac{S^2 - \langle \sigma_{\rm err}^2 \rangle}{\langle F
\rangle^2}}
\end{equation}
where  $\langle F \rangle$ is the mean photon flux, $S$ is the standard deviation of the $N$ flux points, and $\langle \sigma_{\rm err}^2 \rangle$ is the mean-squared error. The error in $F_{\rm var}$ is calculated according to the prescription in Section~2.2 of \cite{Poutanen08},
\begin{equation}
\sigma_{F_{\rm var}}=\sqrt{F_{\rm var}^2+\sqrt{\frac{2\langle \sigma_{\rm err}^2 \rangle^2}{N\langle F\rangle^4}+\frac{4\langle \sigma_{\rm err}^2 \rangle F_{\rm var}^2}{N\langle F \rangle^2}}}-F_{\rm var}
\end{equation}
This prescription is more precise than the method used in \cite{Vaughan2003} when the $\sigma_{\rm err}$ is comparable to or larger than $S$.

The $F_{\rm var}$ values derived from the light curves in Fig.~\ref{fig22} are plotted in Fig.~\ref{fig24}. The values of $F_{\rm var}$ are plotted only for instruments with $S^2>\sigma_{\rm err}^2$. When there is no variability detectable with the sensivity of the instrument, $S^2<\sigma_{\rm err}^2$ might occur (as is the case for \FermiLATc).

The $F_{\rm var}$ is highest at the X-ray band. The values of $F_{\rm var}$ measured by \Swiftc-XRT and \RXTEc-PCA agree well at the 2--10 keV band. We note that \Swiftc-XRT shows a higher $F_{\rm var}$ at the 2--10 keV band than at the 0.3--2 keV band. The uncertainty in the $F_{\rm var}$ values at these two bands is small because the measured X-ray flux variations are very large in comparison to the flux uncertainties (which are smaller than 1\%), and that makes the difference in the measured variability very significant. This difference cannot be attributed to different temporal coverage, as they were observed with the same instrument (and hence the same time). 
%As these two energy ranges were observed with the same instrument (and hence at the same time), we cannot attribute the different $F_{\rm var}$ values to a different temporal coverage. 
%Moreover, the $F_{\rm var}$ measured with  \Swiftc-XRT  and \RXTEc-PCA agree well at the 2--10 keV band.

To study this difference, we calculated the normalized
deviations of the fluxes, $F_{\rm dev}=\left( F - \left< F \right>
\right) / \left< F \right>$ computed with the \Swiftc-XRT light curves
at both energy bands (0.3--2 keV and 2--10 keV). Figure~\ref{fig24b}
shows that the absolute values of $F_{\rm dev}$, $\left| F_{\rm dev}
\right|$, at the 2--10 keV band are always higher than those at the
0.3--2 keV band. This shows that the flux at the 2--10 keV band is
intrinsically more variable than at the 0.3--2 keV band across the
whole temporal range, and hence that the higher $F_{\rm var}$ is not
due to one or a few observations, but rather dominated by a higher
overall relative dispersion at the 2-10 keV flux values during the 13 consecutive days.

The $F_{\rm var}$ at VHE $\gamma$-rays is similar to that at
X-rays. The flux points from VERITAS and Whipple are more concentrated
around their mean values, which yield slightly lower $F_{\rm var}$ than that of MAGIC.  In conclusion, both VHE $\gamma$-rays
and X-rays show higher variability than the flux at the other bands,
which is additional evidence that they have a closer relation to
each other, as reported in several other Mrk~421 flaring episodes \citep[e.g.][]{1999ApJ...526L..81M}.

To better understand the relation between X-rays and VHE
$\gamma$-rays, we examined the correlation between the X-ray energy
flux at the 0.3--2 keV and 2--10 keV bands and the VHE $\gamma$-ray
energy flux above 200 GeV. For this exercise we used the X-ray fluxes
from \Swift and \RXTE and the VHE fluxes from MAGIC and VERITAS. The
VHE photon fluxes given in [cm$^{-2}$ s$^{-1}$] were converted into
energy fluxes reported in [erg cm$^{-2}$ s$^{-1}$] using a power-law
spectrum with index 2.5 above 200 GeV\footnote{The spectral shape of
  the VHE emission of Mrk~421
did vary during the 13-day period considered here. Including these
spectral variations would shift some of the reported $\gamma$-ray energy fluxes by
$\sim$10--15\%, which we considered not essential for this study.}. The top panel in Fig.~\ref{fig24m} shows the VHE $\gamma$-ray flux vs. X-ray flux at the 0.3--2 keV band, and the resulting fits with a linear ($F_{VHE}=k \cdot F_{X-ray}$) and a quadratic ($F_{VHE}=k \cdot F^{2}_{X-ray}$) function. For the fits we used only MAGIC data, which are the VHE observations taken simultaneously or almost simultaneously with the X-ray observations (see Appendix~\ref{sec:simultaneity} for details on simultaneity of the observations). The middle and bottom panels of Fig.~\ref{fig24m} also show the X-ray flux vs. VHE-$\gamma$-ray flux, but using the X-ray flux at the 2--10 keV band measured with \Swiftc~and \RXTEc. Neither a linear nor a quadratic function describes the data perfectly. However, for the 2--10 keV energy range, the VHE to X-ray flux closely follows a linear trend, which it is clearly not the case for the 0.3--2 keV energy range. The physical interpretation of these results is discussed in 
Sect.~\ref{Discussion}.

%%%%%%%%%%%%%%%%%%%%%%%%%%%%%%%%%%%%%%%%%%%%%%%%%%%%%%%%%%%%
%%%%%%%%%%%%%%%%%%%%%%%%%%%%%%%%%%%%%%%%%%%%%%%%%%%%%%%%%%%%
\section{Temporal evolution of the broadband spectral energy distribution} \label{SED_Model}
%%%%%%%%%%%%%%%%%%%%%%%%%%%%%%%%%%%%%%%%%%%%%%%%%%%%%%%%%%%%
%%%%%%%%%%%%%%%%%%%%%%%%%%%%%%%%%%%%%%%%%%%%%%%%%%%%%%%%%%%%

To study this flaring activity, we built 13 successive simultaneous broadband SEDs for 13 consecutive days. 
%% HT141119 %%
%In Sect.~\ref{SED}, we discuss characteristics of the MW data used to produce the SEDs, while in Sects.~\ref{Model1} and \ref{Model2} we study these SEDs within one-zone and two-zone SSC scenarios. 
We study these SEDs within one-zone and two-zone SSC scenarios in Sects.~\ref{Model1} and \ref{Model2}. The characteristics of the MW data are described in Appendix~\ref{sec:sed}. 
Specifically, we investigate whether the temporal evolution of the EED in SSC models can explain the observed variations in the SED during the 13-day period, and hence we try to fix (to their quiescent values) the  model parameters related to the environment, namely the blob radius ($R$), magnetic field ($B$), and Doppler factor ($\delta$). We cannot exclude that other model realizations with a different set of model parameters (e.g., changing the environment parameters, or varying more model parameters) can also provide a satisfactory description of the broadband SEDs, but in this paper we wish to vary as few model parameters as possible to most directly study the evolution of 
the EED, which is the part of the model directly connected to the particle acceleration and cooling mechanisms.

%%%
We applied steady-state SSC models instead of time-dependent models to
the SEDs of each day and estimated physical parameters in the emission
regions, which gives us an estimate of the temporal evolution of these
physical parameters. 
Time-dependent models, as developed 
by \citet[e.g.][]{2002MNRAS.336..721K,2011MNRAS.416.2368C}, are a direct way 
to derive the physical properties of the emission regions, 
but they include many detailed processes,
 such as synchrotron or inverse-Compton cooling of electrons, adiabatic cooling of electrons due to the expansion of an emission blob, 
and the injection of relativistic electrons and its time evolution,  
and therefore are very complex and have an arbitrarily large number of degrees of freedom. 
The snapshot approach with steady-state SSC models allows us to observe 
the time evolution of basic physical parameters averaged over a day in the blobs 
independently of the difficulty associated with time-dependent models. 
The time evolution of the averaged basic parameters observed in this study 
reflects physical mechanisms that are not considered explicitly, but gives us
hints about them. A caveat of this approach is that the SEDs are
observationally determined from short (about one hour) observations
distributed over a relatively long (13 day) period of time, and hence
we cannot exclude that some of the SEDs relate to short-lived active
states that do not necessarily fit in the scheme of a slowly varying activity phase.

%%% This section will be removed (HT141119) %%%
%%%%%%%%%%%%%%%%%%%%%%%%%%%%%%%%%%%%%%%%%%%%%%%%%%%%%%%%%%%%
%\subsection{Characteristics of the Measured Broadband SEDs} \label{SED}
%%%%%%%%%%%%%%%%%%%%%%%%%%%%%%%%%%%%%%%%%%%%%%%%%%%%%%%%%%%%

Given the known multiband variability in the emission of Mrk~421 (and
blazars in general), we paid special attention to organize
observations that were as close in time as possible 
(see Appendix~\ref{sec:simultaneity} for the simultaneity 
of the observations). 
%% HT141119 %% The simultaneity of the observations is depicted in Figs.~\ref{figSim1} and \ref{figSim2}. 
The observations performed with MAGIC, \RXTEc, and
\Swift were scheduled many 
weeks in advance, which resulted in actual
observations occurring always within temporal windows of less
  than two hours. The observations with VERITAS/Whipple were triggered by the high activity detected in 2010 March, and performed typically about seven hours after MAGIC observations because VERITAS and Whipple are located at a different longitude from that of MAGIC. At radio frequencies we have only seven observations during this period, but we neither expected nor detected variability at radio during these short (a few days) timescales. 
Based upon these observations, we show in Appendix~\ref{sec:sed} 13 
consecutive days of SEDs.  Each SED is characterized with a one-zone
and a two-zone SSC model as described in the following two subsections.

The peak luminosities and peak frequencies of the low- and high-energy bumps shift during high activity. In general, the peak frequency and peak luminosity decrease as the flare decays. In addition to the migration in the SED peak positions, the shapes of these SED bumps change. The X-ray and $\gamma$-ray bumps of the SEDs from MJD~55265 and 55266, when Mrk~421 emitted the highest flux, are narrow, and they widen as the flare decays. A quantitative evaluation of the widening of the two SED bumps is reported in Sect.~\ref{Discussion}.

%It is worth noting that the SEDs during the last several nights are very similar to the averaged SED from 2009 reported in \cite{AbdoMrk421}. 
%It is worth noting that the measured SEDs for the last nights of the 13-day periods, the ones reporting the nonflaring (low) activity, are very similar to the averaged SED from 2009 reported in \cite{AbdoMrk421}. 
%Consequently, we decided to use the SED and SSC modeling results from \cite{AbdoMrk421} as a reference for many of the studies/results reported in this paper. 

%%%%%%%%%%%%%%%%%%%%%%%%%%%%%%%%%%%%%%%%%%%%%%%%%%%%%%%%%%%%
\subsection{SED modeling: One-zone SSC model} \label{Model1}
%%%%%%%%%%%%%%%%%%%%%%%%%%%%%%%%%%%%%%%%%%%%%%%%%%%%%%%%%%%%
 
In this SSC model, we assume that emission comes from a single, spherical and homogeneous region in the jet, which is moving relativistically toward us. The one-zone SSC model describes most of the SEDs of high-frequency-peaked BL~Lac objects with the fewest parameters, and hence it is the most widely adopted. 
The emission from radio to X-ray results from synchrotron radiation of
electrons inside a blob of comoving radius $R$, with a Doppler factor
$\delta$. In this emission blob, there is a randomly oriented magnetic
field with uniform strength $B$. The $\gamma$-ray emission is
produced by inverse-Compton scattering of the synchrotron photons with
the same population of electrons that produce them. We
used the numerical code of the SSC model described in
\citet{Takami11}. The algorithm implemented in this code allows us to
very quickly determine the parameters that accurately describe the SED.

%In the SSC model, the emission from radio to X-ray results from synchrotron radiation of electrons inside a blob of comoving radius $R$, with Doppler factor $\delta$. In this emission blob, there is a randomly-oriented magnetic field with uniform strength $B$. The emission of $\gamma$-rays is produced by inverse-Compton scattering of the synchrotron photons with the same population of electrons which produce them. In this work, we use the numerical code of the SSC model described in \citet{Takami11}. 

\begin{table*}
\caption{Integral flux above 200 GeV and parameters of the one-zone SSC model.}\label{OneBlobSSC}
\centering
\begin{tabular}{lllllllll} 
\hline\hline            
  Date            &  MAGIC flux                   &  VERITAS flux                  &  Whipple flux                &  $\gamma_{\rm br1}$&  $\gamma_{\rm br2}$  &  $s_1$ &  $s_2$   &  $n_{\rm e}$             \\
  $[\rm MJD]$     & [$10^{-10}$cm$^{-2}$s$^{-1}$] &  [$10^{-10}$cm$^{-2}$s$^{-1}$] & [$10^{-10}$cm$^{-2}$s$^{-1}$]    &  [$10^{4}$]      &  [$10^{5}$]        &        &            &  [$10^{3} \rm cm^{-3}$]       \\
\hline
55265 & $3.8\pm 0.2$ & $4.0\pm 0.5          $    &  $                  $  & 60. & 6.0 & 2.23 & 2.23  &  1.14   \\
55266 & $4.7\pm 0.2$ & $                    $    &  $                  $  & 66. & 6.6 & 2.23 & 2.23  &  1.16   \\
55267 & $          $ & $4.0\pm 0.5          $    &  $5.3\pm 0.3        $  & 16. & 6.0 & 2.23 & 2.70  &  1.10   \\
55268 & $2.1\pm 0.3$ & $4.0\pm 0.6          $    &  $4.8\pm 0.3        $  & 16. & 6.0 & 2.20 & 2.70  &  0.90   \\
55269 & $3.3\pm 0.3$ & $4.2\pm 0.6          $    &  $4.2\pm 0.3        $  & 12. & 7.0 & 2.20 & 2.70  &  0.95   \\
55270 & $2.3\pm 0.2$ & $2.6\pm 0.4          $    &  $3.0\pm 0.2        $  & 8.0 & 3.9 & 2.20 & 2.70  &  0.90   \\
55271 & $          $ & $3.5\pm 0.4          $    &  $4.1\pm 0.5        $  & 9.0 & 5.0 & 2.20 & 2.70  &  0.90   \\
55272 & $          $ & $2.5\pm 0.4          $    &  $                  $  & 5.0 & 4.0 & 2.20 & 2.50  &  0.90   \\
55273 & $1.5\pm 0.2$ & $2.0\pm 0.4          $    &  $2.5\pm 0.3        $  & 6.0 & 3.9 & 2.20 & 2.70  &  0.90   \\
55274 & $1.0\pm 0.3$ & $1.6\pm 0.3          $    &  $1.9\pm 0.2        $  & 3.5 & 3.9 & 2.20 & 2.70  &  0.90   \\
55275 & $          $ & $                    $    &  $1.8\pm 0.3        $  & 5.0 & 3.9 & 2.20 & 2.70  &  0.85   \\
55276 & $1.6\pm 0.2$ & $                    $    &  $1.5\pm 0.3        $  & 5.7 & 3.9 & 2.20 & 2.70  &  0.90   \\
55277 & $1.2\pm 0.1$ & $                    $    &  $1.4\pm 0.4        $  & 8.0 & 3.9 & 2.20 & 2.70  &  0.70   \\
\hline
\end{tabular}
\tablefoot{VERITAS and Whipple fluxes were measured around seven hours
  after the MAGIC observations. The model parameters that were kept
  constant during the 13-day period are the following ones:  $\gamma_{\rm min}=8 \times 10^{2}$; $\gamma_{\rm max}=1\times 10^{8}$; $s_3=4.70$; $B= 38$~mG; $\log(R{\rm [cm]})=16.72$; $\delta=21$.}
\end{table*}

The one-zone homogeneous SSC scenario with an EED described with a broken power-law function (seven free parameters plus the two parameters defining the edges of the EED) can be formally constrained from the seven characteristic observables that can be obtained from the multi-instrument data covering the two SED bumps, namely the spectral indices below and above the synchrotron peak, the peak frequencies and luminosities of the synchrotron and inverse-Compton bumps, and the variability timescale \citep{1998ApJ...509..608T}. However, in reality, the collected data do not allow us to determine these seven parameters with very good precision (particularly for the variability timescale and the peak frequency of the inverse-Compton bump), which implies some degeneracy in the seven (+two) model parameters, which unavoidably necessitates making some approximations or assumptions.

In previous works related to Mrk~421, it was common to use only one or
two power-law functions (that is, zero or one break) to describe
  the EED. However, such a simple model cannot adequately describe
the broadband SED from the campaign organized in 2009, when Mrk~421
was in its typical nonflaring VHE state \citep{AbdoMrk421}. The SED
from this paper was better sampled (more instruments with higher
sensitivity) than those reported previously, and an additional
break (two additional parameters) was required to properly describe
the shape of the measured synchrotron bump (from 1 eV to 100 keV),
together with the full inverse-Compton bump (from 100 MeV to 10
TeV). Given the similar energy coverage and activity of the source
during many days of the 13-day period considered here, we also used three power-law functions (i.e., two breaks) to parameterize the EED: 
\begin{equation}
\label{e_eq1}
  \frac{dn_{\rm e}}{d\gamma_{\rm e}} = \left\{\begin{array}{ll}
    n_{\rm e}\gamma_{\rm e}^{-s_1} & \quad \textrm{ if }
\gamma_{\rm min}<\gamma_{\rm e}\le\gamma_{\rm br1}\\
    n_{\rm e}\gamma_{\rm e}^{-s_2}\gamma_{\rm br1}^{s_2-s_1}  & \quad \textrm{ if } 
\gamma_{\rm br1}<\gamma_{\rm e}\le\gamma_{\rm br2}\\
    n_{\rm e}\gamma_{\rm e}^{-s_3}e^{-\gamma_{\rm e}/\gamma_{\rm max}}\gamma_{
\rm br1}^{s_2-s_1}\gamma_{\rm br2}^{s_3-s_2}e^{\gamma_{\rm br2}/\gamma_{\rm max}}  &
\quad \textrm{ if } \gamma_{\rm e}>\gamma_{\rm br2}.\\
  \end{array} \right.
\end{equation}
where $n_{\rm e}$ is the number density of electrons, $\gamma_{\rm e}$
is the Lorentz factor of the electrons, $\gamma_{\rm min}$ and
$\gamma_{\rm max}$ define the range of $\gamma_{\rm e}$, $s_1$, $s_2$
and $s_3$ are the indices of the power-law functions, and $\gamma_{\rm
  br1}$ and $\gamma_{\rm br2}$ are the Lorentz factors where the
power-law indices change.  In total, this model has two more free
parameters than the model with a broken power-law EED. The SEDs
from the days with highest activity can be described with an EED
with only one break, but for the nonflaring activity, we need to use an EED with two breaks. The requirement for a more complex parameterization of the EED in the recent works might be due to the better energy coverage (more instruments involved in the campaigns), and better sensitivity to cover the $\gamma$-ray bump. Future observations of Mrk~421 during nonflaring states with as good or better energy coverage will reveal 
whether the two-break EED is always needed, or whether this is something that was required only to describe the 2009 and 2010 data.

\begin{figure*}
\centering
\begin{subfigure}{.57\textwidth}
  \centering
  \includegraphics[width=1.\linewidth]{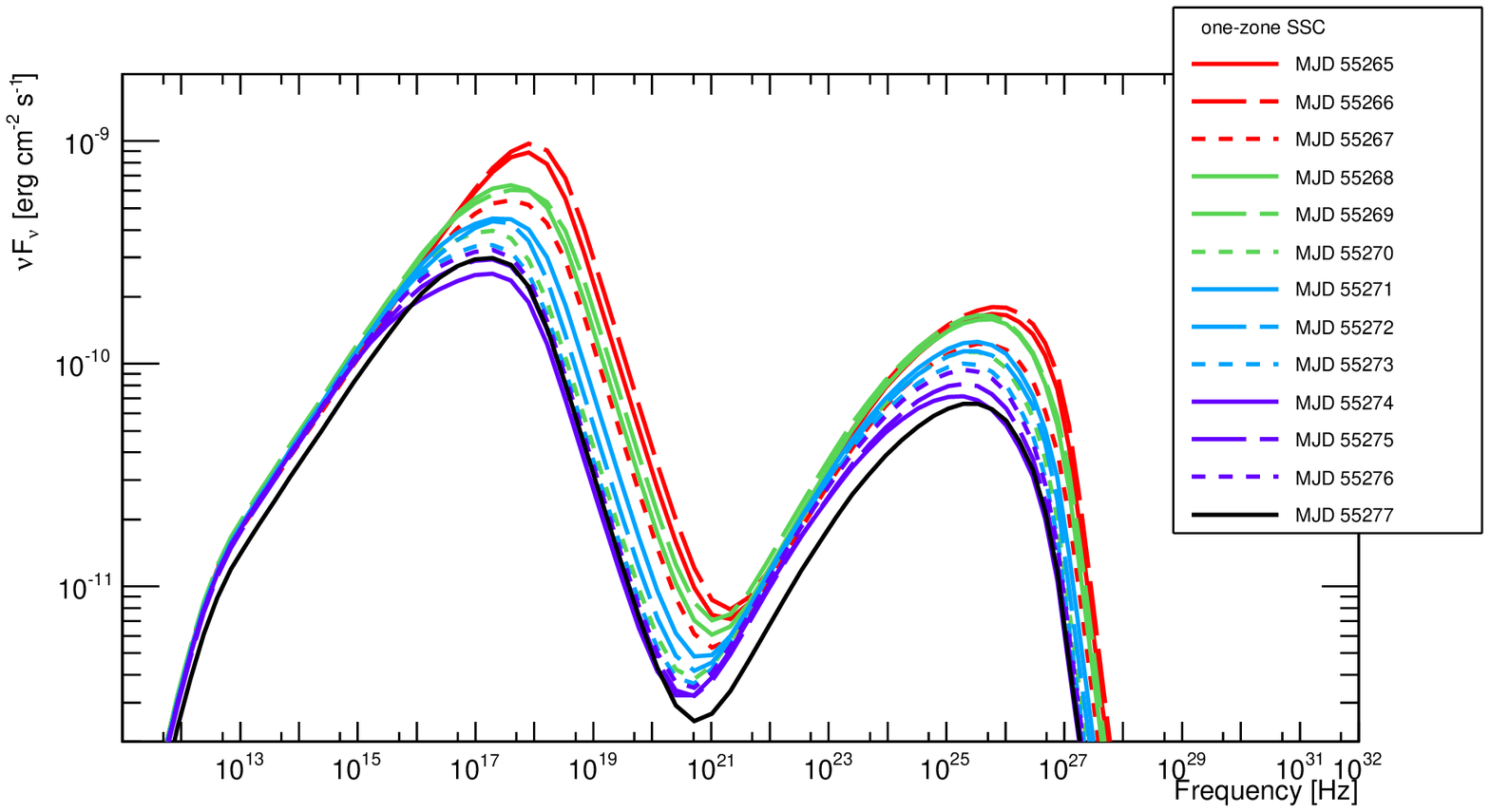}
  \caption{SEDs.}
  \label{fig:models1}
\end{subfigure}
\begin{subfigure}{.42\textwidth}
  \centering
  \includegraphics[width=1.\linewidth]{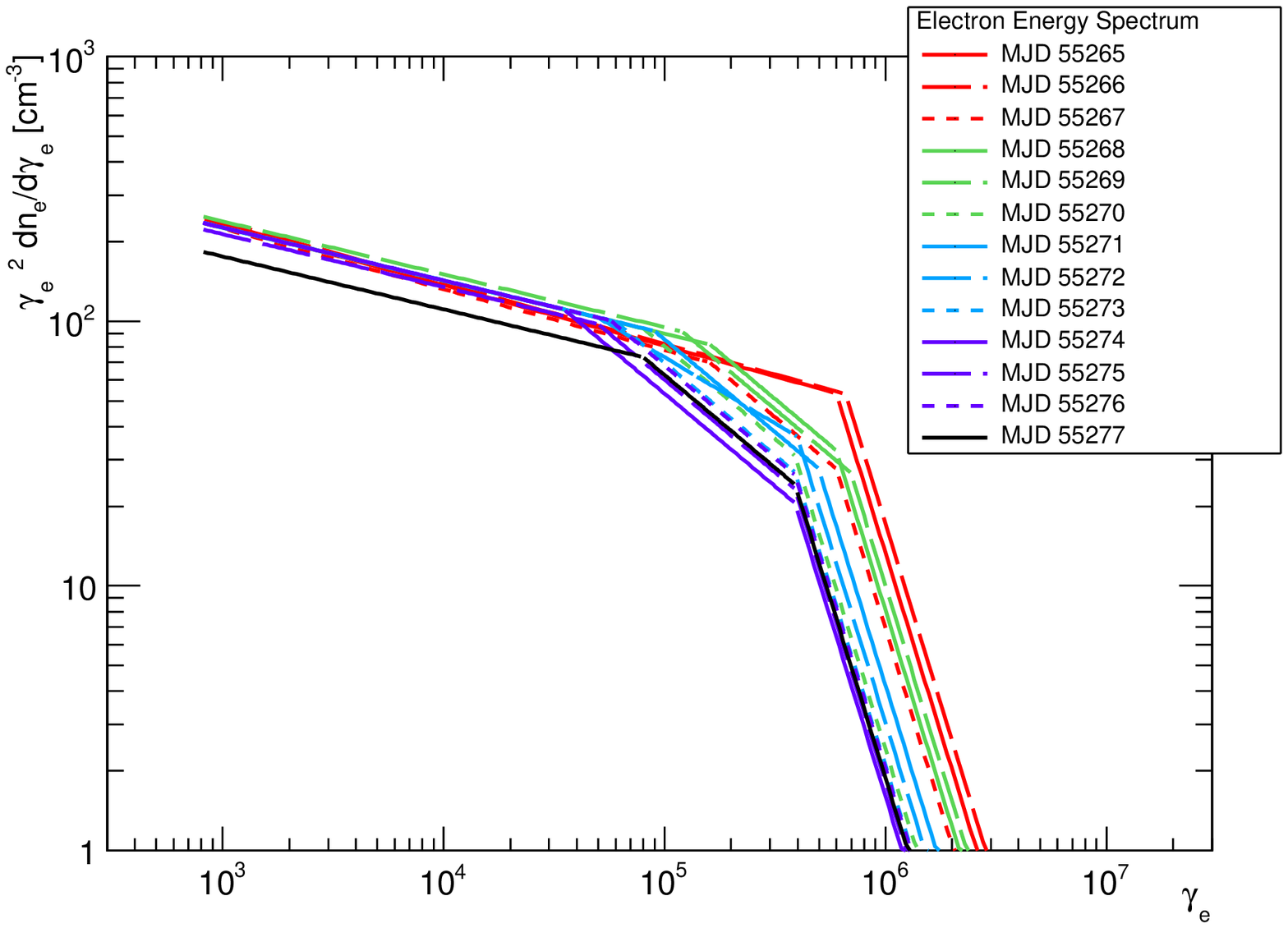}
  \caption{EEDs.}
  \label{fig25}
\end{subfigure}
\caption{One-zone SSC model curves and the related EEDs used to describe the measured SEDs during the 13-day flaring activity. The parameter values are given in Table~\ref{OneBlobSSC}.}
\label{fig:SSCmodel1}
\end{figure*}

Despite the extensive MW data collected in this campaign, there is
still some degeneracy in the choice of the eleven parameter values
required to adjust the SED model to the observational data. Given the
similarities 
between the SEDs of the last few days 
and the SED reported in \cite{AbdoMrk421}, 
%% HT141119 %%
%(e.g. seeFigs.~\ref{fig20} and \ref{fig21}), 
we used the SED model parameter
values from \cite{AbdoMrk421} as a reference for the choice of SSC
parameters to describe the 2010 March broadband observations. In
particular, we wish to test whether the temporal evolution of the EED
can explain the observed variations in the SED during the 13-day
period, and hence we fixed $\gamma_{\rm min}$, $\gamma_{\rm max}$ and the model parameters related to the environment $R$, $B$, and $\delta$ to the values reported in \cite{AbdoMrk421}. The value of the Doppler factor, 21, is higher than the value inferred from VLBA measurements of the blob movement in \cite{Piner10}. This is a common circumstance for VHE sources, which has been dubbed the 
``bulk Lorentz factor crisis'', and requires the radio and TeV emission to be produced in regions with different Lorentz factors \citep{2003ApJ...594L..27G,2005A&A...432..401G,2006ApJ...640..185H}. 
During the adjustment of the model to the measured SED, the VHE and
X-ray data provide the primary constraint because the variability
  is highest in these two energy bands.

The model parameters inferred from the observed SEDs (shown in Appendix~\ref{sec:sed}) 
are reported in Table~\ref{OneBlobSSC}. 
%The broadband SEDs and the model results for each day are shown in Figs.~\ref{fig11} to \ref{fig:SEDs1zone:2}. The resulting model parameters are reported in Table~\ref{OneBlobSSC}. 
Only one break in the EED (instead of two) is sufficient to describe
  the narrow SED bumps on MJD~55265 and 55266, while two breaks are
  necessary to properly describe the wider X-ray and $\gamma$-ray
  bumps from MJD~55267 to MJD~55277, when Mrk~421 shows a somewhat
  lower X-ray and VHE activity. The changes in the SED during the flaring activity are dominated
  by the parameters, 
$n_{\rm e}$, $\gamma_{\rm br1}$, and $\gamma_{\rm br2}$: lower
activity can be parameterized with 
a lower $n_{\rm e}$ and a decrease in the values of the 
%two breaks 
two break Lorentz factors in the EED.
The spectral index $s_2$ is equal to 2.5 for MJD~55272, 
while $s_2 = 2.7$ for the adjacent days. 
The X-ray bump for MJD~55272 (see Fig.~\ref{fig16}) is rather narrow, 
and therefore $s_2$, which affects the SED slope of the lower energy side of the bump, 
needs to be closer to $s_1$ to properly describe the data.
%For MJD~55272, $s_2 = 2.5$, while for the adjacent dates $s_2 = 2.7$. The X-ray bump in Fig.~\ref{fig16} is rather narrow, and $s_2$, which affects the slope of the left side of the bump, needs to be closer to $s_1$ to properly describe the data.}

Given the values of the blob radius and Doppler factor used here, the shortest time of
the flux variation $t_{\rm min} = (1 + z) R / \beta ~c \delta$ is
about one day. This value is reasonable, given the flux variations
measured during the March flaring activity (see Fig.~\ref{fig22}),
but it would not be consistent with the potential intra-night
variability that might have occurred in MJD~55268, as hinted by the
disagreement in the VHE fluxes measured by MAGIC and VERITAS. The
predicted radiative cooling break by synchrotron radiation\footnote{In
  HBLs like Mrk~421, the cooling of the electrons is expected to be
  dominated by the synchrotron emission.}, $\gamma_{\rm c} = 6\pi
m_{\rm e} c^2 / (\sigma_T B^2 R)$, where $m_{\rm e}$ is the electron
mass and $\sigma_T$ is the Thomson cross-section, is $3.2 \times 10^5$
in this model. This formula is derived by equating the timescale of
synchrotron radiation to the timescale of electrons staying in the
blob $\sim$ R / c, on the assumption that the timescale of adiabatic
cooling is much longer than that of synchrotron cooling. This
assumption is reasonable because R is fixed in this study.
The $\gamma_{\rm  br2}$ values in the model range from $3.9\times 10^5$ to $7.0 \times 10^5$, which is comparable to $\gamma_{\rm c}$, hence suggesting that the second break in the EED might be related to the synchrotron cooling break.  
Thus, the decrease of $\gamma_{\rm br2}$ and the weak dependence on $n_e$ implies 
 that the end of a flare is dominated by cooling.
%% HT141119 %%
%This suggests that a flare finishes by cooling. 
However, the change in the power-law index does not match the canonical change expected from synchrotron cooling, $\Delta s = 1$, which is similar to the situation reported in \cite{AbdoMrk421}. The result that $s_3$ is softer than expected can be explained by inhomogeneity of the emission blob, or by a weakening of the electron injection.

In general, the agreement between the one-zone SSC model and the
observational data is quite acceptable,  
%% HT141119 %%
%in Figs.~\ref{fig11} to\ref{fig21}, 
which shows one more the success of the one-zone SSC
model in describing the SEDs of blazars. However, there are several
problems. At the low-energy end of
the VHE spectra, the model is slightly higher than the data for the SEDs
from MJD~55265, 55266, 55268, 55269, and 55273; and 
the model also goes slightly beyond the data in the X-ray bump for MJD~55265
and 55266.  We cannot exclude that these data-model mismatches arise
from the requirement that the EED is the only mechanism responsible
for the blazar variability. For instance, if in addition to changing the model parameters related to the EED, the parameters $B$, $R$ and $\delta$ were varied as well, the relative position of the synchrotron and SSC peak could be modified, possibly achieving better agreement with the data.

%      
% \begin{figure*}
% \includegraphics[width=17cm]{all_models_AA_1blob.eps}
% \caption{One-zone SSC model curves used to describe the measured
%   SEDs during the 13-day flaring activity. The parameter values are given in Table~\ref{OneBlobSSC}.}
% \label{fig:models1}
% \end{figure*} 
% 
% 
% 
% \begin{figure*}
% \sidecaption
% %\includegraphics[width=12cm]{texV_EED_1blob.eps}
% \includegraphics[width=12cm]{EED_1blob.eps}
% \caption{The electron energy distributions used to
%   model the measured SEDs with the one-zone
% SSC scenario. The parameter values are given in Table~\ref{OneBlobSSC}.}
% \label{fig25}
% \end{figure*}

Overall, the temporal evolution of the broadband SEDs can be described by changes in the EED, keeping $\gamma_{\rm min}$, $\gamma_{\rm max}$, and the  model parameters related to the environment (blob radius, magnetic field, and the Doppler factor) constant at the values reported in \cite{AbdoMrk421}. Figures~\ref{fig:models1} and \ref{fig25} depict the one-zone SSC model curves and the parameterized EEDs for the 13 consecutive days. We can divide the whole activity into three periods: MJD~55265-55266 (period 1), MJD~55268-55271 (period 2), and MJD~55272-55277 (period 3), which correspond to a VHE flux of $\sim2$ c.u., $\sim1.5$ c.u., and $\sim0.5$ c.u., respectively. The EEDs of period 1 have one break, while those of period 2 and 3 have two breaks. Moreover, the EEDs of period 1 have a higher electron number density ($n_{\rm e}$) than those of periods 2 and 3. Figure~\ref{fig25} 
shows that the greatest variability occurs above the first break ($\gamma_{\rm br1}$) in the EED.

%%%%%%%%%%%%%%%%%%%%%%%%%%%%%%%%%%%%%%%%%%%%%%%%%%%%%%%%%%%%
\subsection{SED modeling: Two-zone SSC model} \label{Model2}
%%%%%%%%%%%%%%%%%%%%%%%%%%%%%%%%%%%%%%%%%%%%%%%%%%%%%%%%%%%% 

The one-zone SSC model curves reported in the previous section
describe the overall temporal evolution of the low-
and high-energy bumps of the SED during this flaring
activity reasonably well. However, we cannot ignore the model-data mismatches
mentioned in the last section. This was our main motivation for trying
a model with two distinct blobs: one producing the steady emission,
the other producing the temporal evolution of the SED, which is
evident primarily at the X-ray and VHE $\gamma$-ray bands. The two
blobs are assumed to be separated by a long distance and the
individual radiation fields do not interact with each other. We call
these the quiescent blob and the flaring blob. The quiescent blob is described with the parameter values from
the one-zone SSC model reported in Table~\ref{OneBlobSSC} for MJD~55274, which is the SED with the lowest activity
among the 13 consecutive days.
While the EED of the quiescent-state blob is described by three power-law functions, we employ only  two power-law 
functions to describe the EED of the flaring blob:
\begin{equation}
\label{e_eq2}
  \frac{dn_{\rm e}}{d\gamma_{\rm e}} = \left\{\begin{array}{ll}
    n_{\rm e}\gamma_{\rm e}^{-s_1} & \quad \textrm{ if }
\gamma_{\rm min}<\gamma_{\rm e}\le\gamma_{\rm br1}\\
    n_{\rm e}\gamma_{\rm e}^{-s_2}\gamma_{\rm br1}^{s_2-s_1} & \quad \textrm{ if } 
\gamma_{\rm br1}<\gamma_{\rm e}<\gamma_{\rm max}\\
  \end{array} \right.
\end{equation}
where $n_{\rm e}$ is the electron number density, $\gamma_{\rm e}$ is the Lorentz factor of the electrons, $\gamma_{\rm min}$ and $\gamma_{\rm max}$ define the range of $\gamma_{\rm e}$, $s_1$ and $s_2$ are the indices of the power-law function, and $\gamma_{\rm br1}$ is the  Lorentz factor where the power-law index changes.

\begin{figure*}
\sidecaption
\includegraphics[width=12cm]{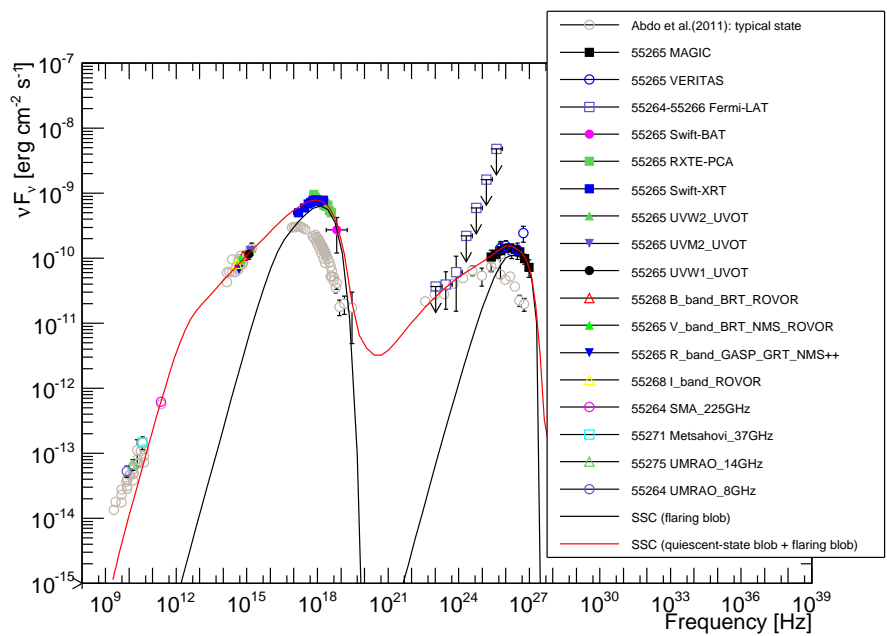}
\caption{Largely simultaneous broadband SED of Mrk~421 on MJD 55265. The correspondence between markers and instruments is given in the legend. The full names of the instruments can be found in Table~\ref{TableWithInstruments}. Because of space limitations, 
R-band instruments other than GASP, GRT, and NMS are denoted with the 
symbol "++".   Whenever a simultaneous observation is not available, the fluxes from the closest date are reported, and their observation time in MJD is reported next to the instrument name in the legend.  The red curve depicts the two-zone SSC model matching the SED data, while the black line shows the contribution of the flaring blob. The gray circles depict the averaged SED from the 2009 MW campaign reported in \cite{AbdoMrk421}, which is a good representation of the nonflaring (typical) SED of Mrk~421. \label{fig11b}}
\end{figure*}

In the overall process of adjusting the model to the 13 measured SEDs,
we used a flaring blob size $R$ about one order of
magnitude smaller than the quiescent blob, which naturally allows
faster variability. The size of the blob was kept constant, while
the other parameters were allowed to change to describe the
characteristics of the flare evolution.
%In particular, since the changes in the SED mostly occur at the X-ray and VHE range, the peak frequencies of the two SED bumps produced by the flaring blob should be higher than those produced by the quiescent-state blob. 

\begin{table*}
\caption{Parameters for the flaring blob in the two-zone SSC model.}\label{TwoBlobSSC}
\centering
\begin{tabular}{lllll} 
\hline\hline          
 Date          &  $\gamma_{\rm min}$  &  $\gamma_{\rm br1}$ & $n_{\rm e}$        & $B$         \\
  $[\rm MJD]$  &  [$10^{4}$]          &  [$10^{5}$]         & [$10^{3}\rm cm^{-3}$]& [$\rm mG$]\\
\hline
55265  &   3.0  & 3.0    &  5.0  &   105 \\
55266  &   3.0  & 3.0    &  6.0  &   100 \\
55267  &   2.5  & 1.1    &  5.9  &   100 \\
55268  &   5.3  & 1.8    &  5.6  &   100 \\
55269  &   3.0  & 2.3    &  5.2  &    90 \\
55270  &   3.5  & 0.8    &  6.0  &    75 \\
55271  &   3.5  & 1.2    &  6.5  &    75 \\
55272  &   3.5  & 2.0    &  3.0  &    75 \\
55273  &   3.5  & 0.5    &  4.0  &    75 \\
55274  &   - -  & - -    &   - - &    - -\\
55275  &   3.5  & 0.5    &  5.0  &    60 \\
55276  &   3.5  & 1.0    &  3.0  &    60 \\
55277  &   3.5  & 0.8    &  2.5  &    60 \\
\hline
\end{tabular}
\tablefoot{ 
 The model parameters that were kept
  constant during the 13-day period are the following ones:
  $\gamma_{\rm max}=6\times 10^{5}$; $s_1=2.0$; $s_2=3.0$; $\log(R{\rm
    [cm]})=15.51$; $\delta=35$.  The quiescent blob is parameterized with the parameter values from
the one-zone SSC model reported in Table~\ref{OneBlobSSC} for
MJD~55274.  We refer to Table~\ref{OneBlobSSC} for the $\gamma$-ray flux
above 200 GeV measured with MAGIC, VERITAS and Whipple.
}
\end{table*}

% \begin{figure*}
% \includegraphics[width=17cm]{all_models_AA_2blob_2.eps}
% \caption{All the two-zone SSC models used to describe the measured SEDs in the flaring activity. The parameter values are given in Table~\ref{TwoBlobSSC}.}
% \label{fig:models2}
% \end{figure*}
% 
% \begin{figure*}
% \sidecaption
% %\includegraphics[width=12cm]{deabsorb_EED_2blob.eps}
% %\includegraphics[width=12cm]{EED_2blob.eps}
% \includegraphics[width=12cm]{EED_2blob_b.eps}
% \caption{The electron energy distributions of the flaring blob used to
%   model the measured SEDs with the two-zone
% SSC scenario. The parameter values are given in Table~\ref{TwoBlobSSC}.}
% \label{fig25b}
% \end{figure*} 

\begin{figure*}
\centering
\begin{subfigure}{.57\textwidth}
  \centering
  \includegraphics[width=1.\linewidth]{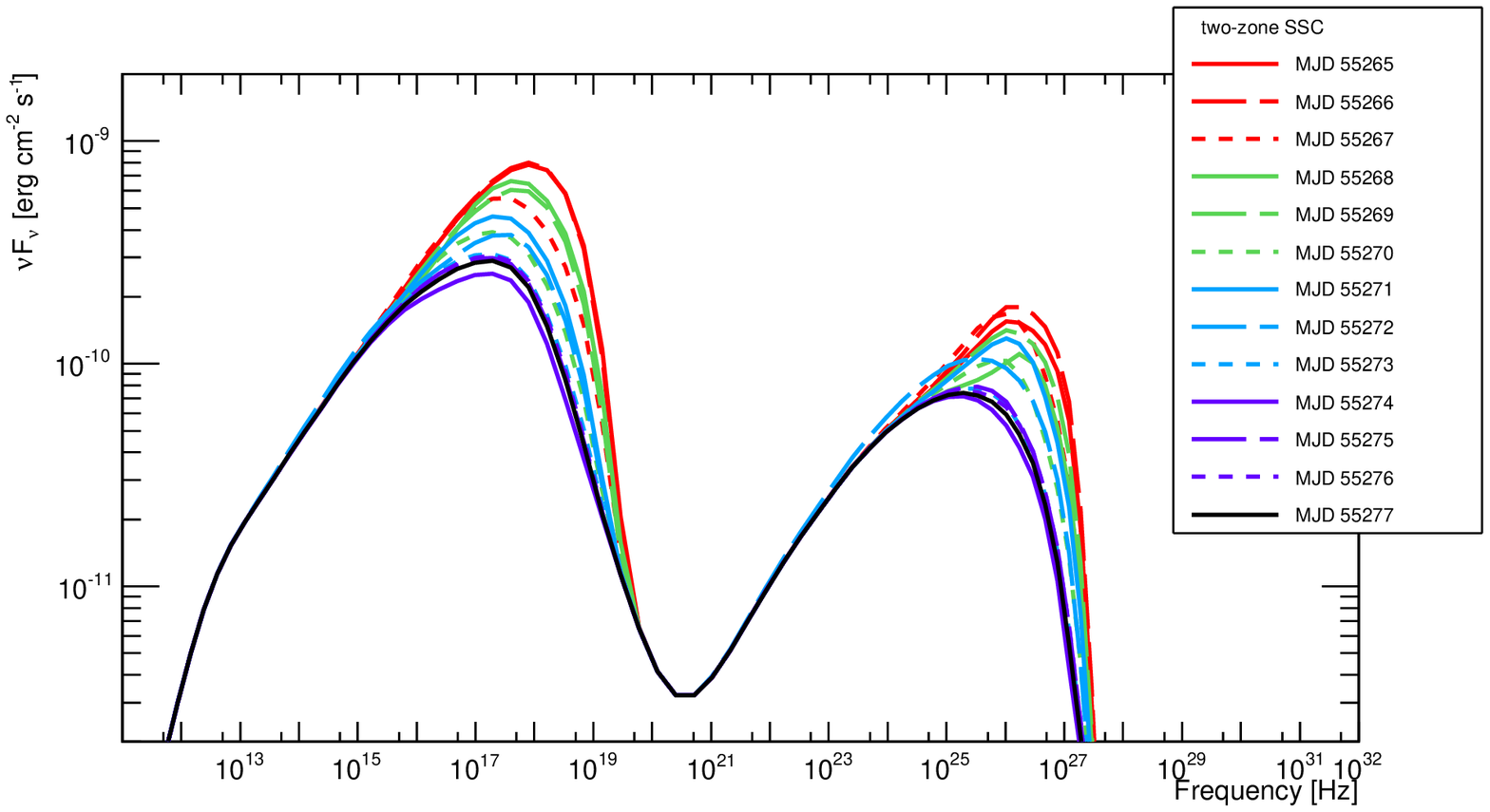}
  \caption{SEDs.}
  \label{fig:models2}
\end{subfigure}
\begin{subfigure}{.42\textwidth}
  \centering
  \includegraphics[width=1.\linewidth]{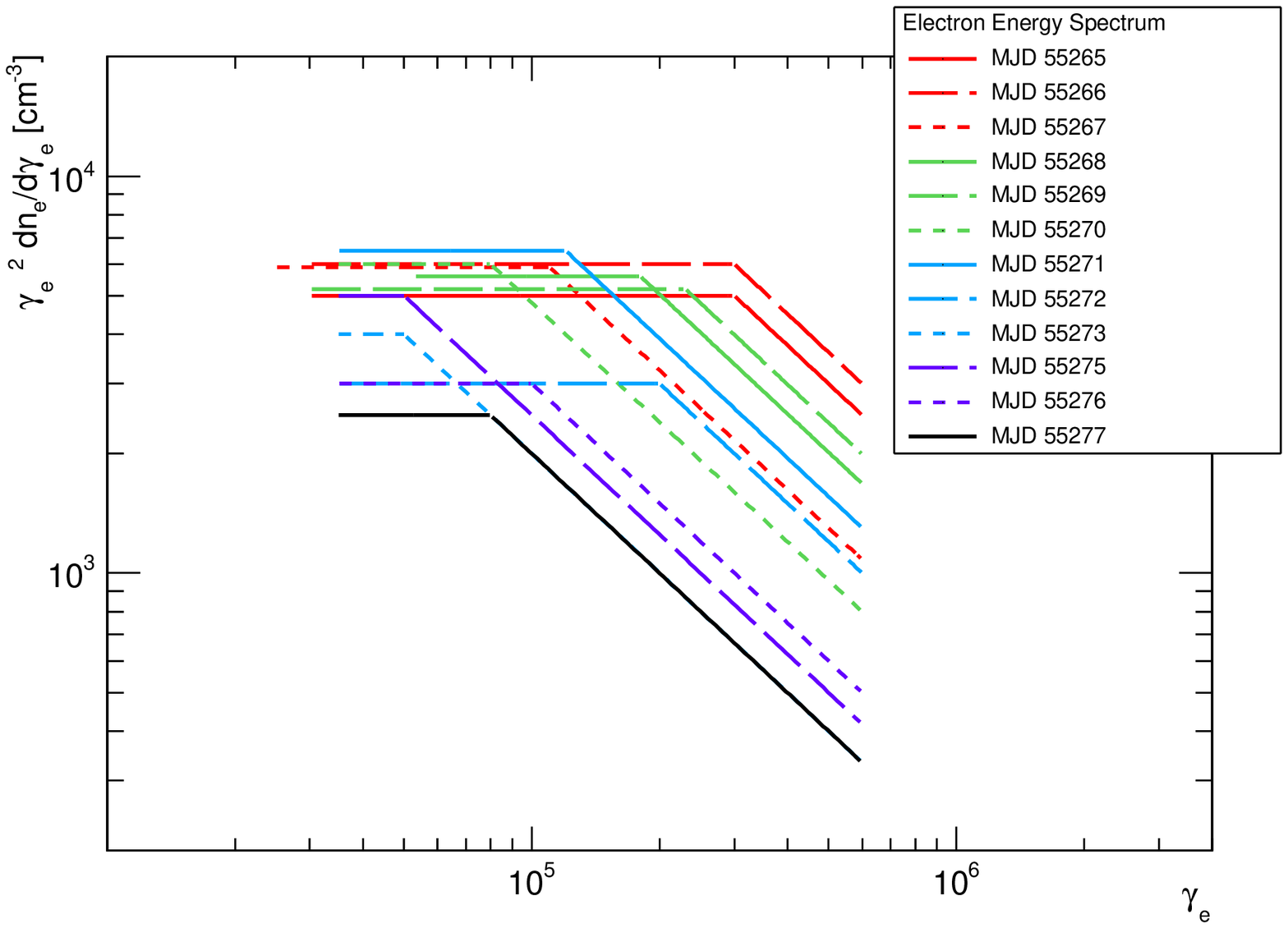}
  \caption{EEDs.}
  \label{fig25b}
\end{subfigure}
\caption{Two-zone SSC model curves (sum of the emission from the quiescent and the flaring blobs) and the related EEDs from the flaring blob used to describe the measured SEDs during the 13-day flaring activity. The parameter values are given in Table~\ref{TwoBlobSSC}.} \label{fig:SSCmodel2}
\end{figure*}

Figure~\ref{fig11b} depicts the two-zone model curve adjusted to
  the broadband SED from MJD~55265. It is worth noting that the
  contribution from the flaring blob is relevant only at the X-ray and
VHE bands. The model curves related to the remaining 12 consecutive
SEDs are shown in 
Appendix~\ref{sec:sed} (Figs.~\ref{fig:SEDs2zone:1} and \ref{fig:SEDs2zone:2}),
%% HT141119 %%
%Figs.~\ref{fig:SEDs2zone:1} and \ref{fig:SEDs2zone:2},}
and Table~\ref{TwoBlobSSC} reports the two-zone SSC model parameters that adequately describe the measured SEDs. Except for the magnetic field, which decreases during the decay of the flare, the other model parameters related to the environment 
%% HT141119 %%
%, namely the blob radius and the Doppler factor, 
remain constant. The changes occur in the three model parameters $n_{\rm e}$, $\gamma_{\rm min}$, and $\gamma_{\rm br1}$, while $s_1$, $s_2$, $\gamma_{\rm max}$ can be kept constant for all the 13 SEDs. With this two-zone SSC model, the shortest variability timescale $t_{\rm min}$ is about one hour, which is comparable to the length of our single-instrument observations, during which we did not measure significant variability. This shortest variability timescale would be consistent with the 
potential intra-night VHE variability on MJD~55268. The predicted synchrotron cooling break $\gamma_{\rm c}$ for the flaring blob is $7 \times 10^5$ for MJD~55265. For this day, the parameter $\gamma_{\rm br1}$ for the flaring blob is $3 \times 10^5$, with a change in the EED power-law index of 1, which is the canonical change for synchrotron cooling. During the following three days $\gamma_{\rm c}/\gamma_{\rm br1} \lapp 8$, and after MJD~55269 $\gamma_{\rm c}/\gamma_{\rm br1}$ is much larger, which means that the break in the EED of the flaring blob is intrinsic to the acceleration mechanism, and cannot be directly related to the synchrotron cooling during these days.

The flaring blob is characterized by an EED with a very high $\gamma_{\rm min}$ ($>3 \times 10^4$), which means that it lacks low-energy electrons, and so does not contribute to the radio/optical emission. This is necessary for improving (with respect to the one-zone SSC model from Sect.~\ref{Model1}) the description of the very narrow peaks at the X-ray and the $\gamma$-ray bumps occurring on some days (e.g. MJD~55265 and 55266).

Figures~\ref{fig:models2} and \ref{fig25b} depict the two-zone SSC model curves and the parameterized EEDs for the 13 consecutive days. In this case, by construction, all the SED variations occur at the X-ray and the VHE bands, and the SED peaks are narrower than those from the one-zone SSC scenario. Overall, the decay of the flaring activity is dominated by a reduction in $n_{\rm e}$ and $\gamma_{\rm br1}$. The magnetic field also varies with time (not shown in this plot, see Table~\ref{TwoBlobSSC}); lower activity is related to lower values of $B$.

The two-zone SSC model is described by 20 parameters, the
one-zone SSC model by 11. However, after fixing the parameters of the
quiescent-state blob, we only needed to change the values of four
parameters ($\gamma_{\rm min}$, $\gamma_{\rm br1}$, $n_{\rm e}$, and
$B$) in the flaring blob, while in the one-zone SSC model we had to
change five parameters ($\gamma_{\rm br1}$, $\gamma_{\rm br2}$, $s_1$,
$s_2$, $n_{\rm e}$) to describe the SEDs during these 13 consecutive
days (see Sect.~\ref{Model1}). 
Therefore, once the parameters of the quiescent blob are fixed, the two-zone SSC model describes the measured temporal evolution of the broadband SED with one free parameter less than the one-zone SSC model.

%%%%%%%%%%%%%%%%%%%%%%%%%%%%%%%%%%%%%%%%%%%%%%%%%%%%%%%%%%%%
%%%%%%%%%%%%%%%%%%%%%%%%%%%%%%%%%%%%%%%%%%%%%%%%%%%%%%%%%%%%
\section{Discussion} \label{Discussion}
%%%%%%%%%%%%%%%%%%%%%%%%%%%%%%%%%%%%%%%%%%%%%%%%%%%%%%%%%%%%
%%%%%%%%%%%%%%%%%%%%%%%%%%%%%%%%%%%%%%%%%%%%%%%%%%%%%%%%%%%%

The broadband SEDs during this flaring episode, resolved on timescales
of one day, allows for an unprecedented characterization of the time
evolution of the radio to $\gamma$-ray emission of Mrk~421. 
We find that both the one-zone SSC and the two- zone SSC models can
describe the daily SEDs by varying five and four model parameters, mostly related to the EED. This shows that the particle acceleration and cooling mechanism producing the EED could be the main mechanism responsible for the broadband SED variations during the flaring episodes in blazars.
 
In this theoretical framework, the two-zone SSC model provides 
better data-model agreement at the peaks of the low- and high-energy SED bumps.
Additionally, the two-zone SSC scenario presented here naturally
provides shorter timescales (one hour vs. one day) for variability at the
X-ray and VHE bands, as the correlated variability at X-ray and VHE bands without any variation at the optical and radio bands. Because low-energy electrons are absent, the peak frequency of the $\gamma$-ray bump becomes sensitive to $\gamma_{\rm min}$ as a result of the strong Klein-Nishina effect, which provides a rather independent channel to adjust the $\gamma$-ray bump for the flaring state. On the other hand, the X-ray bump is more sensitive to the magnetic field and $\gamma_{\rm br1}$. Hence this phenomenological scenario of two distinct zones (quiescent+flaring) allows for more flexibility in the locations and shapes of the two bumps than in the one-zone SSC model, while still varying fewer
parameters. This was particularly useful to adequately describe the
evolution of the width of the two SED bumps. 
%% HT 141119 %%
%mentioned in Sect.~\ref{SED}. 
We can quantify this effect by computing the widths
of the bumps as the full width at half maximum (FWHM) in the
logarithmic scale, $\log(\nu_2/\nu_1)$, where  $\nu_1$ and $\nu_2$ are
the frequencies at which the energy flux is half of that at the peak
position. The widths of the SED bumps for the 13 consecutive days are
reported in Table~\ref{peak2}, showing that both the synchrotron and
inverse-Compton peak widths increase from $\log(\nu_2/\nu_1)\sim$2 to $\log(\nu_2/\nu_1)\sim$3 during the decay of the flare, which means that the width of the two bumps (in logarithmic scale) is about 50\% greater during the nonflaring (low) activity. 

\begin{sidewaystable}
\caption{Peak positions and widths of the synchrotron and inverse-Compton bumps derived from the two-zone SSC model parameters reported in Table~\ref{TwoBlobSSC}.}\label{peak2}
\centering
\begin{tabular}{ccccccccccc} 
\hline
\hline 
Date   & $\nu^{\rm syn}_{\rm peak}$ & $(\nu F_{\nu})^{\rm syn}_{\rm peak}$ & $\nu^{\rm syn}_1$ & $\nu^{\rm syn}_2$ & $\log(\nu^{\rm syn}_2/\nu^{\rm syn}_1)$ & $\nu^{\rm ic}_{\rm peak}$ & $(\nu F_{\nu})^{\rm ic}_{\rm peak}$ & $\nu^{\rm ic}_1$ & $\nu^{\rm ic}_2$ & $\log(\nu^{\rm ic}_2/\nu^{\rm ic}_1)$ \\
 - -   &[$10^{17}$]& [$10^{-10}$]            &[$10^{15}$]&[$10^{18}$]& - - &[$10^{25}$]& [$10^{-11}$]          &[$10^{23}$]&[$10^{26}$]& - - \\
$[\rm MJD]$ & [Hz] & [erg cm$^{-2}$s$^{-1}$] & [Hz]      & [Hz]      & - - & [Hz]      &[erg cm$^{-2}$s$^{-1}$]& [Hz] & [Hz] & - - \\
\hline

\hline
55265 & 8.1 & 7.9 & 34. & 6.1 & 2.3 & 10. & 15. & 60. & 9.5 & 2.2  \\
55266 & 8.1 & 8.0 & 34. & 5.9 & 2.2 & 10. & 18. & 94. & 9.9 & 2.0  \\
55267 & 4.0 & 5.5 & 11. & 3.3 & 2.5 & 10. & 17. & 56. & 5.1 & 2.0  \\
55268 & 4.0 & 6.6 & 30. & 4.5 & 2.2 & 17. & 11. & 16. & 7.3 & 2.7  \\
55269 & 4.0 & 6.1 & 1.9 & 4.5 & 2.4 & 10. & 14. & 42. & 7.8 & 2.3  \\
55270 & 2.0 & 3.9 & 5.7 & 2.3 & 2.6 & 6.0 & 10. & 11. & 4.3 & 2.6  \\
55271 & 2.0 & 4.6 & 9.0 & 2.6 & 2.5 & 1.0 & 13. & 30. & 5.4 & 2.3  \\
55272 & 4.0 & 3.8 & 4.9 & 2.8 & 2.8 & 3.4 & 11. & 7.4 & 4.5 & 2.8  \\
55273 & 2.0 & 3.1 & 3.1 & 1.9 & 2.8 & 1.9 & 7.7 & 3.9 & 3.0 & 2.9  \\
55274 & 2.0 & 2.5 & 1.8 & 1.6 & 2.9 & 1.9 & 7.1 & 3.0 & 2.4 & 2.9  \\
55275 & 2.0 & 3.0 & 2.8 & 1.8 & 2.8 & 3.4 & 7.9 & 4.2 & 3.0 & 2.9  \\
55276 & 2.0 & 3.1 & 3.1 & 1.8 & 2.8 & 1.9 & 7.5 & 3.6 & 3.2 & 2.9  \\
55277 & 2.0 & 2.9 & 2.7 & 1.7 & 2.8 & 1.9 & 7.4 & 3.4 & 2.8 & 2.9  \\
\hline
\end{tabular}
\tablefoot{$\nu^{\rm syn}_{\rm peak}$: the peak frequency of the synchrotron bump; $(\nu F_{\nu})^{\rm syn}_{\rm peak}$: the peak energy flux of the synchrotron bump; $\nu^{\rm ic}_{\rm peak}$: the peak frequency of the inverse-Compton bump; $(\nu F_{\nu})^{\rm ic}_{\rm peak}$: the peak energy flux of the inverse-Compton bump. For each bump in the SED, the value of $(\nu F_{\nu})_{\rm peak}/2$ determines the two frequencies ($\nu_1$ and $\nu_2$) that are used to quantify the width of the bump in the logarithmic scale $\log(\nu_2/\nu_1)$.}
\end{sidewaystable}

\begin{figure*}
\centering
\begin{subfigure}{.49\textwidth}
  \centering
  \includegraphics[width=1.\linewidth]{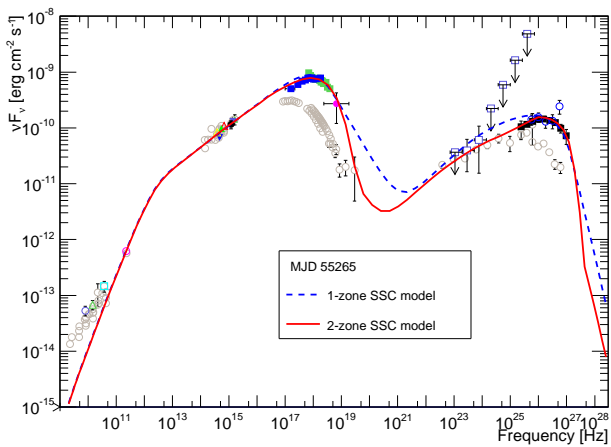}
  \caption{MJD~55265.}
  \label{fig:2SSC65}
\end{subfigure}
\begin{subfigure}{.49\textwidth}
  \centering
  \includegraphics[width=1.\linewidth]{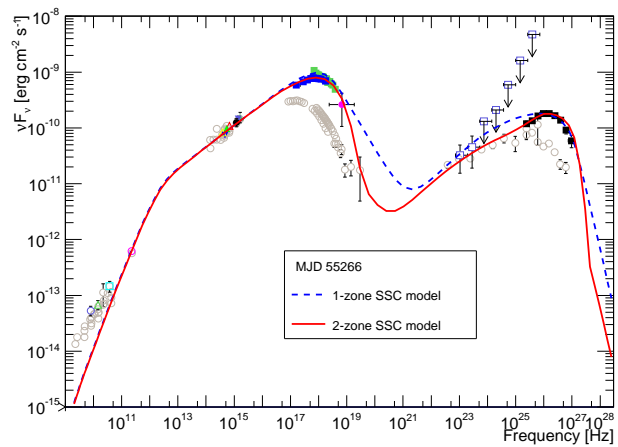}
  \caption{MJD~55266.}
  \label{fig:2SSC66}
\end{subfigure}
\caption{Broadband SEDs from MJD~55265 and 55266 (the two days with
  the highest activity) with the one-zone and two-zone model curves described in sections 4.2 and 4.3. We refer to Figs.~\ref{fig11} and \ref{fig12} for details of the data points.}
\label{fig:2SSC}
\end{figure*}

The additional flexibility of the two-zone SSC model 
%% HT141119 %%
%(in comparison tothe one-zone SSC model) 
helps to improve the agreement of the model
SEDs with the data from MJD~55265, 55266, 55268, 55269, and 55273. 
%% HT141119 %%
%(compare Figs.~\ref{fig11}, \ref{fig12}, \ref{fig13}, \ref{fig14}, and \ref{fig17} with Figs.~\ref{fig11b}, \ref{fig12b}, \ref{fig13b}, \ref{fig14b}, and \ref{fig17b}). 
The largest data-model differences
occur for the first two days, which are the days with the highest
activity and the narrowest low- and high-energy
bumps. Figures~\ref{fig:2SSC65} and \ref{fig:2SSC66} compare the
data-model agreement for these two days. Note the better agreement of
the two-zone SSC model curves with the X-ray data points and,
especially, the $\gamma$-ray data points. The agreement can be
quantified using $\chi ^2$ on the broadband SEDs, after excluding the
radio data, which are considered as upper limits for the models. In
total, we have 50 and 51 data points for MJD~55265 and MJD~55266,
respectively. With a one-zone SSC model we obtain a $\chi ^2$
of $4.0 \times 10^{3}$ for MJD~55265 and $3.6 \times 10^{3}$ for
MJD~55266 , while we obtain $1.2 \times 10^{3}$ for MJD~55265 and
$0.7 \times 10^{3}$ for MJD~55266 with the two-zone SSC model,
which shows that the agreement of the model with the data is better
for the latter theoretical scenario. An F-test on the obtained $\chi
^2$ values, and assuming conservatively that the one-zone model has 11
free parameters and the two-zone model has 20 free parameters (hence
not considering that many of these parameters are kept constant)
rejects the one-zone model in favor of the two-zone model for the given set of model parameters with a
$p$-value lower than $10^{-5}$. 
If one considers that many model parameters are kept constant, the
rejection of the reported one-zone model in favor of the reported
two-zone model would be even clearer. The reduced $\chi ^2$ for
  all cases is well above 1, which shows that none of the models
  describe the observations perfectly well. Both models oversimplify the complexity in the
  blazar jets, and hence we do not intend to explain the data at the
  percent level.

It is worth noting that the EED of the flaring blob is constrained to a very narrow range of energies, namely $\gamma_{\rm min}$--$\gamma_{\rm max}$ $\sim 3 \times 10^4$--$6 \times 10^5$. One theoretical possibility to produce such a narrow EED is stochastic particle acceleration via scattering by magnetic inhomogeneities in the jet, namely second-order Fermi acceleration \citep[e.g.,][]{Stawarz2013ApJ681p1725,Lefa2011ApJ740p64,2014ApJ...780...64A}. The spectrum in this model is localized around a characteristic Lorentz factor $\gamma_{ch}$ determined by the power spectrum of magnetic turbulence $q$ and the cooling timescale of electrons, with a shape proportional to $\gamma_e^2 \exp \left[ - (\gamma_e / \gamma_{ch})^{3-q} \right]$ \citep[e.g.,][]{Schlickeiser1985AA143p431}. Such a spectrum can realize the narrow peaks of synchrotron radiation and inverse-Compton scattering that we measured for Mrk~421 during the 2010 March flare.

The treatment made with the one- and two-zone homogeneous
SSC models is a simplification of the problem. For instance,
relativistic travel within a jet can change the properties of a blob
(e.g. expansion of the size $R$ of the emitting region, and decrease
in the magnetic field $B$). This issue has been discussed in several
papers \citep[e.g][ for the case of 1ES
1959+650]{2008ApJ...679.1029T}. The fact that we can explain the
temporal evolution of the SED during 13 consecutive days without
changing the  model parameters related to the environment could be
interpreted as meaning that the blazar emission region is not traveling
relativistically, but rather is stationary in one or several regions of the jet where there is a standing shock. Such standing shocks could be produced, for instance, by recollimation in the jet, and the particles would be accelerated as the jet flows or the superluminal knots cross it \citep{1997MNRAS.288..833K,2004ApJ...613..725S,2014ApJ...780...87M}. The Lorentz factor of the plasma, as it flows through the standing shock, would be the Lorentz 
factor that would lead to the Doppler factor (depending on the angle) used in the model.

This MW campaign reveals that the correlation between the X-ray
flux at the $2$--$10$ keV band and the VHE $\gamma$-ray flux above
$200$ GeV shows an approximately linear trend (see Fig.~\ref{fig24m}
middle and bottom panels), while the correlation between X-ray flux at
the $0.3$-$2$ keV band and the VHE $\gamma$-ray flux is equally close
to both a linear and quadratic trend (see Fig.~\ref{fig24m} top
panel). This is an interesting result because the 0.3--2 keV band
reports the synchrotron emission below or at the low-energy
(synchrotron) peak of the SED, while the 2--10 keV band reports the
emission at or above the low-energy peak. During the Mrk~421 flaring
activity observed in 2001, it was also noted that the VHE-to-X-ray
(above 2 keV) correlation was linear when considering day timescales,
but the correlation was quadratic when considering few-hour long variability \citep[see][]{fossati08}.
A quadratic (or more-than-quadratic) correlation between X-ray and VHE
$\gamma$-ray fluxes in the decaying phase is hard to
explain with conventional SSC models \citep{Katarzynski05}. 
During the flaring activity observed in 2010 March, we do not detect any significant intra-night variability, which might be due to the shorter (about one hour) duration of the observations (in comparison to the many-hour long observations reported in \cite{fossati08}), or perhaps due to the lower X-ray and VHE activity (in contrast to that of 2001).

The almost linear correlation at $2$-$10$ keV X-rays can be explained as follows: In the framework of the one-zone SSC model, the SED peaks at $\gamma$-ray frequencies are produced by the smaller cross-section in the Klein-Nishina regime, rather than by the breaks $\gamma_{{\rm br},1/2}$ in the EED. Therefore, the $\gamma$-ray emission with energies above the SED peak energy is affected by the lower Klein-Nishina cross-section and is dominated by inverse-Compton scattering off infrared-to-optical photons. Since these target photons are produced by the synchrotron radiation of electrons with a Lorentz factor well below $\gamma_{{\rm br} 1}$, whose density is almost constant during this decaying phase (see Fig.~\ref{fig25}), the density of target photons is almost constant. Thus, the change in the number density of electrons above $\gamma_{{\rm br} 2}$ is directly reflected in the $\gamma$-ray flux, resulting in the almost linear correlation between X-ray and $\gamma$-ray fluxes. A similar mechanism also works in the two-zone SSC model in each blob. In a flaring blob, $\gamma$-ray SED peaks originate from the Klein-Nishina effect. Therefore, $\gamma$-rays with energies above the SED peak result from inverse-Compton scattering of electrons off photons below the SED peak at the X-ray band as well as in the one-zone SSC model. Thus, the almost linear relation is realized in both the quiescent and flaring blobs, and hence it is also realized in the total spectra.

The correlation between X-rays and $\gamma$-rays was analyzed with a great level of detail in \cite{Katarzynski05}, where the evolution of several quantities such as the number density of electrons, magnetic fields, and the size of the emission region, are simply parameterized to study their contribution to the index of the correlation. Evolution of the emission region volume is a possibility to naturally explain the reduction of the electron number density in the emission region. In the results presented here we fixed the size $R$ to properly study the evolution of the electron spectrum with the steady SSC models at each moment. Further studies of the temporal broadband emission evolution involving such additional parameters will be performed elsewhere.

The SED model results described in Sects.~\ref{Model1} and
\ref{Model2} allow for an estimate of several physical properties of
Mrk~421 during the flaring activity from 2010 March: the total
electron number density $N_{\rm e}$, mean electron Lorentz factor
$\langle \gamma_{\rm e} \rangle$, the jet power carried by electrons
$L_{\rm e}$, the jet power carried by the magnetic field $L_{\rm B}$,
the ratio of comoving electron and magnetic field energy densities
$U'_{\rm e}/U'_{\rm B}=L_{\rm e}/L_{\rm B}$, the synchrotron
luminosity $L_{\rm syn}$ (integrated from $10^{9.5}$ Hz to $10^{20.5}$
Hz), the inverse-Compton luminosity $L_{\rm IC}$ (integrated from
$10^{20.5}$ Hz to $10^{28}$ Hz), and the total photon luminosity from
the SSC model $L_{\rm ph}=L_{\rm syn}+L_{\rm IC}$. We can also compute
the jet power carried by protons $L_{\rm p}$ assuming one proton per
electron ($N_{\rm p} = N_{\rm e}$). The total jet power is $L_{\rm
  jet}=L_{\rm p}+L_{\rm e}+L_{\rm B}$. 
We follow the prescriptions given in \cite{2008MNRAS.385..283C}. Specifically, the following formulae are used:
\begin{equation}
N_{\rm e}=\int_{\gamma_{\rm min}}^{\gamma_{\rm max}} \frac{dn_{\rm e}}{d\gamma_{\rm e}} d\gamma_{\rm e},
\label{eq:electron_number}
\end{equation}
\begin{equation}
\langle \gamma_{\rm e} \rangle = \frac{\int_{\gamma_{\rm min}}^{\gamma_{\rm max}} \gamma_{\rm e}\frac{dn_{\rm e}}{d\gamma_{\rm e}} d\gamma_{\rm e}}{N_{\rm e}},
\end{equation}
\begin{equation}
L_{\rm e}=\pi R^2 \Gamma^2 \beta c N_{\rm e} \langle \gamma_{\rm e} \rangle m_{\rm e} c^2,
\end{equation}
\begin{equation}
L_{\rm p}=\pi R^2 \Gamma^2 \beta c \cdot N_{\rm e}  \cdot m_{\rm p} c^2,
\end{equation}
\begin{equation}
L_{\rm B}=\frac{1}{8} R^2 \Gamma^2 \beta c B^2,
\end{equation}
\begin{equation}
L_{\rm ph}=\int \frac{\pi D_{\rm L}^2  F_{\rm \nu}}{\Gamma^2} \frac{(1+z)d\nu}{\delta},
\label{eq:photon_luminosity}
\end{equation}
where $\Gamma \sim \delta$, $\beta = \frac{v}{c} =
\sqrt{1-\frac{1}{\Gamma^2}} \sim 1 - \frac{1}{2\Gamma^2}  $, $D_{\rm
  L}$=134 Mpc 
%% HT141119 %%
%(derived from $H_0$ = 71 km s$^{-1}$ Mpc$^{-1}$, $\Omega_M$ = 0.27, $\Omega_{\Lambda}$ = 0.73). 
In the jet power
calculation, only one side is considered, differently to what was done
in \cite{2008ApJ...686..181F}, who used a two-sided jet. 
%% HT141119 %%
%Table~\ref{allL1} reports the calculated values for the one-zone SSC model described in Sect.~\ref{Model1}, while Table~\ref{allL2} reports the calculated values for the flaring blob and the sum of the two blobs in the two-zone SSC model described in Sect.~\ref{Model2}. 
The details of these quantities derived with the SSC model parameters are tabulated 
in Appendix~\ref{sec:phys_parameters}.

In both the one-zone and two-zone model, the electron luminosity $L_{\rm e}$ and magnetic 
luminosity $L_{\rm B}$ are more than one order of magnitude away from equipartition, 
which
was reported in \cite{AbdoMrk421,2011ApJ...733...14M,2012A&A...542A.100A}. In addition, we found that the ratio
$L_{\rm e}/L_{\rm B}$ does not vary much during the 13-day period considered here.

In the two-zone model, the total power $L_{\rm p} + L_{\rm e} + L_{\rm B}$ of the flaring blob is 
about one order of magnitude smaller than that of the quiescent-state blob 
($\sim 10^{43}$ erg s$^{-1}$ vs. $\sim 10^{44}$ erg s$^{-1}$) even though 
$\langle \gamma_{\rm e} \rangle$ is 20 -- 30 times higher. This is caused by the smaller size 
of the flaring blob, in spite of its stronger magnetic field and higher electron 
density. Nevertheless, the flaring blob is responsible for about half of
the photon luminosity $L_{\rm ph} (= L_{\rm syn} + L_{\rm IC})$ of the quiescent-state blob during the highest X-ray/VHE $\gamma$-ray activity. This indicates that 
the radiative efficiency of electrons is high in the flaring blob as a result of the strong 
magnetic field $B$ and high electron number density $n_{\rm e}$. Since the contribution 
of the flaring blob to the total photon luminosity decreases with the
decline of the X-ray/VHE activity, the total photon luminosity in the two-zone model does not change 
substantially during the 13-day period with the VHE flux going from $\sim 2$ c.u. 
down to $\sim 0.5$ c.u., remaining at about $(3 - 5) \times 10^{42}$ erg s$^{-1}$. 
On the other hand, the variation of the total photon luminosity in the one-zone model is from 
$9 \times 10^{42}$ erg s$^{-1}$ to $3 \times 10^{42}$ erg s$^{-1}$, and hence, in terms of 
jet energetics, the production of the measured X-ray/VHE flaring activity is more 
demanding in the one-zone scenario than in the two-zone scenario.

%%%%%%%%%%%%%%%%%%%%%%%%%%%%%%%%%%%%%%%%%%%%%%%%%%%%%%%%%%%%
%%%%%%%%%%%%%%%%%%%%%%%%%%%%%%%%%%%%%%%%%%%%%%%%%%%%%%%%%%%%
\section{Conclusion} \label{Conclusion}
%%%%%%%%%%%%%%%%%%%%%%%%%%%%%%%%%%%%%%%%%%%%%%%%%%%%%%%%%%%%
%%%%%%%%%%%%%%%%%%%%%%%%%%%%%%%%%%%%%%%%%%%%%%%%%%%%%%%%%%%%

We 
have reported 
the MW observations of the decaying phase of a Mrk~421 flare from 2010 March, and characterized it with two leptonic scenarios: a one-zone SSC model, and a two-zone SSC model where one zone is responsible for the quiescent emission, while the other (smaller) zone, which is spatially separated from the former one, contributes to the daily-variable emission occurring mostly at X-rays and VHE $\gamma$-rays. 
We found
that flux variability is noticeable at the X-ray and VHE
  $\gamma$-ray bands, while it is minor or not significant in the
  other bands.  These observations revealed 
an almost linear correlation between the X-ray flux at the 2--10 keV band and the VHE $\gamma$-ray flux above 200 GeV, consistent with the $\gamma$-rays being produced by inverse-Compton scattering in the Klein-Nishina regime in the framework of SSC models.

The broadband SEDs during this flaring episode, resolved on timescales 
of one day, allowed 
for an unprecedented characterization of the time
evolution of the radio to $\gamma$-ray emission of Mrk~421. Such a
detailed study has not been performed on Mrk~421 or any other
blazar before. 
Both the one-zone SSC and the two-zone SSC models can
describe the daily SEDs via the variation of only five and four model
parameters respectively, under the hypothesis that the variability is associated
mostly with the underlying particle population. This shows that
blazar variability might be dominated by the acceleration and cooling
mechanisms that produce the EED. For both cases (one-zone and
two-zone SSC models), an EED parameterized by two power-law functions
is sufficient to describe the emission during the very high states
(MJD~55265 and 55266), but an EED with three power-law functions is needed
during the somewhat lower blazar activity.

We also found
that the two-zone SSC model describes the measured SED data at the peaks of the low- and high-energy
bumps better, although the reported one-zone SSC model could be further improved by the variation of the parameters related to the emitting region itself, in addition to the parameters related to the particle population.
The two-zone SSC scenario presented here naturally provides shorter
timescales (one hour vs. one day) for variability at the X-ray and VHE
bands, as well as lack of correlation between the
radio/optical/GeV emission and the variability in the
X-ray/VHE bands. Within this two-zone SSC scenario, the EED of the flaring blob is constrained to a very narrow range of energies, namely $\gamma_{\rm min}$--$\gamma_{\rm max}$ $\sim 3 \times 10^4$--$6 \times 10^5$, which could be produced through stochastic particle acceleration via scattering by magnetic inhomogeneities in the jet.

%$\gamma_{\rm min}$ &  $\gamma_{\rm max}$ &  $\gamma_{\rm br1}$&  $\gamma_{\rm br2}$  &  $s_1$ &  $s_2$ &  $s_3$  &  $n_{\rm e}$                &  $B $          &  $log(R)$             &  $\delta$ 

%total electron number density $N_{\rm e}$          & $\langle \gamma_{\rm e} \rangle$  & $L_{\rm e}$             & $L_{\rm p}$             & $L_{\rm B}$             & $U^'{\rm e}/U^'{\rm B}$ & $L_{\rm syn}$           & $L_{\rm IC}$            & $L_{\rm ph}$     & $^{sum}L_{\rm e}$             & $^{sum}L_{\rm p}$             & $^{sum}L_{\rm B}$             & $L_{\rm syn}$          & $^{sum}L_{\rm IC}$            & $^{sum}L_{\rm ph}$

%x_erg=SSCfreq/keV2Hz*keV2erg*1.03/Doppler;
%y_e10_s_1=SSCSEDdeabsorb/x_erg/(1.e20.)*pi*(LD_pc*pc2cm)*(LD_pc*pc2cm)/pow(Doppler,2.);
%one-sided jet

\begin{acknowledgements}
The authors thank the anonymous referee for providing a very
  detailed and constructive list of remarks that helped us to improve the manuscript.\\

The MAGIC collaboration would like to thank the Instituto de Astrof\'isica de Canarias for the excellent working conditions at the Observatorio del Roque de los Muchachos in La Palma. The financial support of the German BMBF and MPG, the Italian INFN and INAF,  the Swiss National Fund SNF, the ERDF under the Spanish MINECO, and the Japanese JSPS and MEXT is gratefully acknowledged. This work was also supported by the Centro de Excelencia Severo Ochoa SEV-2012-0234, CPAN CSD2007-00042, and MultiDark CSD2009-00064 projects of the Spanish Consolider-Ingenio 2010 programme, by grant 268740 of the Academy of Finland, by the Croatian Science Foundation (HrZZ) Project 09/176 and the University of Rijeka Project 13.12.1.3.02, by the DFG Collaborative Research Centers SFB823/C4 and SFB876/C3, and by the Polish MNiSzW grant 745/N-HESS-MAGIC/2010/0.\\

The VERITAS collaboration acknowledges supports from the grants from the U.S. Department of Energy Office of Science, the U.S. National Science Foundation and the Smithsonian Institution, by NSERC in Canada, by Science Foundation Ireland (SFI 10/RFP/AST2748) and by STFC in the U.K. We acknowledge the excellent work of the technical support staff at the Fred Lawrence Whipple Observatory and at the collaborating institutions in the construction and operation of the instrument.\\

The \textit{Fermi} LAT collaboration acknowledges generous ongoing support from a number of agencies and institutes that have supported both the development and the operation of the LAT as well as scientific data analysis. These include the National Aeronautics and Space Administration and the Department of Energy in the United States, the Commissariat \`a l'Energie Atomique and the Centre National de la Recherche Scientifique / Institut National de Physique Nucl\'eaire et de Physique des Particules in France, the Agenzia Spaziale Italiana and the Istituto Nazionale di Fisica Nucleare in Italy, the Ministry of Education, Culture, Sports, Science and Technology (MEXT), High Energy Accelerator Research Organization (KEK) and Japan Aerospace Exploration Agency (JAXA) in Japan, and the K.~A.~Wallenberg Foundation, the Swedish Research Council and the Swedish National Space Board in Sweden. Additional support for science analysis during the operations phase is gratefully acknowledged from the Istituto Nazionale 
di Astrofisica in Italy and the Centre National d'\'Etudes Spatiales in France.\\

The research at Boston University was funded in part by NASA Fermi Guest Investigator grant NNX11AQ03G. The Submillimeter Array is a joint project between the Smithsonian Astrophysical Observatory and the Academia Sinica Institute of Astronomy and Astrophysics and is funded by the Smithsonian Institution and the Academia Sinica. 

The OVRO 40-m monitoring program is supported in part by NASA grants NNX08AW31G and NNX11A043G, and NSF grants AST-0808050 and AST-1109911. 
The Mets\"ahovi team acknowledges the support from the Academy of Finland to our observing projects (numbers 212656, 210338, 121148, and others).
This work was partly supported by Russian RFBR grant 12-02-00452 and St.Petersburg University research grants 6.0.163.2010, 6.38.71.2012.
The Abastumani Observatory team acknowledges financial support by the Georgian National Science Foundation through grant GNSF/ST07/4-180 and by
the Shota Rustaveli National Science Foundation through the grant
FR/577/6-320/13.
We acknowledge the use of public data from the Swift and RXTE data archive.

\end{acknowledgements}

%-------------------------------------------------------------------

\clearpage

%%%%%%%%%%%%%%%%%%%%%%%%%%%%%%%%%%%%%%%%%%%%%%%%%%%%%%%%%%%%
%%%%%%%%%%%%%%%%%%%%%%%%%%%%%%%%%%%%%%%%%%%%%%%%%%%%%%%%%%%%
%%%%%%%%%%%%%%%%%%%%%%%%%%%%%%%%%%%%%%%%%%%%%%%%%%%%%%%%%%%%
\begin{appendix}
%%%%%%%%%%%%%%%%%%%%%%%%%%%%%%%%%%%%%%%%%%%%%%%%%%%%%%%%%%%%
%%%%%%%%%%%%%%%%%%%%%%%%%%%%%%%%%%%%%%%%%%%%%%%%%%%%%%%%%%%%
%%%%%%%%%%%%%%%%%%%%%%%%%%%%%%%%%%%%%%%%%%%%%%%%%%%%%%%%%%%%

%%%%%%%%%%%%%%%%%%%%%%%%%%%%%%%%%%%%%%%%%%%%%%%%%%%%%%%%%%%%
%%%%%%%%%%%%%%%%%%%%%%%%%%%%%%%%%%%%%%%%%%%%%%%%%%%%%%%%%%%%
\section{Simultaneity in the multi-instrument observations} 
\label{sec:simultaneity}
%%%%%%%%%%%%%%%%%%%%%%%%%%%%%%%%%%%%%%%%%%%%%%%%%%%%%%%%%%%%
%%%%%%%%%%%%%%%%%%%%%%%%%%%%%%%%%%%%%%%%%%%%%%%%%%%%%%%%%%%%

%{\bf This section displays the simultaneity in the multi-instrument data collected during the 13-day period considered in this paper, namely from  2010 March 10 (MJD~55265) to 2010 March 22 (MJD~55277). }

The energy coverage as a function of the time for
the daily multi-instrument observations from 2010 March 10 (MJD~55265) to
2010 March 22 (MJD~55277) is depicted in Figs.~\ref{figSim1} and
\ref{figSim2}, which show that most of the observations used to determine the SEDs
reported in Appendix~\ref{sec:sed} occur within less than 2 hours.

\begin{figure*}
\centering
\includegraphics[width=17cm]{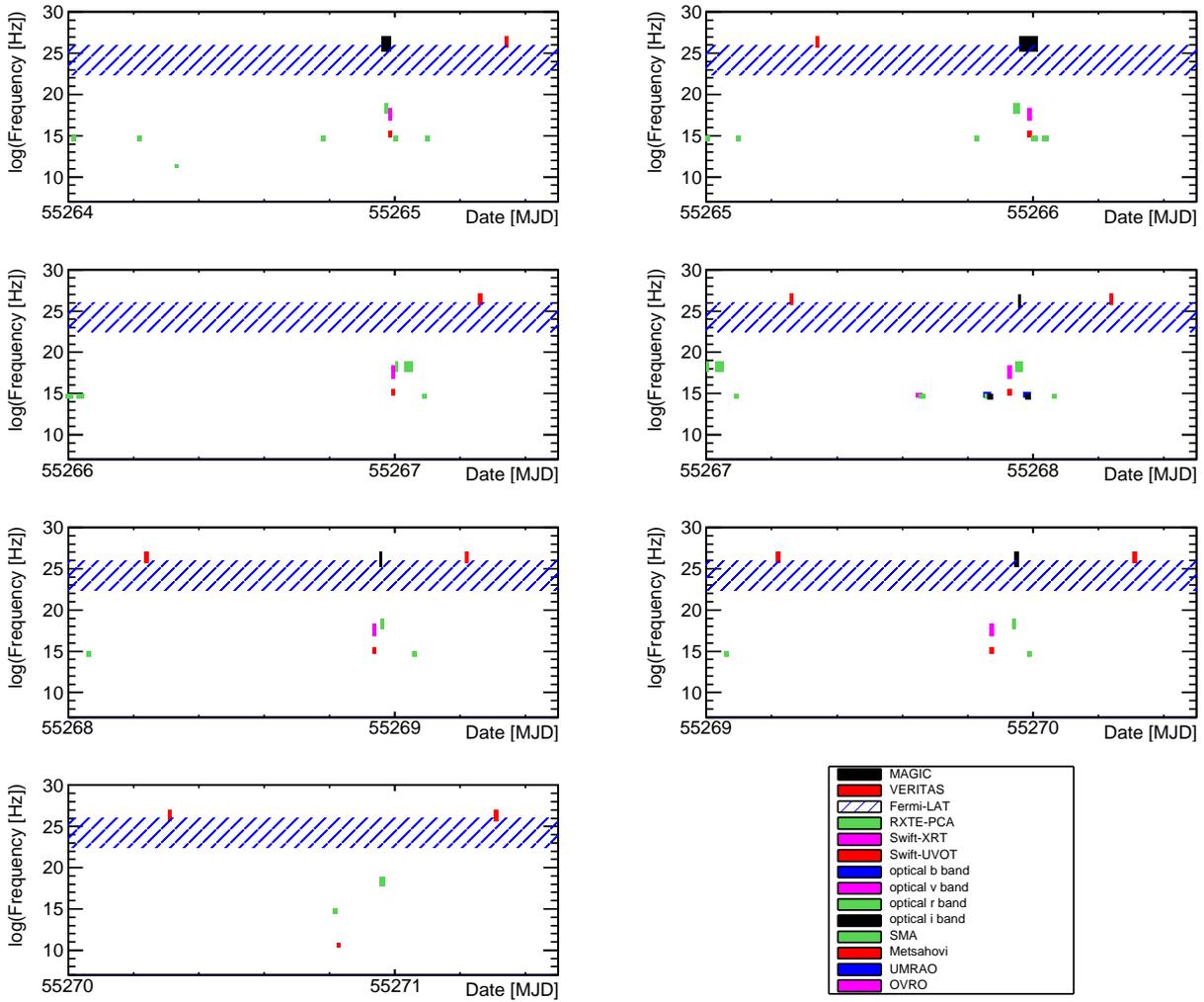}
\caption{Temporal and energy coverage during the flaring activity from
  2010 March 10 (MJD~55265) to 2010 March 16 (MJD~55271). 
\FermiLAT data were accumulated during two-day time intervals to ensure significant detections of Mrk\,421, and is depicted here with
a blue band. For better visibility of the observations at UV, optical, and radio band, where the observation time is usually short and the covered frequency band is narrow, an additional 20 minutes in time and half a decade in frequency are included when displaying the results. The names of all the optical instruments are listed in Table~\ref{TableWithInstruments}.}
\label{figSim1}
\end{figure*}

\begin{figure*}
\centering
\includegraphics[width=17cm]{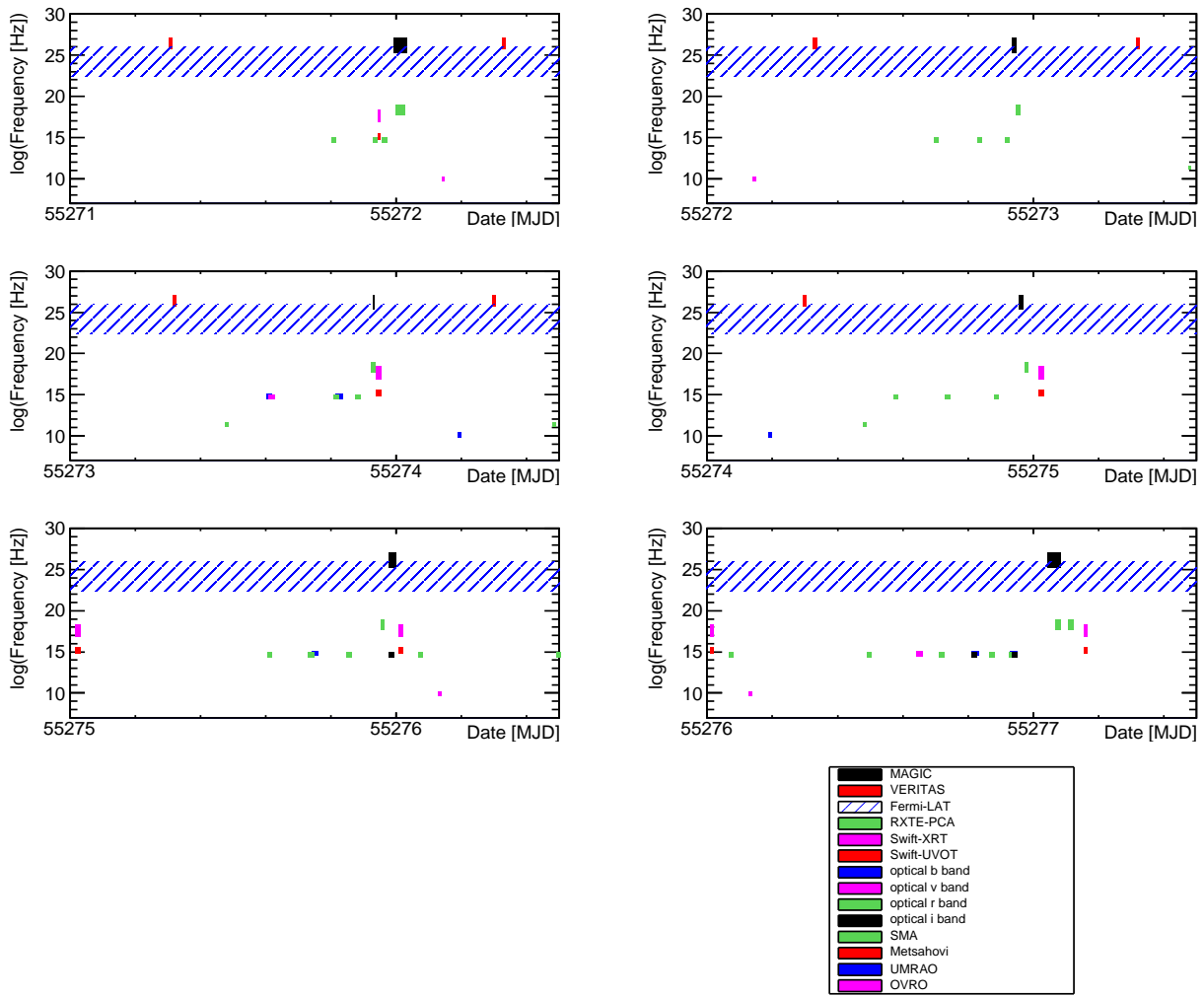}
\caption{Temporal and energy coverage during the flaring activity from 2010 March 17 (MJD~55272) to 2010 March 22 (MJD~55277). See the caption of Fig.~\ref{figSim1} for further details.
\label{figSim2}}
\end{figure*}

\clearpage

%%%%%%%%%%%%%%%%%%%%%%%%%%%%%%%%%%%%%%%%%%%%%%%%%%%%%%%%%%%%
%%%%%%%%%%%%%%%%%%%%%%%%%%%%%%%%%%%%%%%%%%%%%%%%%%%%%%%%%%%%
\section{Broadband SEDs for the 13 consecutive days} \label{sec:sed}
%%%%%%%%%%%%%%%%%%%%%%%%%%%%%%%%%%%%%%%%%%%%%%%%%%%%%%%%%%%%
%%%%%%%%%%%%%%%%%%%%%%%%%%%%%%%%%%%%%%%%%%%%%%%%%%%%%%%%%%%%

%% HT 141119 %%
%{\bf This section reports the broadband SEDs for each single day from the 13-day period considered in this paper. The one-zone SSC model curves are overlaid in Figs.~\ref{fig11}, \ref{fig:SEDs1zone:1} and \ref{fig:SEDs1zone:2}, while the two-zone SSC model curves are depicted in Fig.~\ref{fig11b} (in the main text) and Figs.~\ref{fig:SEDs2zone:1} and \ref{fig:SEDs2zone:2}.}

\begin{figure*}
\sidecaption
\includegraphics[width=12cm]{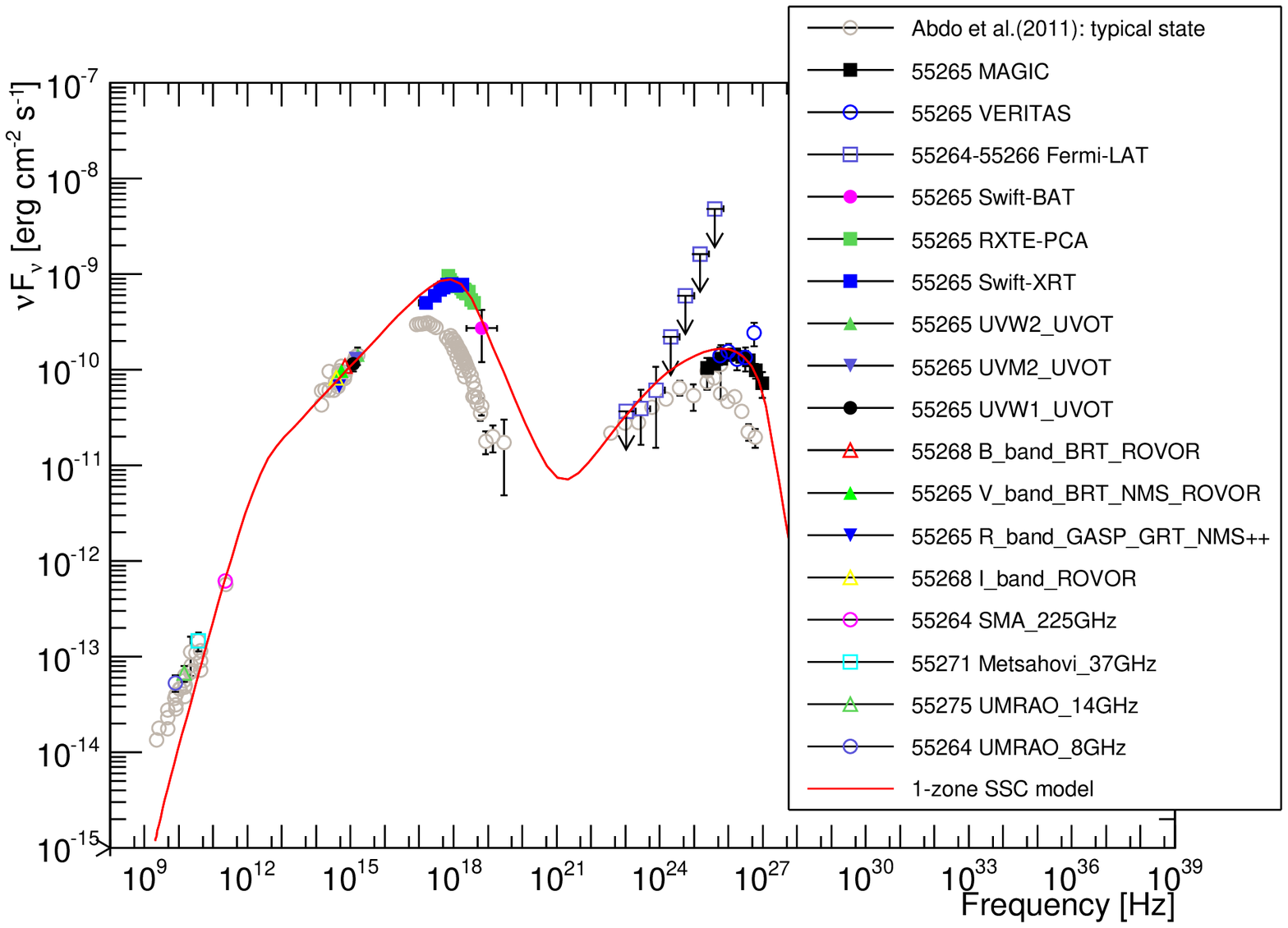}
\caption{Largely simultaneous broadband SED of Mrk~421 on MJD
  55265. The correspondence between markers and instruments is given
  in the legend. The full names of the instruments can be found in
  Table~\ref{TableWithInstruments}. Because of space limitations, 
R-band instruments other than GASP, GRT, and NMS are denoted with the 
symbol "++". Whenever a simultaneous observation is not available, the fluxes from the closest date are reported, and their observation time in MJD is reported next to the instrument name in the legend. The red curve depicts the one-zone SSC model matching the data. The gray circles depict the average SED from the 2009 MW campaign reported in \cite{AbdoMrk421}, which is a good representation of the nonflaring (typical) SED of Mrk~421. 
\label{fig11}}
\end{figure*}

\begin{figure*}
\centering
\begin{subfigure}{.49\textwidth}
  \centering
  \includegraphics[width=1.\linewidth]{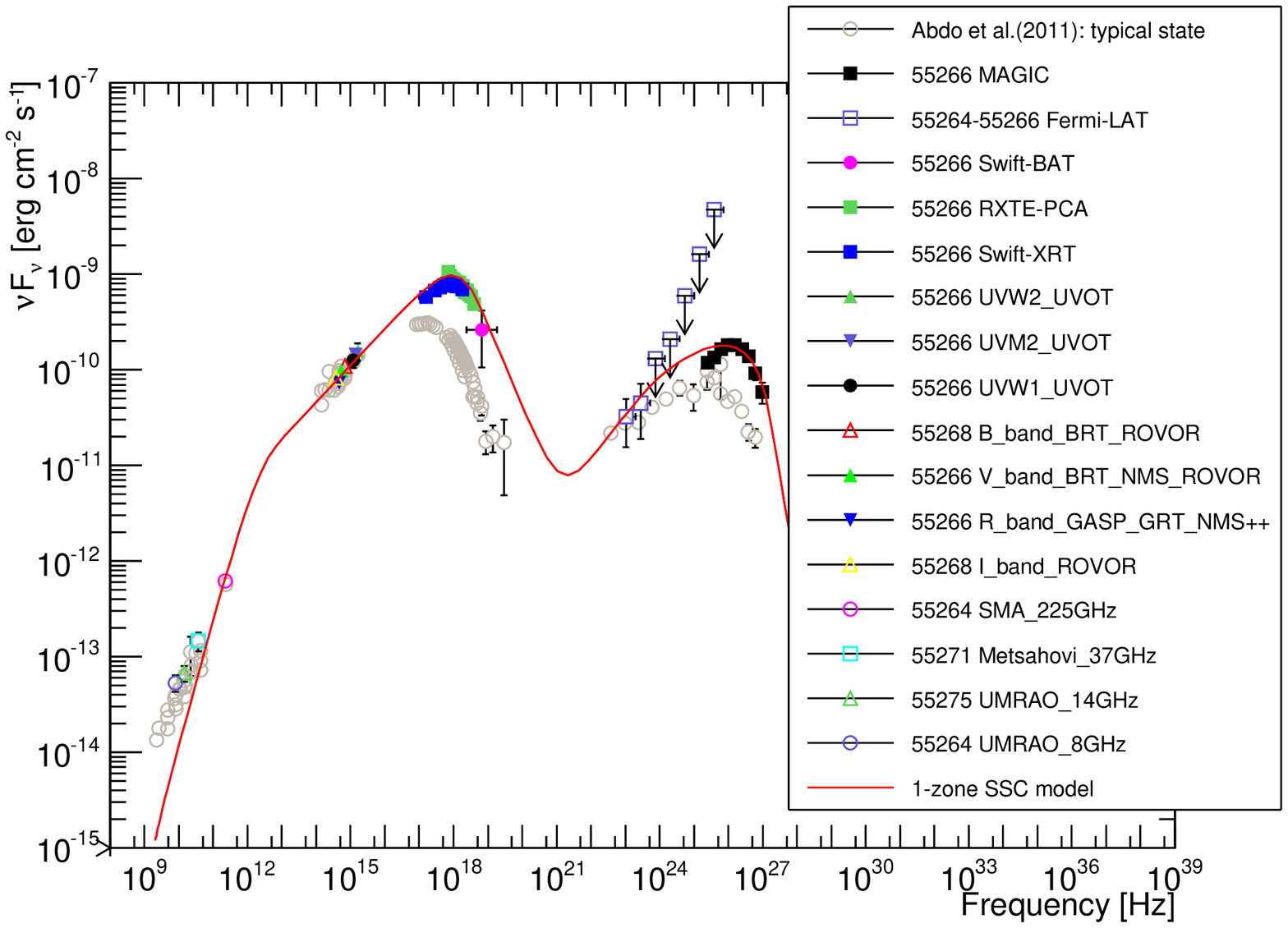}
  \caption{MJD~55266.}
  \label{fig12}
\end{subfigure}
\begin{subfigure}{.49\textwidth}
  \centering
  \includegraphics[width=1.\linewidth]{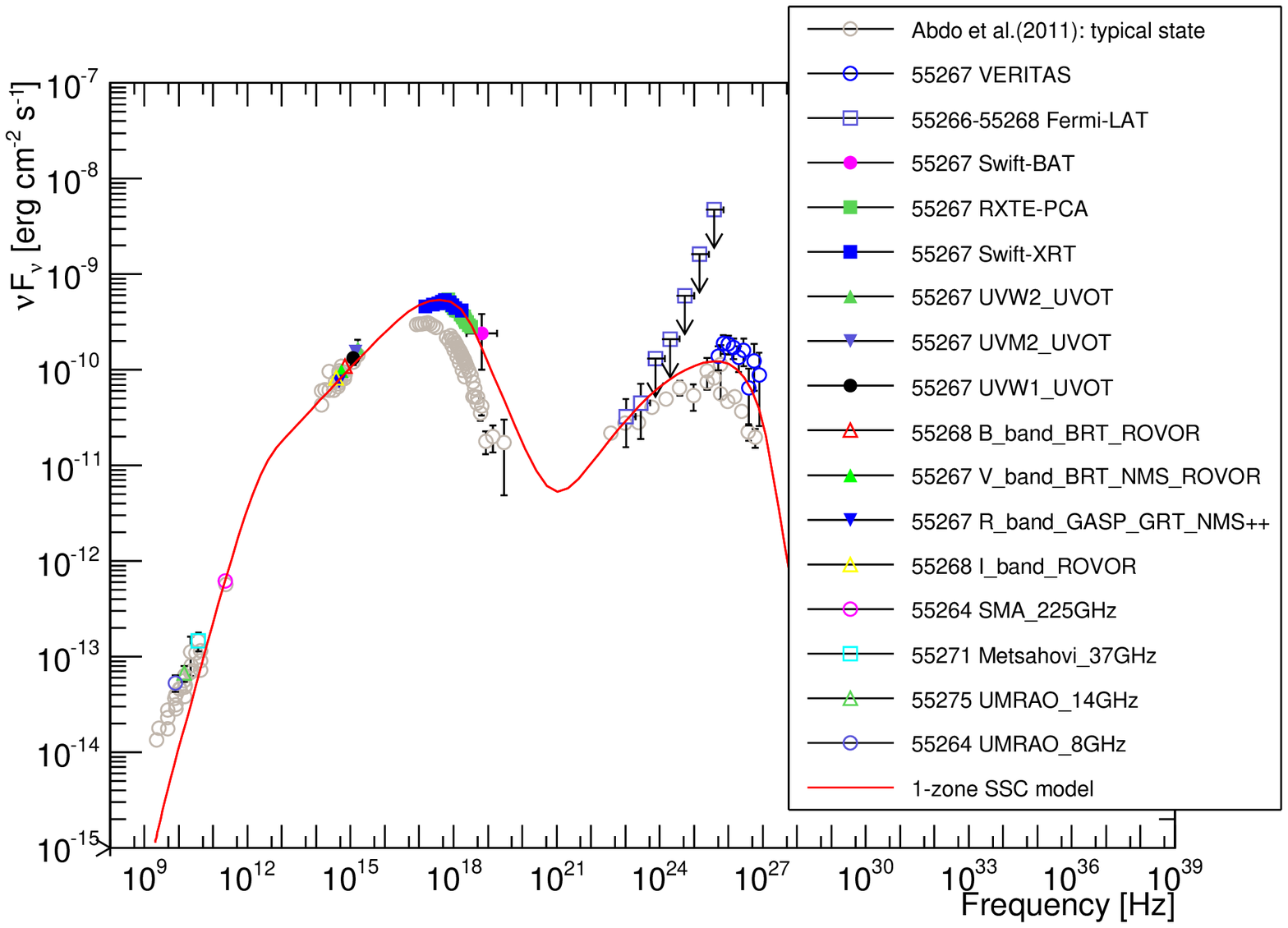}
  \caption{MJD~55267.}
  \label{fig12a}
\end{subfigure}

\begin{subfigure}{.49\textwidth}
  \centering
  \includegraphics[width=1.\linewidth]{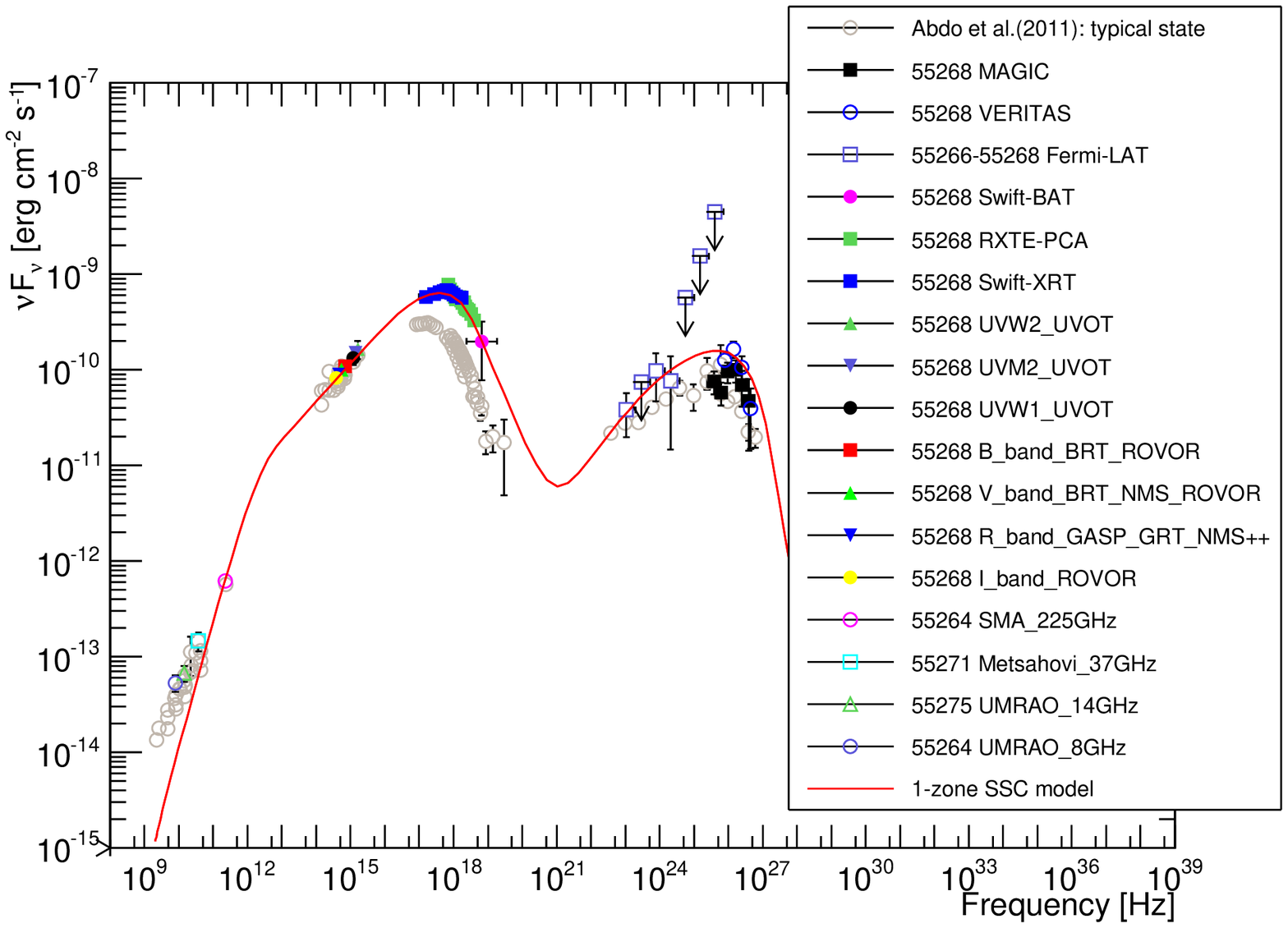}
  \caption{MJD~55268.}
  \label{fig13}
\end{subfigure}
\begin{subfigure}{.49\textwidth}
  \centering
  \includegraphics[width=1.\linewidth]{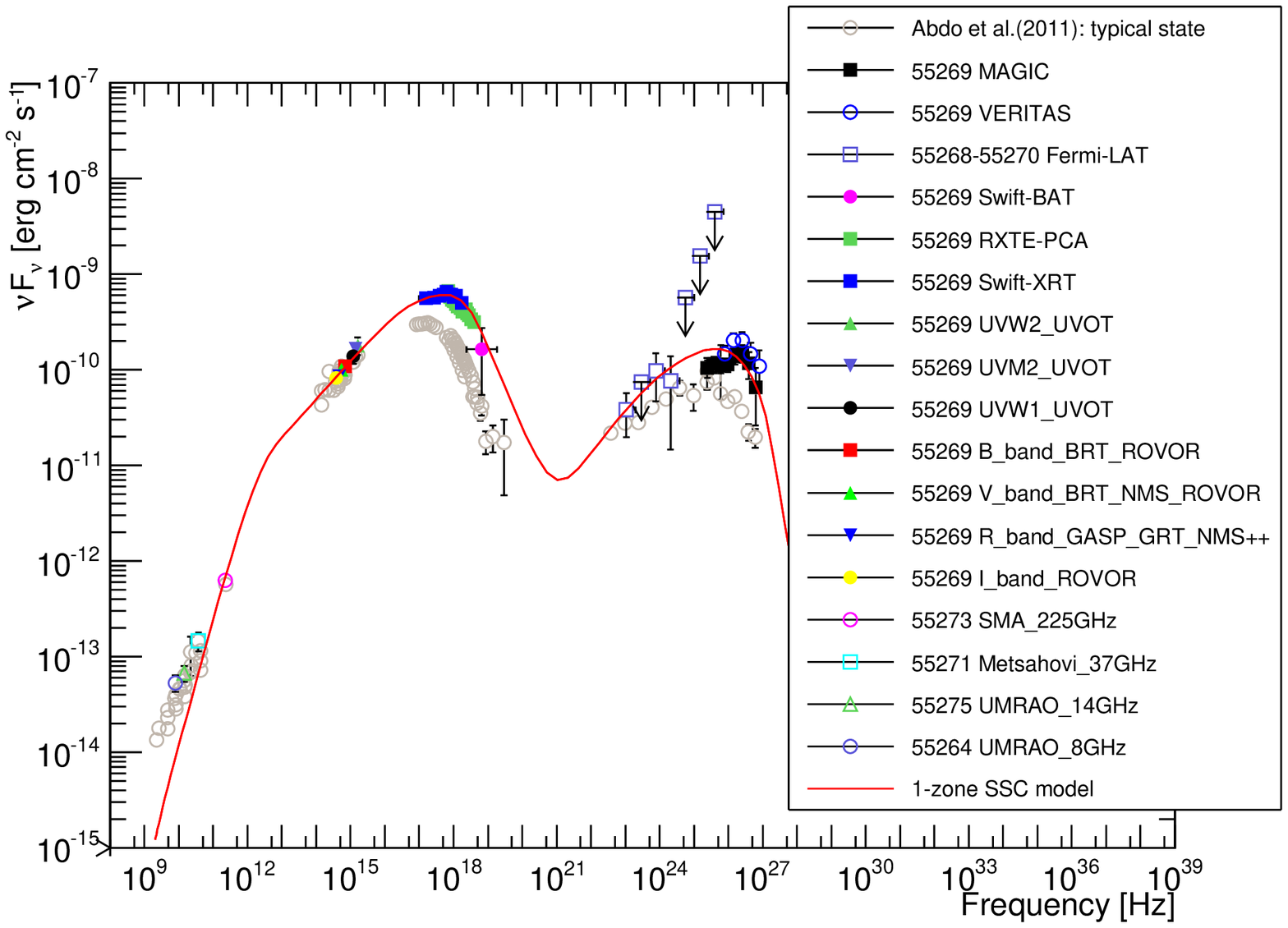}
  \caption{MJD~55269.}
  \label{fig14}
\end{subfigure}

\begin{subfigure}{.49\textwidth}
  \centering
  \includegraphics[width=1.\linewidth]{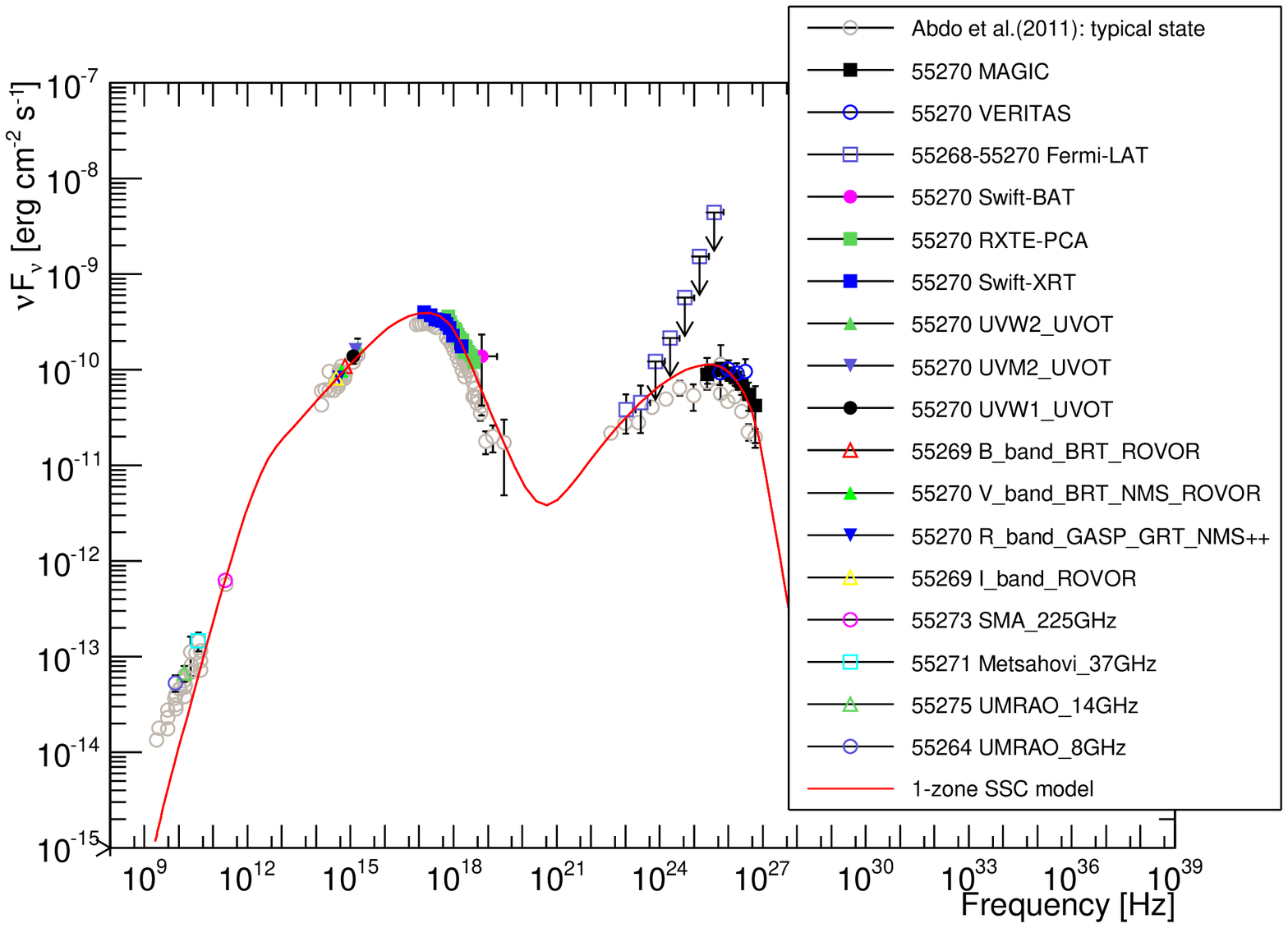}
  \caption{MJD~55270.}
  \label{fig15}
\end{subfigure}
\begin{subfigure}{.49\textwidth}
  \centering
  \includegraphics[width=1.\linewidth]{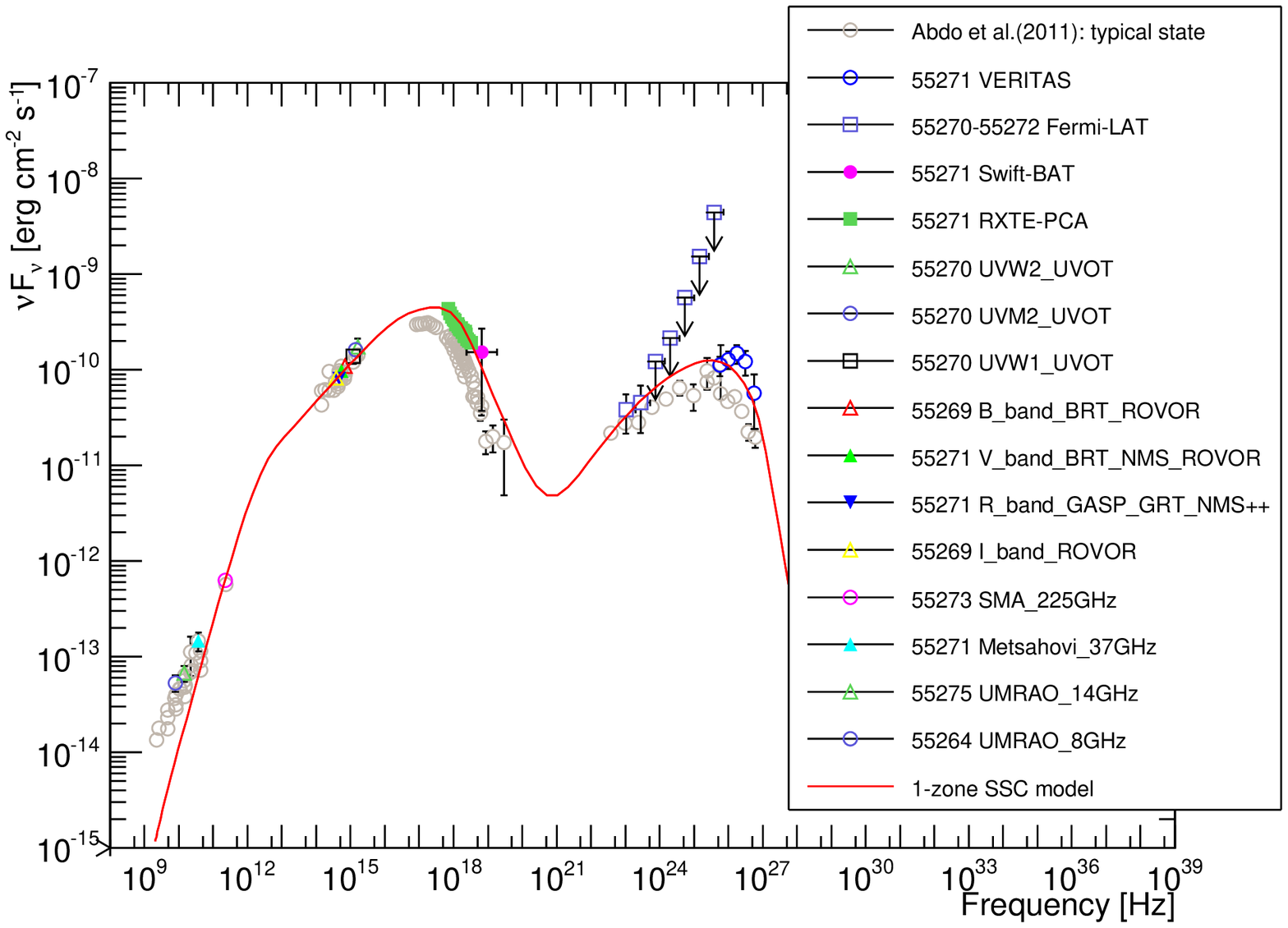}
  \caption{MJD~55271.}
  \label{fig15a}
\end{subfigure}
\caption{Simultaneous broadband SEDs and their one-zone SSC model
fits. See caption of Fig.~\ref{fig11} for further details.}
\label{fig:SEDs1zone:1}
\end{figure*}

\begin{figure*}
\centering
\begin{subfigure}{.49\textwidth}
  \centering
  \includegraphics[width=1.\linewidth]{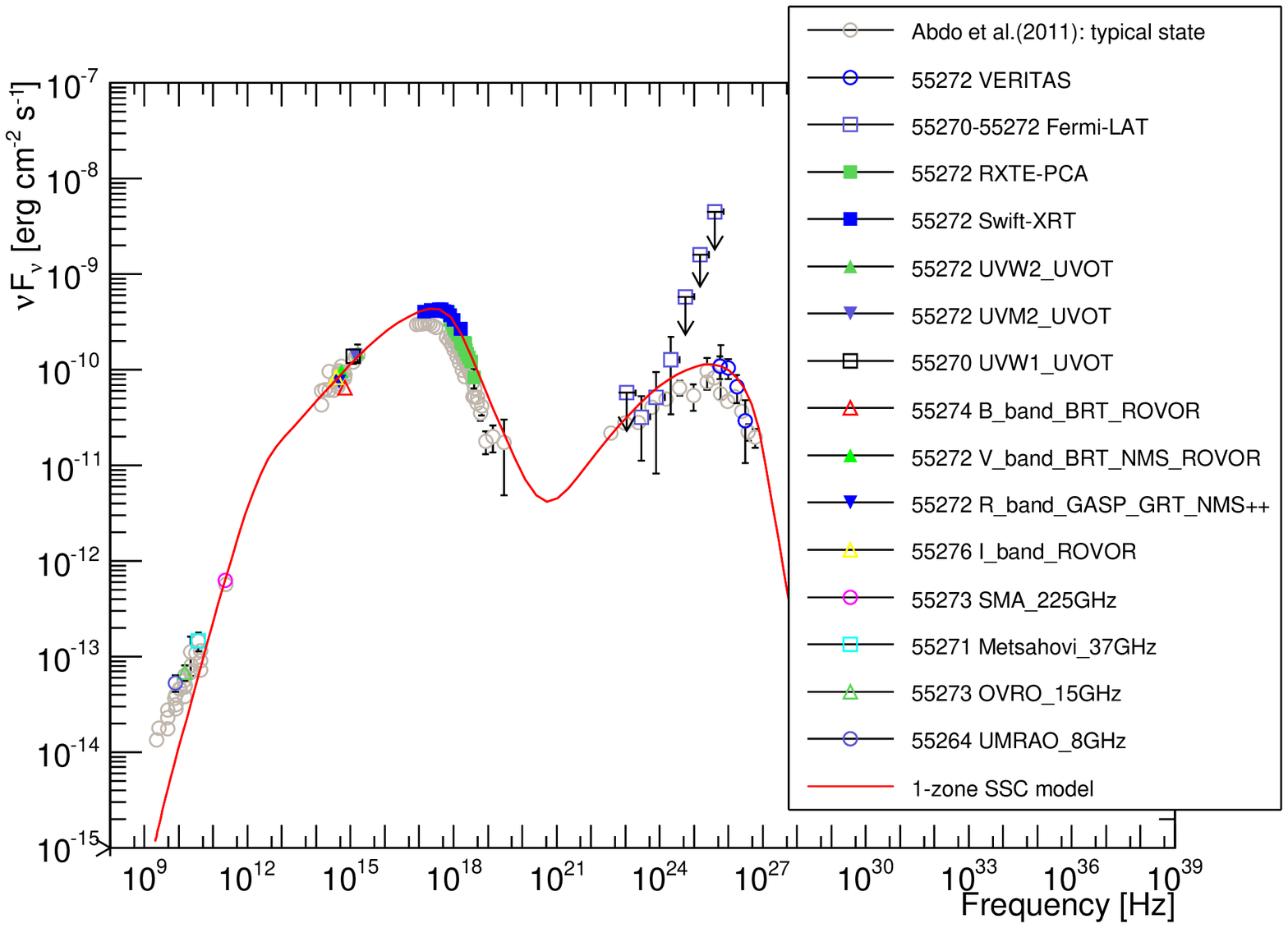}
  \caption{MJD~55272.}
  \label{fig16}
\end{subfigure}
\begin{subfigure}{.49\textwidth}
  \centering
  \includegraphics[width=1.\linewidth]{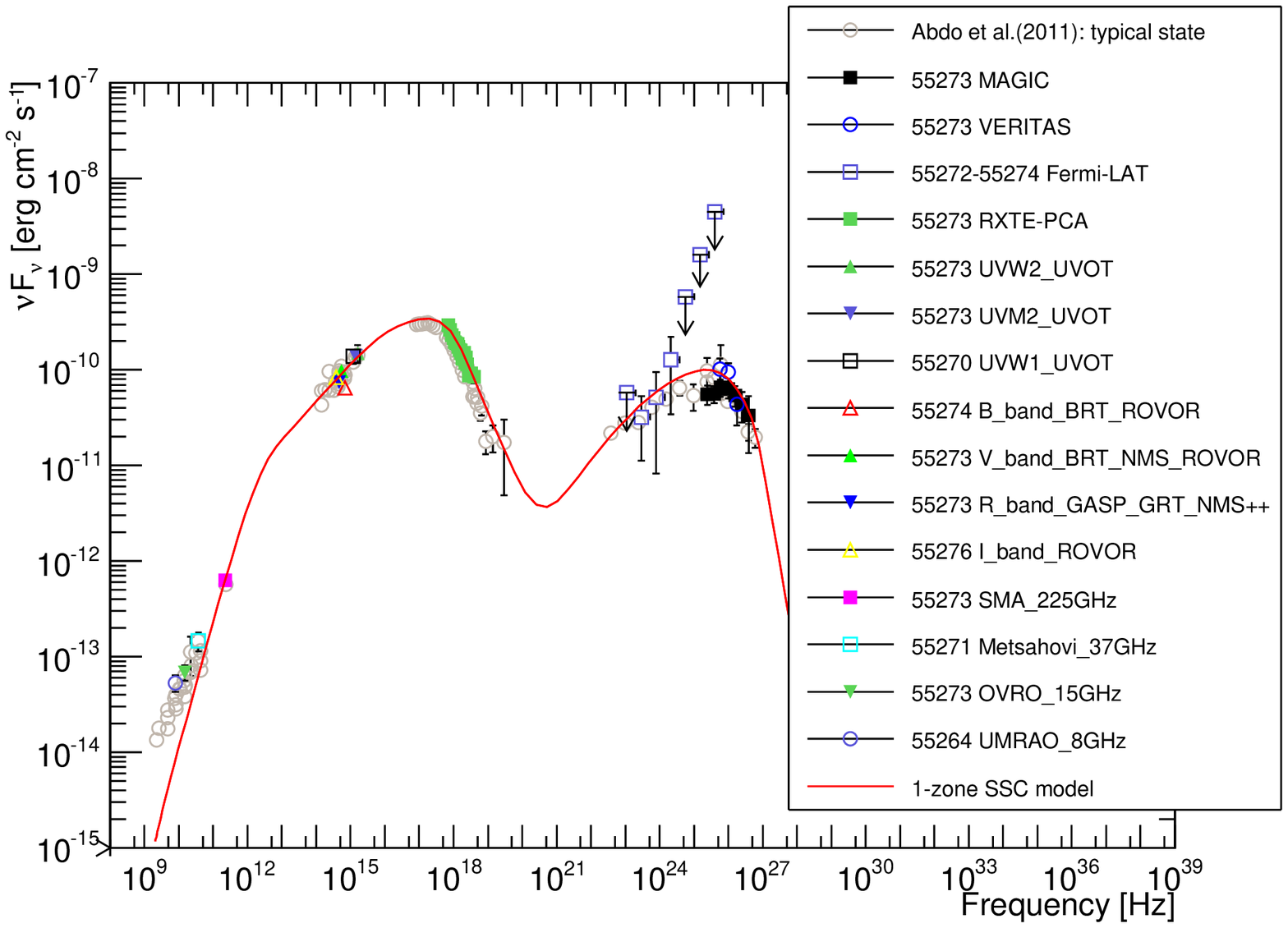}
  \caption{MJD~55273.}
  \label{fig17}
\end{subfigure}

\begin{subfigure}{.49\textwidth}
  \centering
  \includegraphics[width=1.\linewidth]{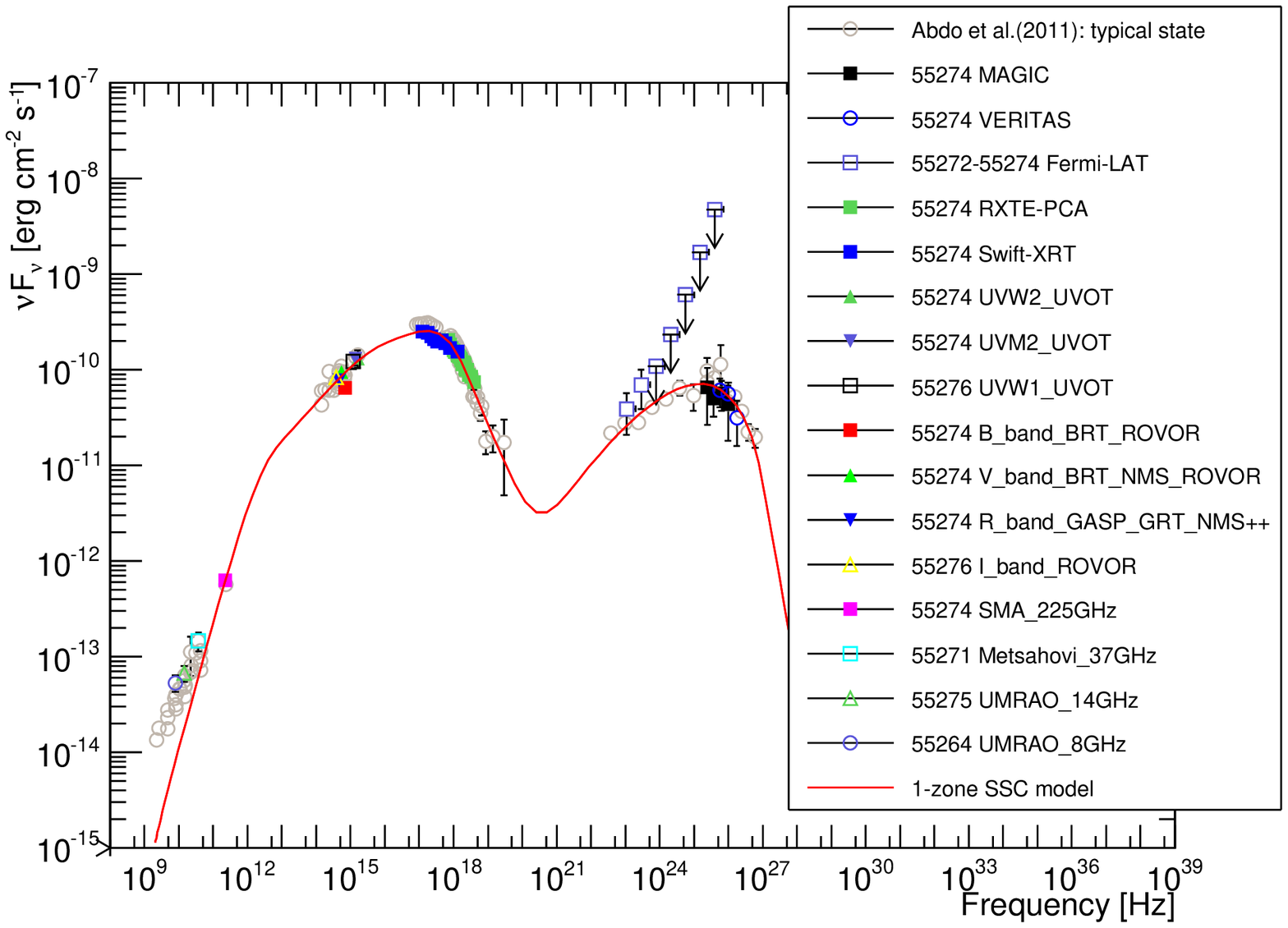}
  \caption{MJD~55274.}
  \label{fig18}
\end{subfigure}
\begin{subfigure}{.49\textwidth}
  \centering
  \includegraphics[width=1.\linewidth]{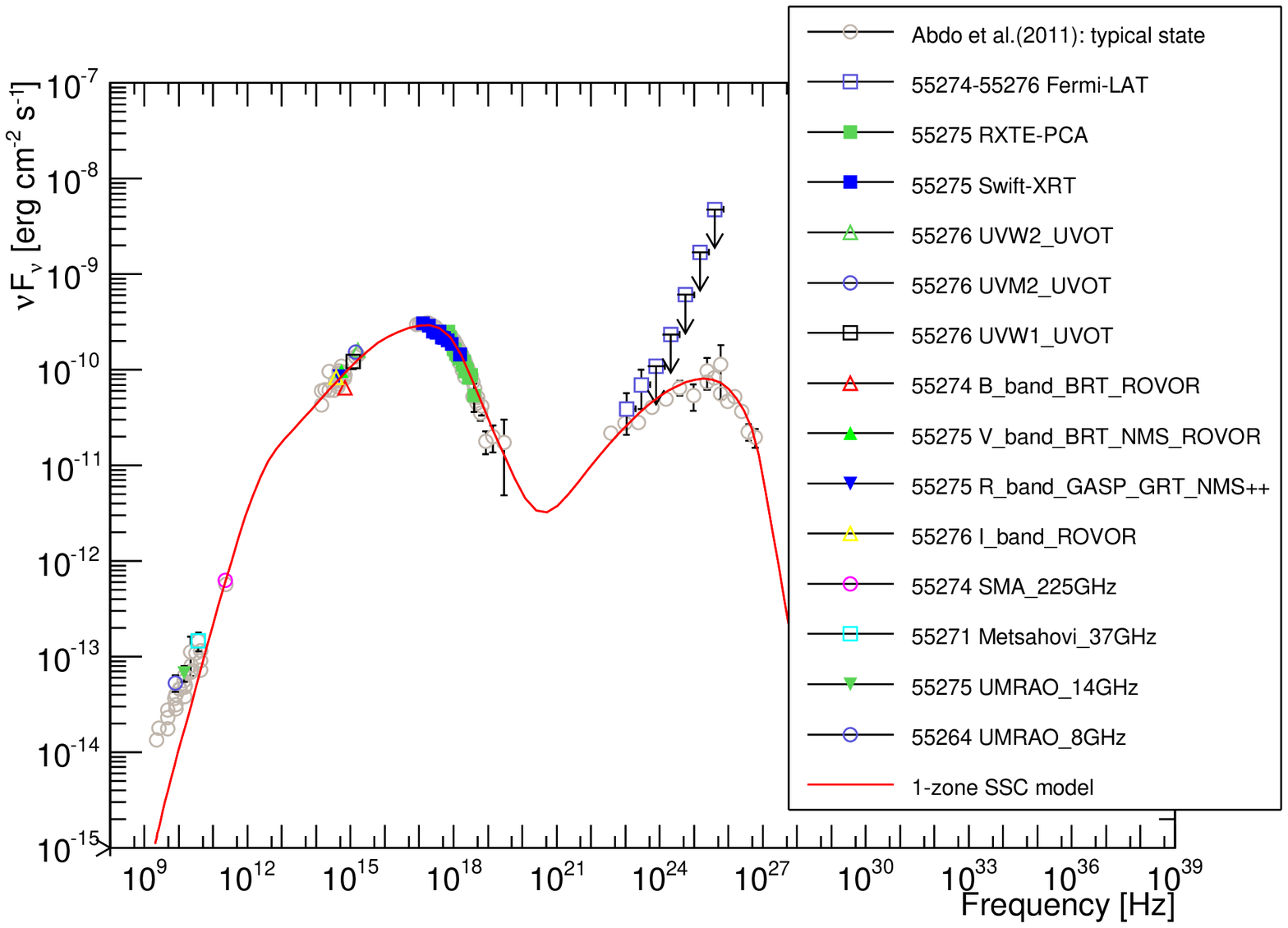}
  \caption{MJD~55275.}
  \label{fig18a}
\end{subfigure}

\begin{subfigure}{.49\textwidth}
  \centering
  \includegraphics[width=1.\linewidth]{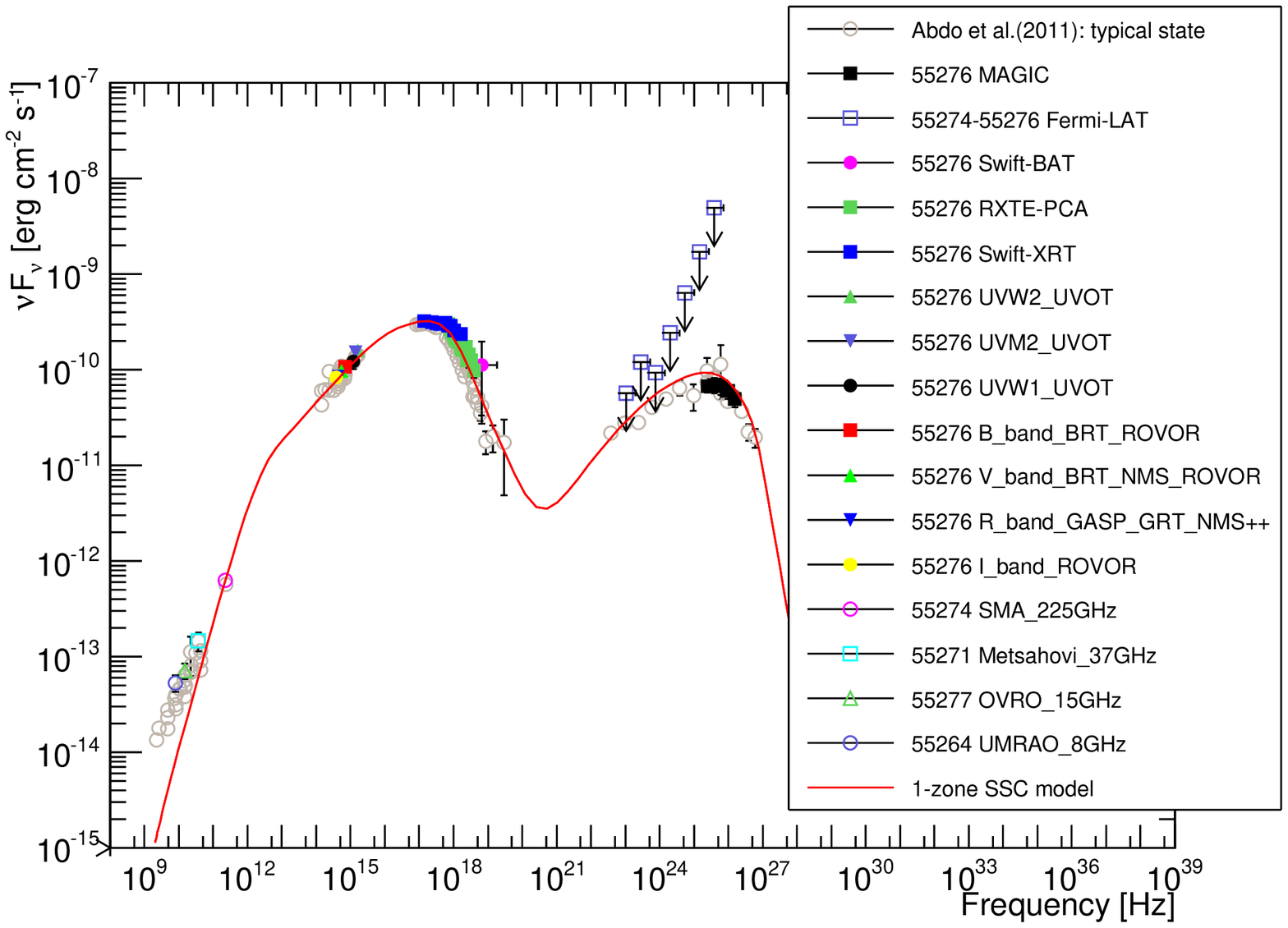}
  \caption{MJD~55276.}
  \label{fig20}
\end{subfigure}
\begin{subfigure}{.49\textwidth}
  \centering
  \includegraphics[width=1.\linewidth]{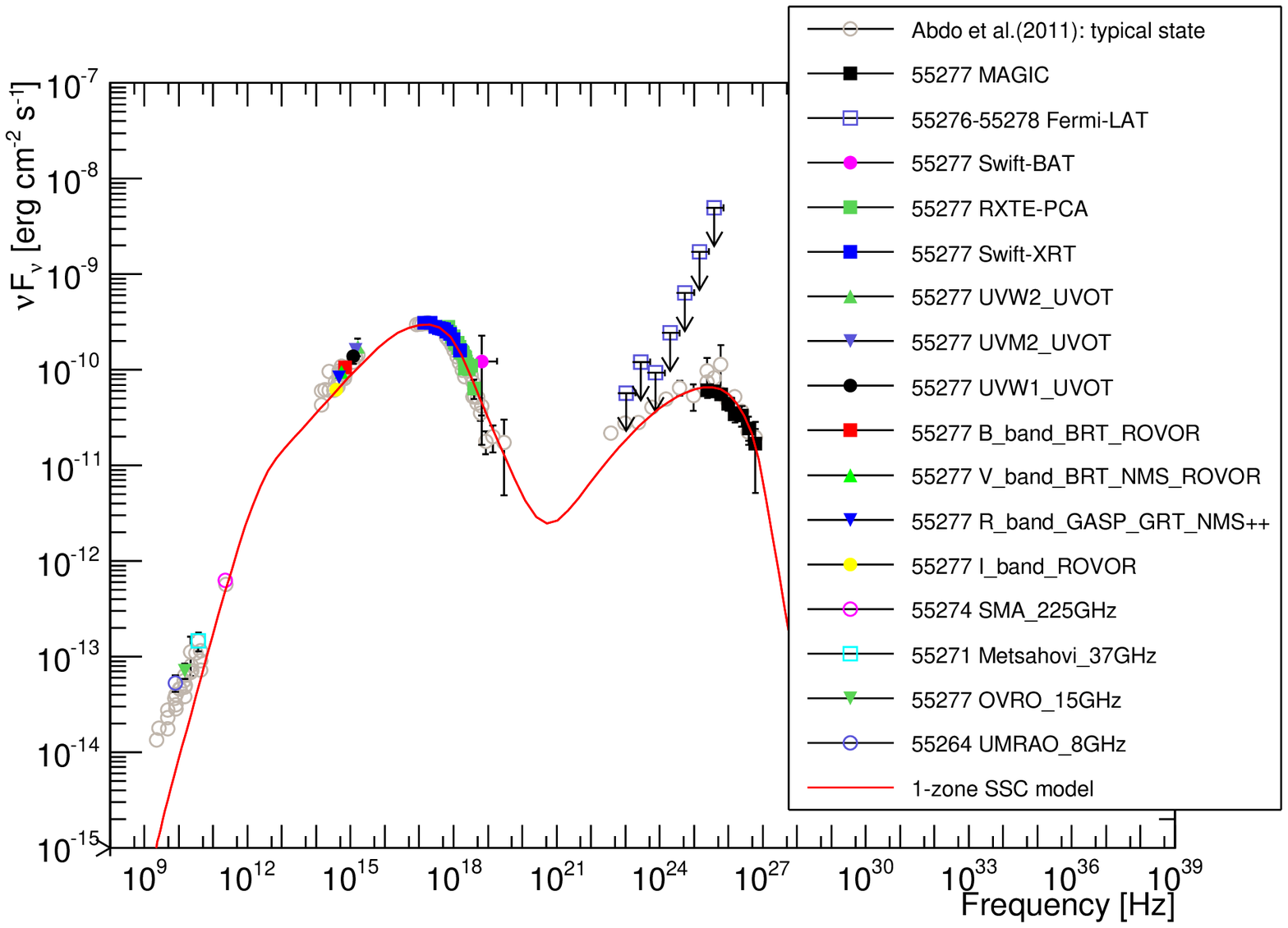}
  \caption{MJD~55277.}
  \label{fig21}
\end{subfigure}
\caption{Simultaneous broadband SEDs and their one-zone SSC model
fits. See caption of Fig.~\ref{fig11} for further details.}
\label{fig:SEDs1zone:2}
\end{figure*}

\begin{figure*}
\centering
\begin{subfigure}{.49\textwidth}
  \centering
  \includegraphics[width=1.\linewidth]{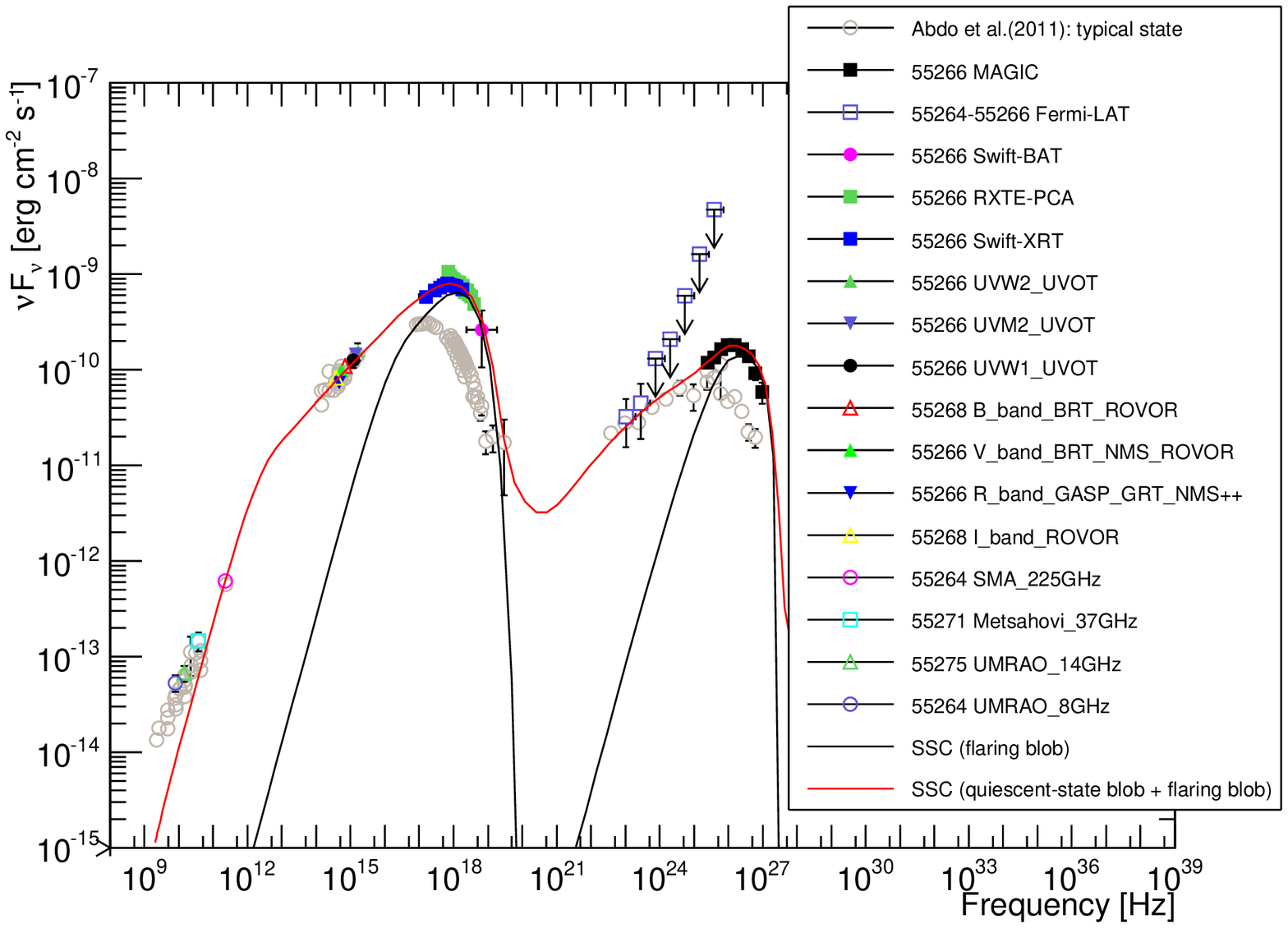}
  \caption{MJD~55266.}
  \label{fig12b}
\end{subfigure}
\begin{subfigure}{.49\textwidth}
  \centering
  \includegraphics[width=1.\linewidth]{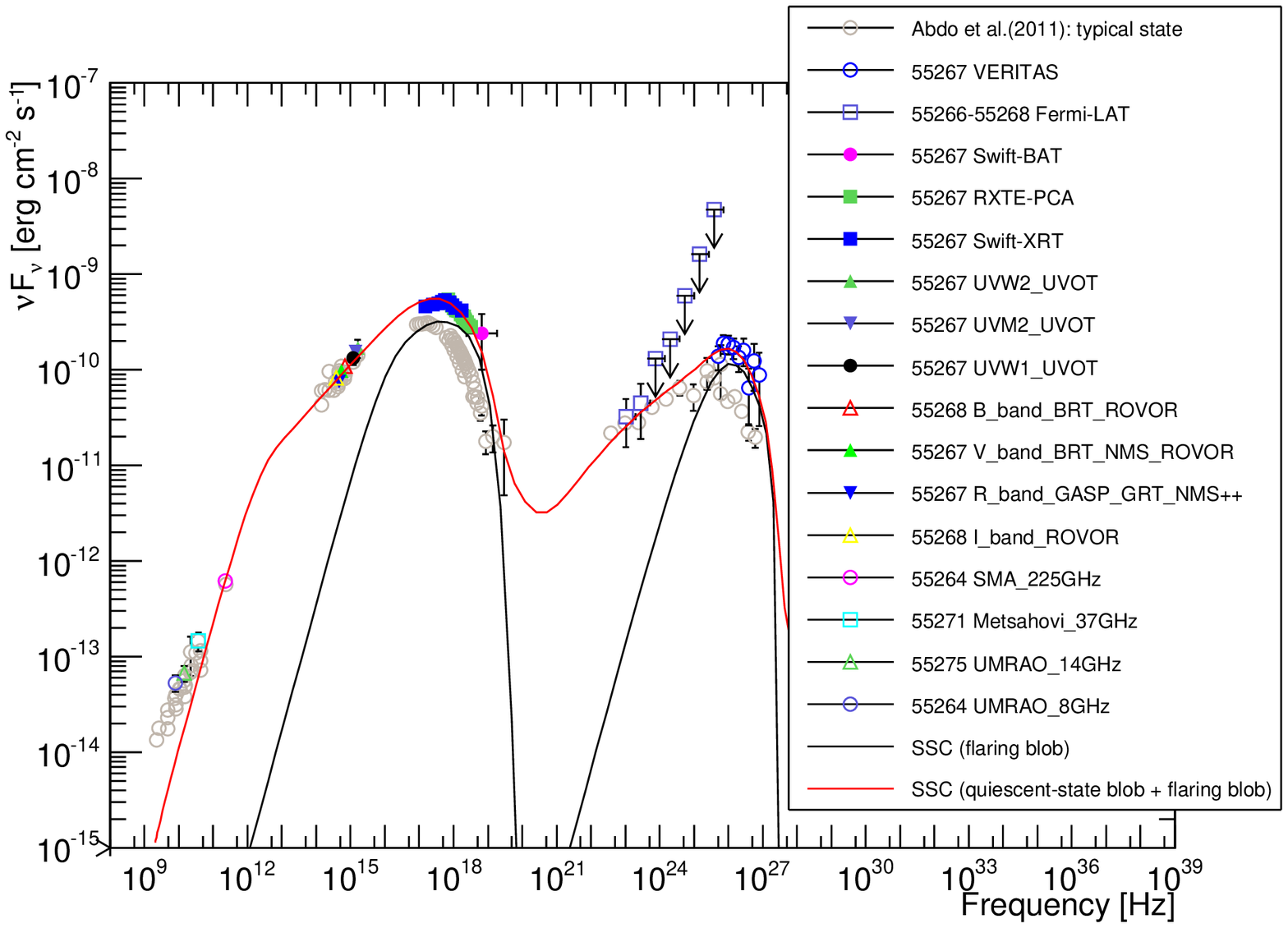}
  \caption{MJD~55267.}
  \label{fig12bb}
\end{subfigure}

\begin{subfigure}{.49\textwidth}
  \centering
  \includegraphics[width=1.\linewidth]{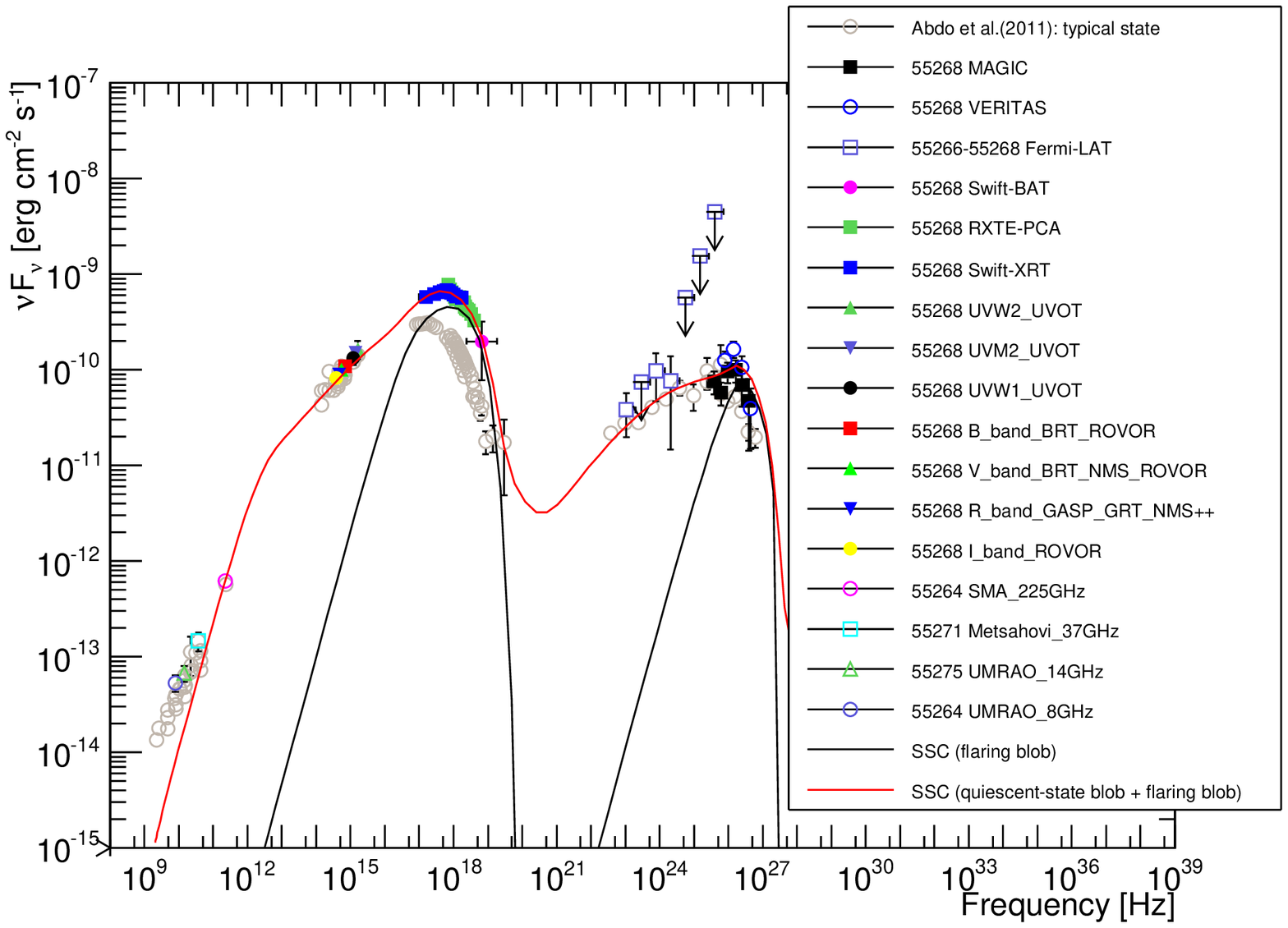}
  \caption{MJD~55268}
  \label{fig13b}
\end{subfigure}
\begin{subfigure}{.49\textwidth}
  \centering
  \includegraphics[width=1.\linewidth]{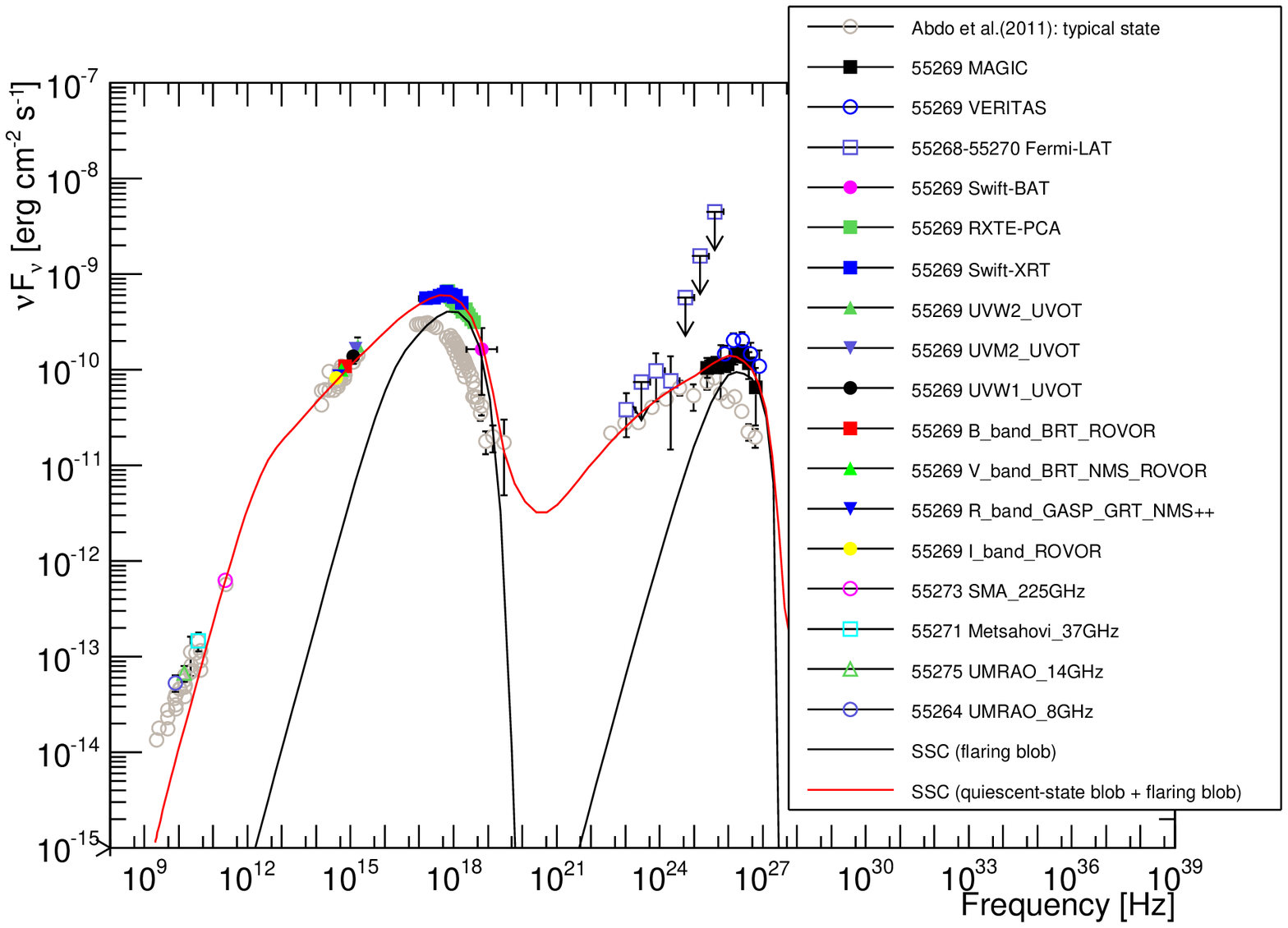}
  \caption{MJD~55269.}
  \label{fig14b}
\end{subfigure}

\begin{subfigure}{.49\textwidth}
  \centering
  \includegraphics[width=1.\linewidth]{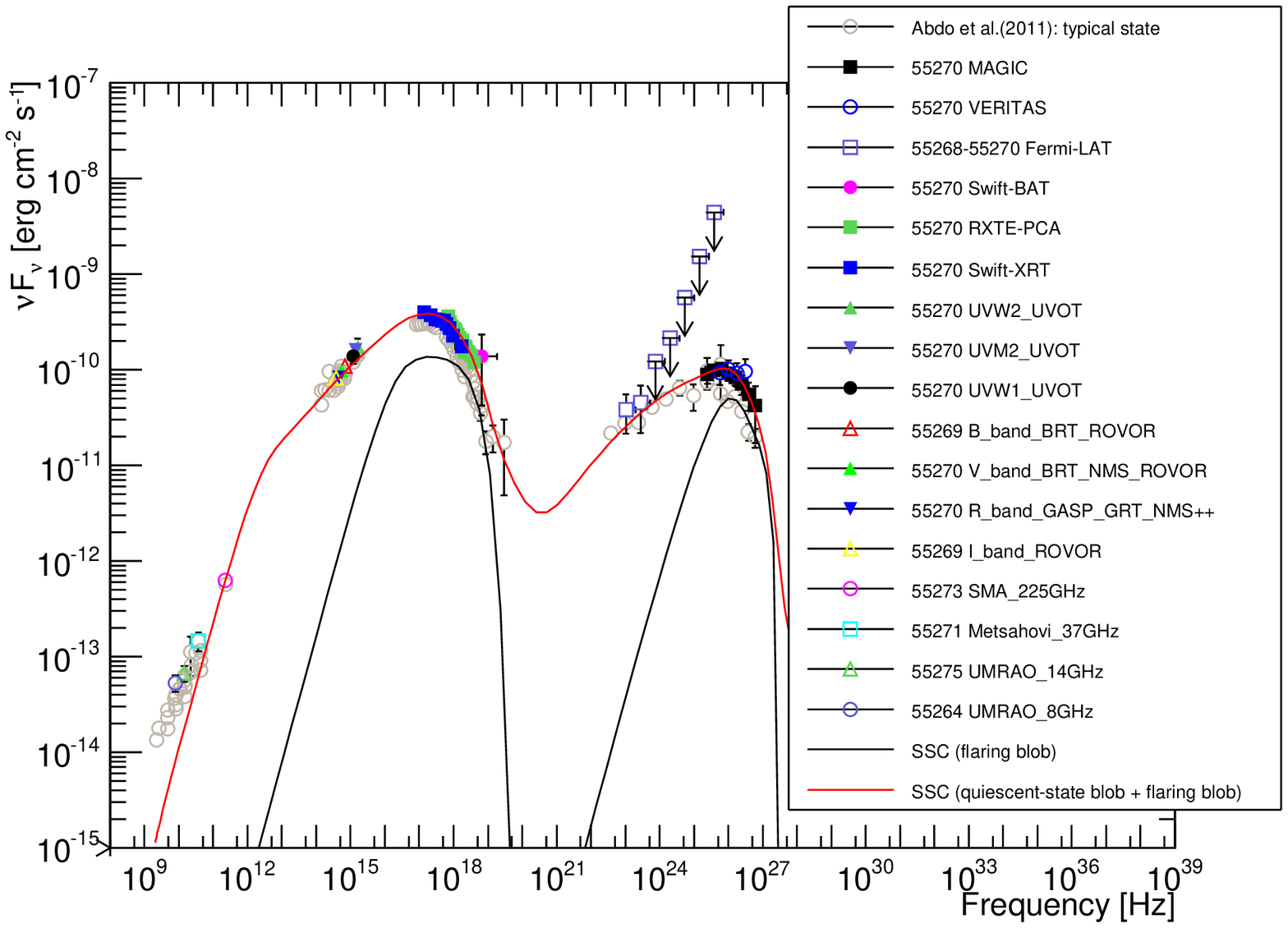}
  \caption{MJD~55270.}
  \label{fig15b}
\end{subfigure}
\begin{subfigure}{.49\textwidth}
  \centering
  \includegraphics[width=1.\linewidth]{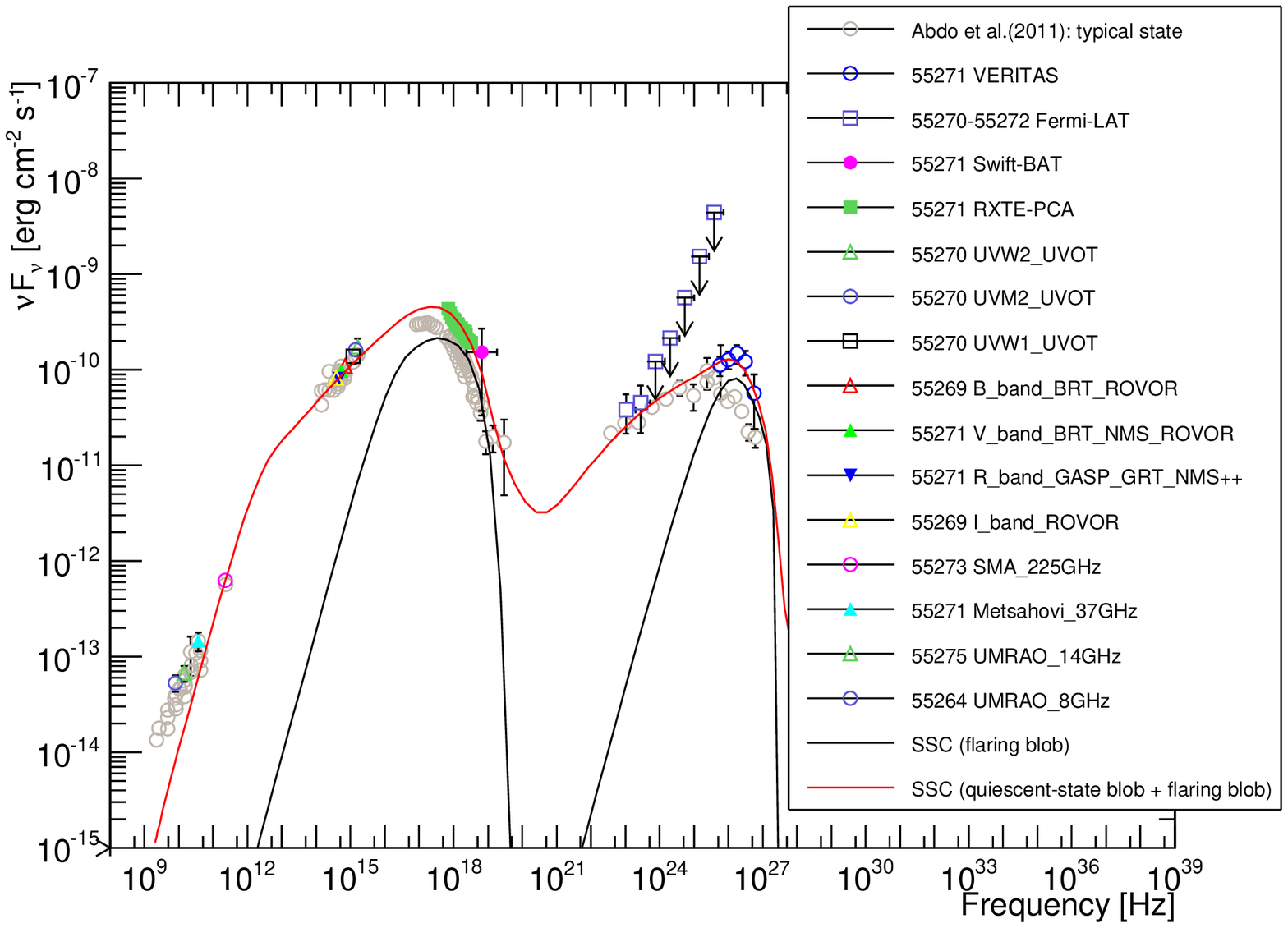}
  \caption{MJD~55271.}
  \label{fig15bb}
\end{subfigure}
\caption{Simultaneous broadband SEDs and their two-zone SSC model fits. See caption of Fig.~\ref{fig11b} for further details.}
\label{fig:SEDs2zone:1}
\end{figure*}

\begin{figure*}
\centering
\begin{subfigure}{.49\textwidth}
  \centering
  \includegraphics[width=1.\linewidth]{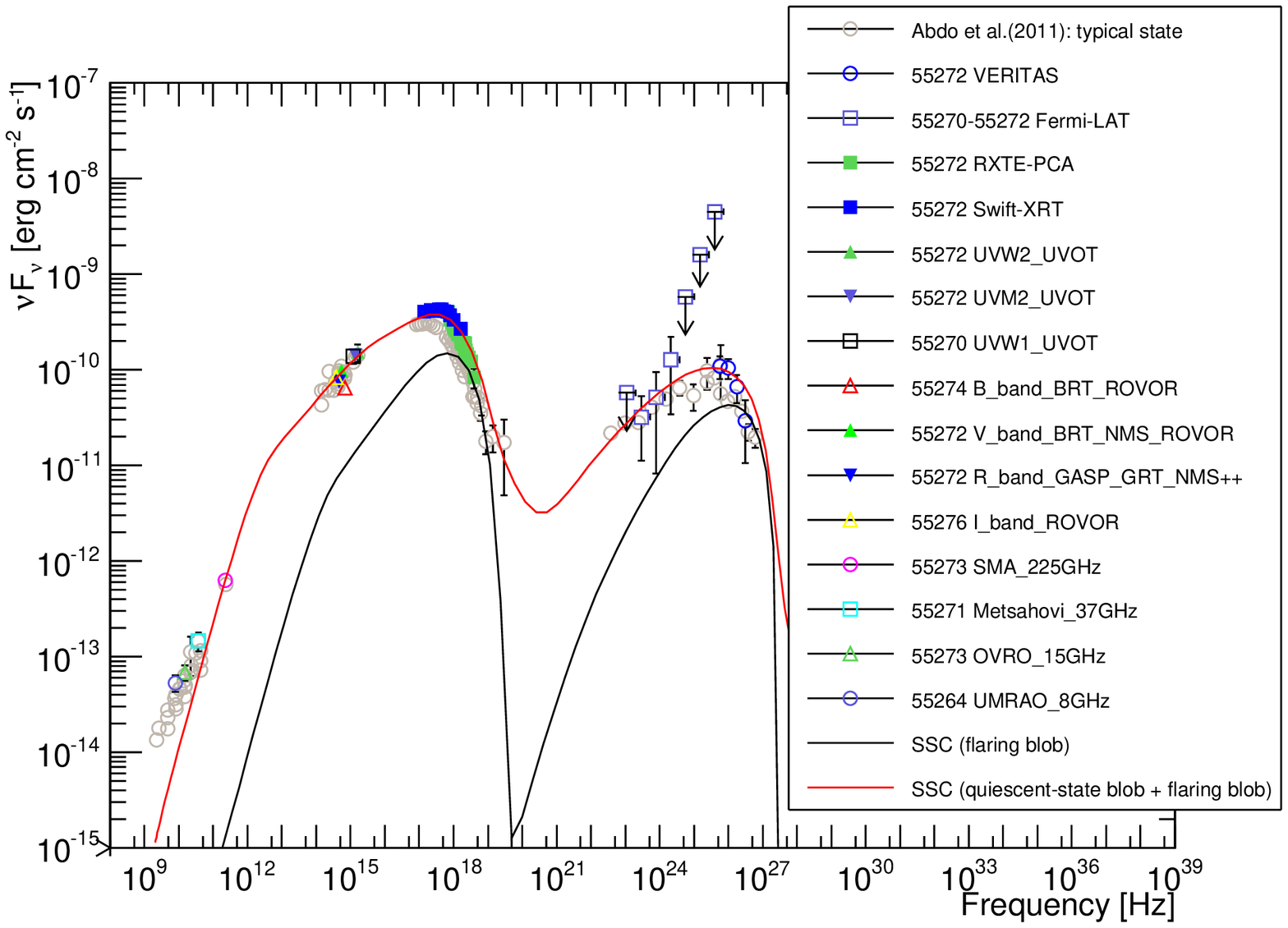}
  \caption{MJD~55272}
  \label{fig15bbb}
\end{subfigure}
\begin{subfigure}{.49\textwidth}
  \centering
  \includegraphics[width=1.\linewidth]{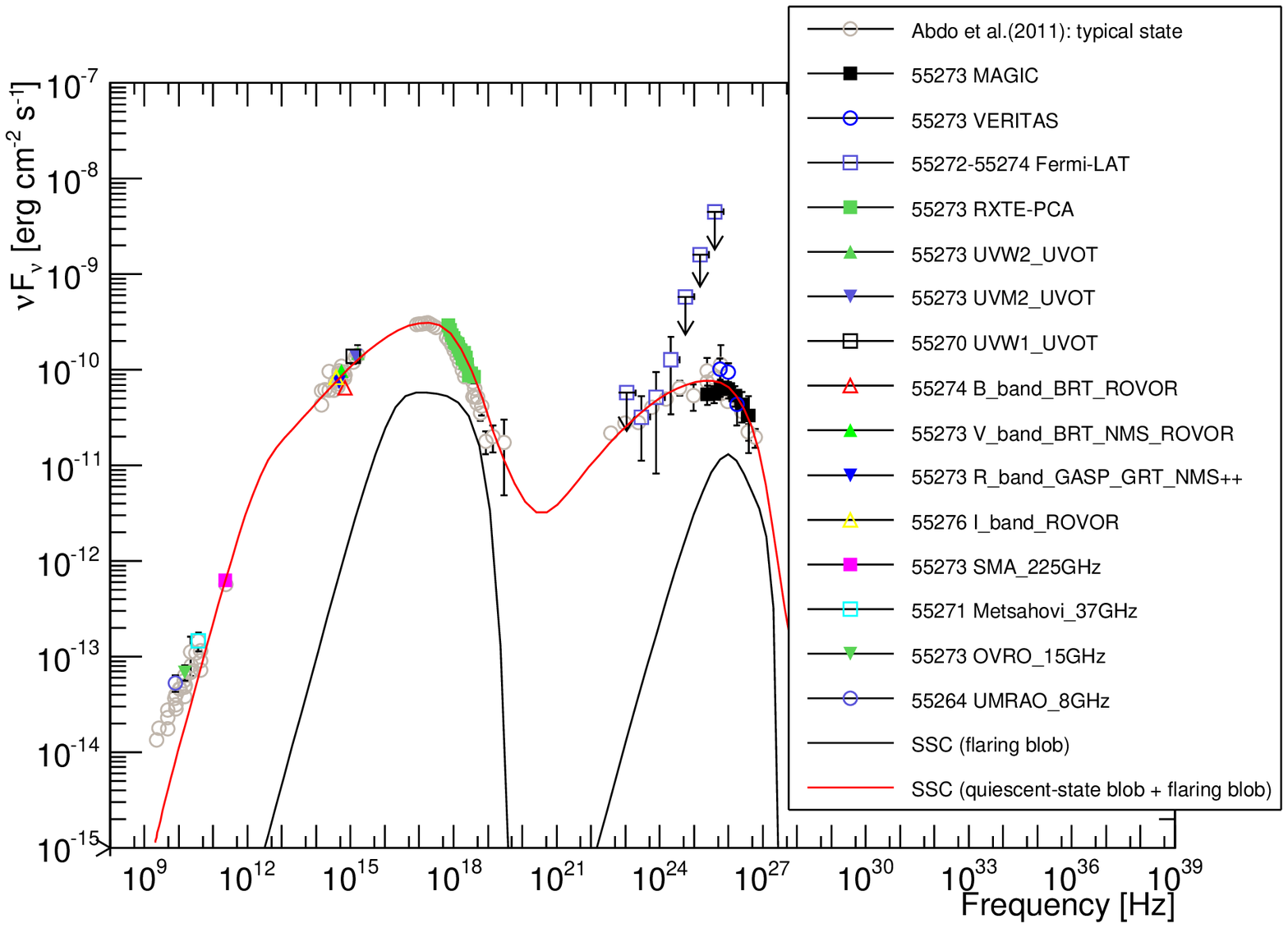}
  \caption{MJD~55273.}
  \label{fig17b}
\end{subfigure}

\begin{subfigure}{.49\textwidth}
  \centering
  \includegraphics[width=1.\linewidth]{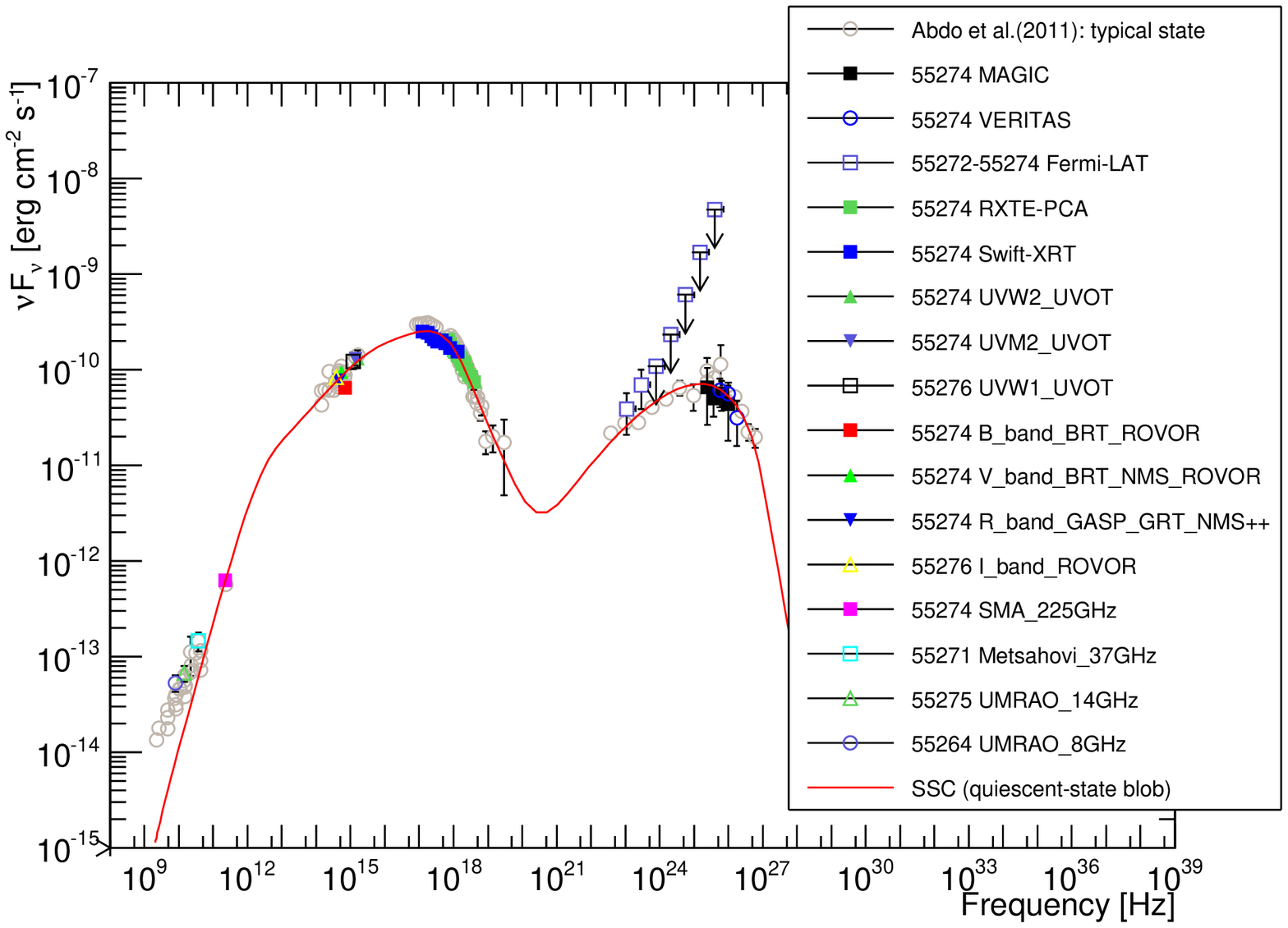}
  \caption{MJD~55274.}
  \label{fig18b}
\end{subfigure}
\begin{subfigure}{.49\textwidth}
  \centering
  \includegraphics[width=1.\linewidth]{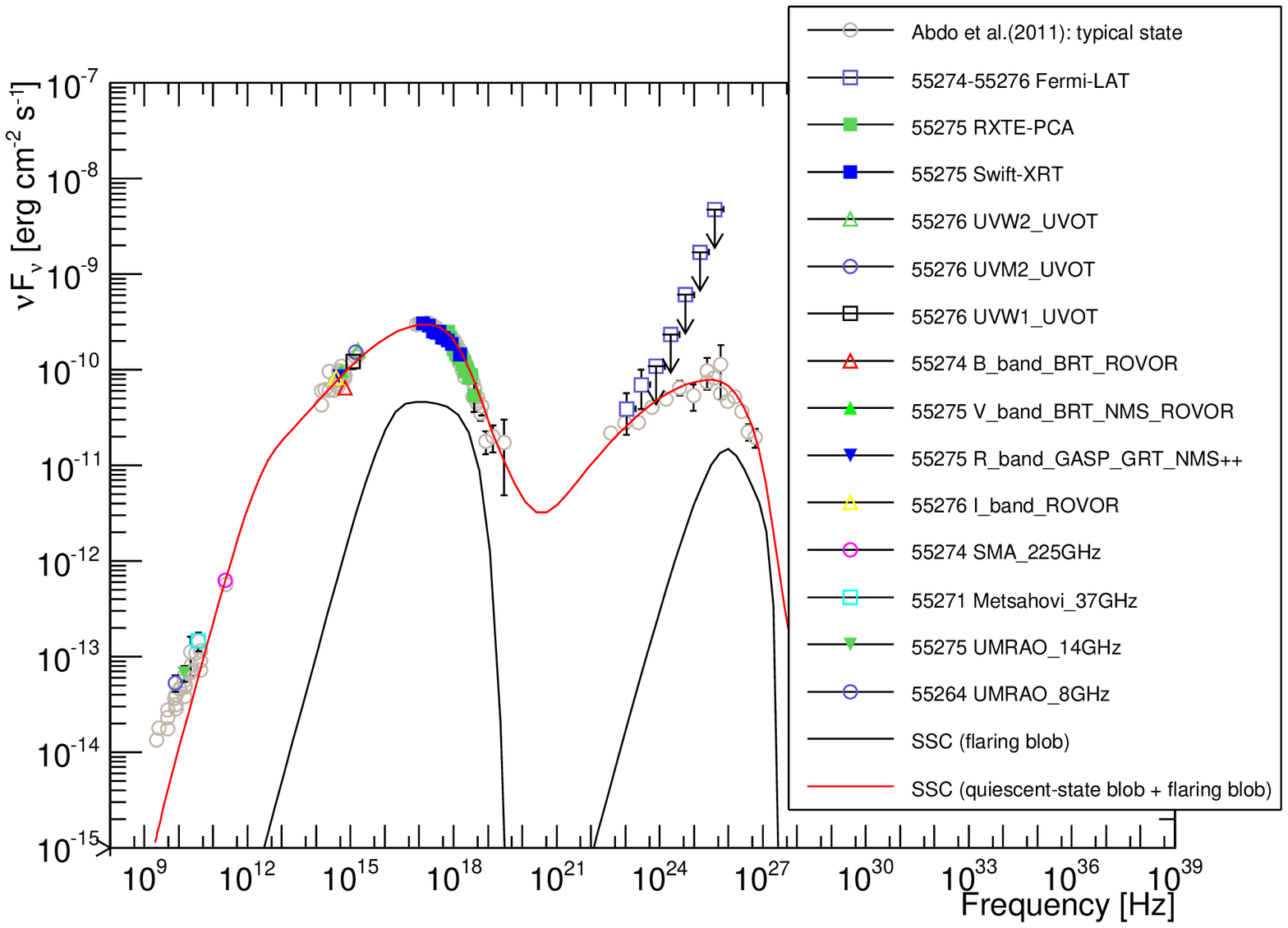}
  \caption{MJD~55275.}
  \label{fig18bb}
\end{subfigure}

\begin{subfigure}{.49\textwidth}
  \centering
  \includegraphics[width=1.\linewidth]{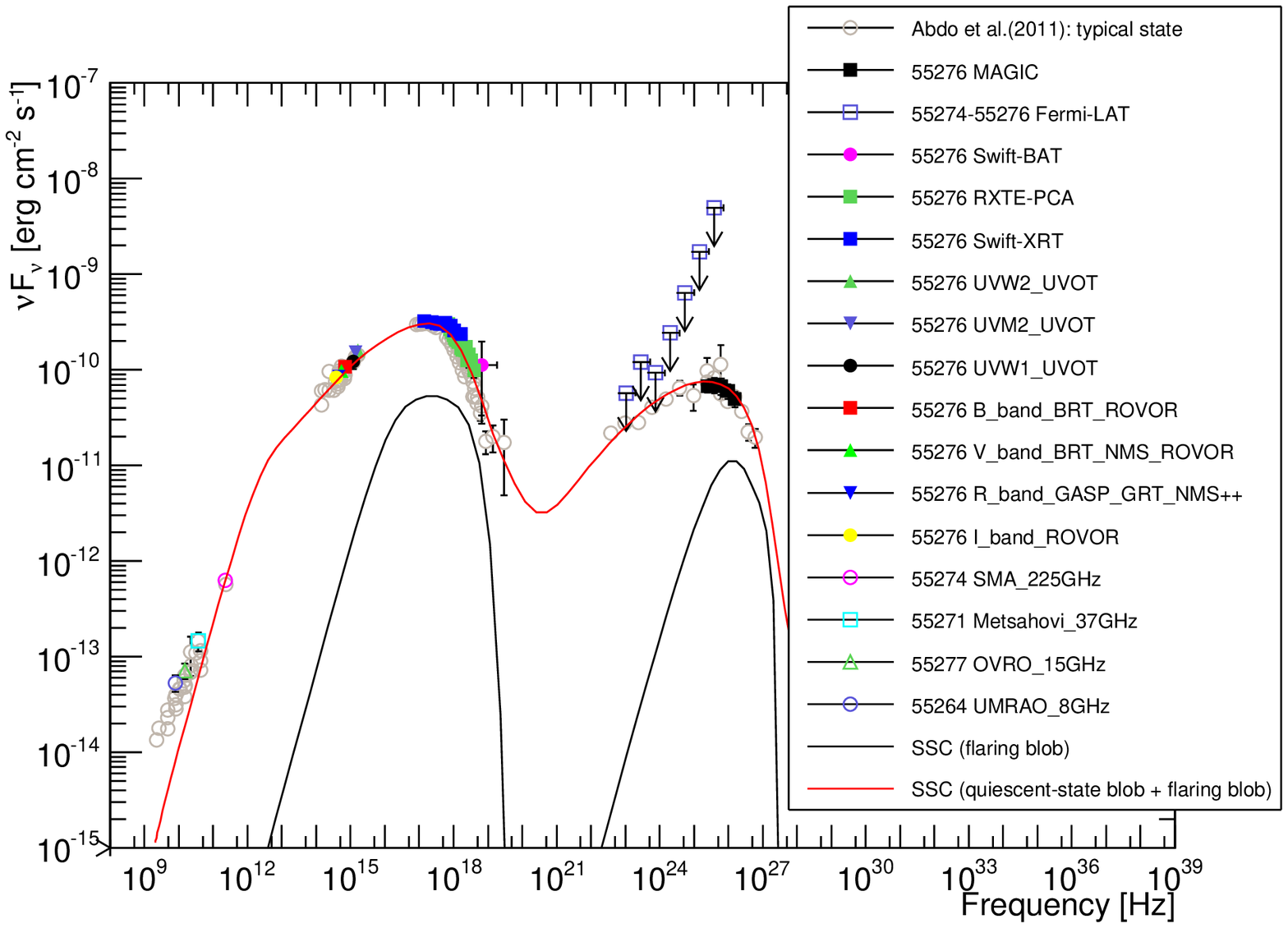}
  \caption{MJD~55276.}
  \label{fig20b}
\end{subfigure}
\begin{subfigure}{.49\textwidth}
  \centering
  \includegraphics[width=1.\linewidth]{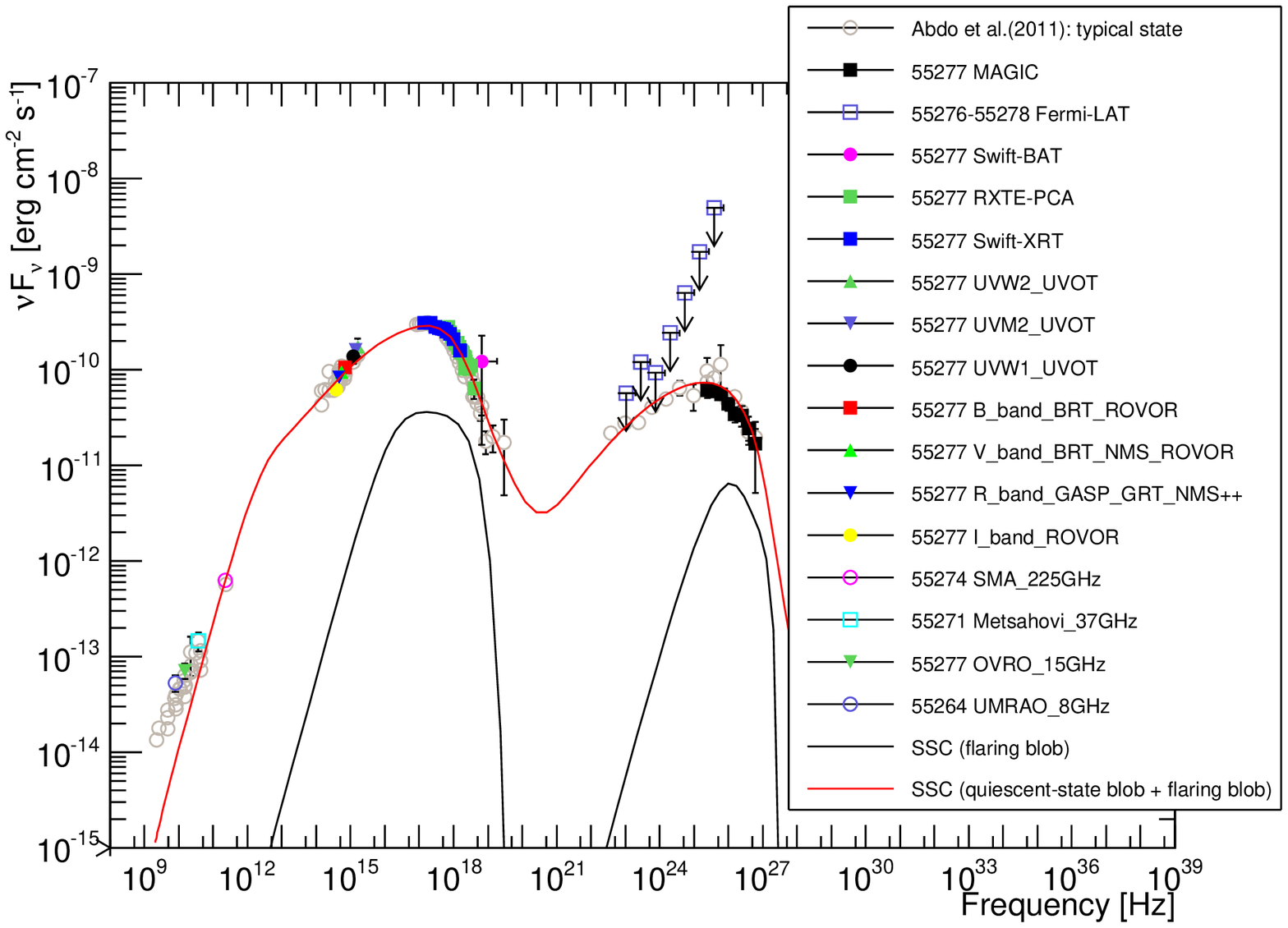}
  \caption{MJD~55277.}
  \label{fig21b}
\end{subfigure}
\caption{Simultaneous broadband SEDs and their two-zone SSC model fits. See caption of Fig.~\ref{fig11b} for further details. The emission of the quiescent blob was set to the one describing the SED from MJD~55274, which is the lowest SED among all the 13 dates considered in this paper.  Consequently, there is no flaring blob emission for MJD~55274.}
\label{fig:SEDs2zone:2}
\end{figure*}

The measured SEDs for these 13 consecutive days are shown in Figs.~\ref{fig11} to \ref{fig:SEDs2zone:2} with one-zone SSC model curves (from Figs.~\ref{fig11} to Fig.~\ref{fig:SEDs1zone:2}) and two-zone SSC model curves (Figs.~\ref{fig:SEDs2zone:1} and \ref{fig:SEDs2zone:2}). The SED with a two-zone SSC model curve measured on the first day (MJD 55265) was shown in Fig.~\ref{fig11b} in the main text. For comparison, the average SED from the 2009 MW campaign \citep{AbdoMrk421} is shown in all the figures, which is a good representation of the SED of Mrk~421 during its nonflaring (typical) state. The details of the models and the characterization of the SED evolution were discussed in Sects.~\ref{Model1} and \ref{Model2} in the main text.

The actual MJD date for each data entry is given in the legend of each
figure. For optical bands, the reported SED data points correspond to
the averaged values (host-galaxy subtracted) for the specified
observing night. As reported in Sect.~\ref{LightCurves}, the
variability at the optical band is small and occurs on timescales
of several days. Therefore, if there was no instrument observing at a
particular optical energy band, then the nearest observation was used, and the corresponding MJD date is described
in the legend of the figure.

Although Mrk\,421 is cosmologically nearby, at a redshift of 0.03, the absorption of $\gamma$-rays by the extragalactic background light (EBL) is not negligible at TeV energies. The VHE spectra are corrected (de-absorbed) with the EBL model provided by \cite{franc08}, where $e^{-\tau_{\gamma\gamma}} = 0.58$ at 4~TeV. At this energy, which is roughly the highest energy bin in the VHE spectra, most models provide $0.5 < e^{-\tau_{\gamma\gamma}} < 0.6$, such as models from \cite{kneiske04}, \cite{finke10}, and \cite{Dominguez11}, which means that the results are not sensitive to the particular published EBL model that we selected.

\clearpage

%%%%%%%%%%%%%%%%%%%%%%%%%%%%%%%%%%%%%%%%%%%%%%%%%%%%%%%%%%%%
%%%%%%%%%%%%%%%%%%%%%%%%%%%%%%%%%%%%%%%%%%%%%%%%%%%%%%%%%%%%
\section{Physical parameters derived from one-zone and two-zone SSC scenarios}
\label{sec:phys_parameters}
%%%%%%%%%%%%%%%%%%%%%%%%%%%%%%%%%%%%%%%%%%%%%%%%%%%%%%%%%%%%
%%%%%%%%%%%%%%%%%%%%%%%%%%%%%%%%%%%%%%%%%%%%%%%%%%%%%%%%%%%%

Physical parameters inferred from spectral modeling 
are tabulated in Table~\ref{allL1} for the one-zone SSC model 
and in Table~\ref{allL2} for the two-zone SSC model. 
The definition of these quantities are provided by Eqs.~\ref{eq:electron_number} to \ref{eq:photon_luminosity}. 

\begin{sidewaystable}
\caption{Jet powers and luminosities derived with the parameters from the one-zone SSC model reported in Table~\ref{OneBlobSSC}.}\label{allL1}
\centering
\begin{tabular}{ccccccccccc} 
\hline\hline 
Date        & $N_{\rm e}$  & $\langle \gamma_{\rm e} \rangle$  & $L_{\rm e}$   & $L_{\rm p}$  & $L_{\rm B}$  & $U'_{\rm e}/U'_{\rm B}$   & $L_{\rm jet}$  & $L_{\rm syn}$ & $L_{\rm IC}$  & $L_{\rm ph}$ \\
  - - -     & [$10^{-1}$] &  [$10^{3}$]     & [$10^{43}$] & [$10^{43}$] & [$10^{42}$] &   [$10^{1}$]  & [$10^{44}$]      & [$10^{42}$]& [$10^{41}$] & [$10^{42}$]\\
$[\rm MJD]$ & [cm$^{-3}$] &  - - - & [erg s$^{-1}$] & [erg s$^{-1}$] & [erg s$^{-1}$] &    - - -   & [erg s$^{-1}$]  & [erg s$^{-1}$]& [erg s$^{-1}$] & [erg s$^{-1}$]                     \\
\hline
55265 & 2.5 & 3.4 & 7.8 & 4.2 & 6.5 & 1.2 & 1.3 & 6.6 & 14. & 8.1  \\
55266 & 2.5 & 3.4 & 8.0 & 4.3 & 6.5 & 1.2 & 1.3 & 7.2 & 16. & 8.8  \\
55267 & 2.4 & 3.3 & 7.3 & 4.0 & 6.5 & 1.1 & 1.2 & 4.6 & 11. & 5.7  \\
55268 & 2.5 & 3.5 & 7.9 & 4.2 & 6.5 & 1.2 & 1.3 & 5.4 & 14. & 6.7  \\
55269 & 2.6 & 3.4 & 8.2 & 4.4 & 6.5 & 1.3 & 1.3 & 5.5 & 14. & 6.9  \\
55270 & 2.5 & 3.3 & 7.5 & 4.1 & 6.5 & 1.2 & 1.2 & 3.5 & 9.8 & 4.5  \\
55271 & 2.5 & 3.4 & 7.6 & 4.1 & 6.5 & 1.2 & 1.2 & 4.0 & 11. & 5.1  \\
55272 & 2.5 & 3.3 & 7.5 & 4.1 & 6.5 & 1.1 & 1.2 & 3.7 & 10. & 4.7  \\
55273 & 2.5 & 3.2 & 7.3 & 4.1 & 6.5 & 1.1 & 1.2 & 3.1 & 8.7 & 4.0  \\
55274 & 2.5 & 3.1 & 7.0 & 4.1 & 6.5 & 1.1 & 1.2 & 2.5 & 6.5 & 3.1  \\
55275 & 2.3 & 3.2 & 6.8 & 3.9 & 6.5 & 1.1 & 1.1 & 2.8 & 7.2 & 3.5  \\
55276 & 2.5 & 3.2 & 7.3 & 4.1 & 6.5 & 1.1 & 1.2 & 3.0 & 8.2 & 3.8  \\
55277 & 1.9 & 3.3 & 5.8 & 3.2 & 6.5 & .90 & .97 & 2.6 & 5.7 & 3.2  \\
\hline
\end{tabular}
\tablefoot{
$N_{\rm e}$: total electron number density; $\langle \gamma_{\rm e} \rangle$: mean electron Lorentz factor; $L_{\rm e}$: jet power carried by electrons; $L_{\rm p}$:the jet power carried by protons; $L_{\rm B}$: jet power carried by the magnetic field; $U'_{\rm e}/U'_{\rm B}$: the ratio of comoving electron and magnetic field energy densities; $L_{\rm jet}$: total jet power; $L_{\rm syn}$: the synchrotron luminosity; $L_{\rm IC}$: inverse-Compton luminosity; $L_{\rm ph}$: total photon luminosity from the SSC model. See the calculation explanation in Sect.~\ref{Discussion}.
}
\end{sidewaystable}

\begin{sidewaystable}
\caption{Jet powers and luminosities derived with the parameters from the two-zone SSC model reported in Table~\ref{TwoBlobSSC}.}\label{allL2}
\centering
\begin{tabular}{ccccccccccc|ccccccc} 
\hline\hline 
Date        & $N_{\rm e}$  & $\langle \gamma_{\rm e} \rangle$  & $L_{\rm e}$   & $L_{\rm p}$  & $L_{\rm B}$  & $U'_{\rm e}/U'_{\rm B}$   & $L_{\rm jet}$  & $L_{\rm syn}$ & $L_{\rm IC}$  & $L_{\rm ph}$ & $^{\rm sum}L_{\rm e}$  & $^{\rm sum}L_{\rm p}$  & $^{\rm sum}L_{\rm B}$ &  $^{\rm sum}L_{\rm jet}$ & $^{\rm sum}L_{\rm syn}$ & $^{\rm sum}L_{\rm IC}$ & $^{\rm sum}L_{\rm ph}$ \\
 - - -       & [$10^{-1}$] &  [$10^{4}$]& [$10^{43}$] & [$10^{41}$] & [$10^{41}$] &    [$10^{1}$]  &    [$10^{43}$]       & [$10^{41}$]& [$10^{40}$] & [$10^{41}$]                       & [$10^{43}$] & [$10^{43}$] & [$10^{42}$] & [$10^{44}$]  & [$10^{42}$]& [$10^{41}$] & [$10^{42}$]\\
\tiny $[\rm MJD]$ &\tiny [cm$^{-3}$] &\tiny  - - - &\tiny [erg s$^{-1}$] &\tiny [erg s$^{-1}$] &\tiny [erg s$^{-1}$] &\tiny    - - -   &\tiny [erg s$^{-1}$]  &\tiny [erg s$^{-1}$]&\tiny [erg s$^{-1}$] &\tiny [erg s$^{-1}$]                       &\tiny [erg s$^{-1}$] &\tiny [erg s$^{-1}$] &\tiny [erg s$^{-1}$]   &\tiny [erg s$^{-1}$]&\tiny [erg s$^{-1}$] &\tiny [erg s$^{-1}$] &\tiny [erg s$^{-1}$]\\
\hline

\hline
  \multicolumn{11}{c|}{the quiescent blob} & \multicolumn{7}{c}{     } \\ % 
\hline
 - - & 2.5 & .31 & 7.0 & 410 &65. & 1.1 & 12. & 25. & 65. & 31. &     &     &     &     &     &     &     \\
%- - & 2.5 & 0.31 & 7.0 & 1400 & 65. & 1.1 & 9.0 & 22. & 58. & 28. &     &     &     &     &     &     &     \\
\hline

\hline
  \multicolumn{11}{c|}{the flaring blob} &  \multicolumn{7}{c}{the quiescent blob + the flaring blob} \\
\hline
% 55265 & 1.6 & 10. & 1.6 & 9.6 & 5.3 & 3.1 & 1.7 & 12. & 16. & 13. & 8.6 & 1.4 & 7.0 & 1.1 & 3.4 & 7.4 & 4.1 \\
% 55266 & 1.9 & 10. & 2.0 & 11. & 4.8 & 4.1 & 2.0 & 12. & 20. & 14. & 9.0 & 1.4 & 7.0 & 1.1 & 3.4 & 7.8 & 4.2 \\
% 55267 & 2.1 & 7.0 & 1.4 & 13. & 4.8 & 3.0 & 1.5 & 7.0 & 16. & 8.7 & 8.4 & 1.4 & 7.0 & 1.1 & 2.9 & 7.4 & 3.7 \\
% 55268 & .90 & 14. & 1.2 & 5.5 & 4.8 & 2.5 & 1.3 & 8.5 & 7.9 & 9.3 & 8.2 & 1.4 & 7.0 & 1.0 & 3.0 & 6.6 & 3.7 \\
% 55269 & 1.6 & 9.7 & 1.6 & 9.8 & 3.9 & 4.0 & 1.6 & 7.8 & 14. & 9.1 & 8.6 & 1.4 & 6.9 & 1.1 & 3.0 & 7.2 & 3.7 \\
% 55270 & 1.3 & 8.2 & 1.1 & 8.1 & 2.7 & 4.0 & 1.1 & 3.1 & 6.5 & 3.7 & 8.1 & 1.4 & 6.8 & 1.0 & 2.5 & 6.4 & 3.2 \\
% 55271 & 1.6 & 9.1 & 1.4 & 9.6 & 2.7 & 5.3 & 1.5 & 4.5 & 10. & 5.5 & 8.4 & 1.4 & 6.8 & 1.0 & 2.6 & 6.8 & 3.4 \\
% 55272 & .78 & 11. & .81 & 4.7 & 2.7 & 3.0 & .84 & 3.1 & 8.8 & 4.0 & 7.8 & 1.4 & 6.8 & .98 & 2.5 & 6.7 & 3.2 \\
% 55273 & .74 & 7.3 & .54 & 4.5 & 2.7 & 2.0 & .57 & 1.4 & 1.7 & 1.5 & 7.5 & 1.4 & 6.8 & .96 & 2.3 & 6.0 & 3.0 \\
% 55274 & - - & - - & - - & - - & - - & - - & - - & - - & - - & - - & 7.0 & 1.4 & 6.5 & .90 & 2.2 & 5.8 & 2.8 \\
% 55275 & .93 & 7.3 & .67 & 5.6 & 1.7 & 3.9 & .69 & 1.1 & 2.0 & 1.3 & 7.7 & 1.4 & 6.7 & .97 & 2.3 & 6.0 & 2.9 \\
% 55276 & .71 & 8.7 & .61 & 4.3 & 1.7 & 3.5 & .63 & 1.2 & 1.5 & 1.3 & 7.6 & 1.4 & 6.7 & .96 & 2.3 & 5.9 & 2.9 \\                          
% 55277 & .56 & 8.2 & .45 & 3.4 & 1.7 & 2.6 & .47 & .82 & .85 & .90 & 7.5 & 1.4 & 6.7 & .95 & 2.3 & 5.9 & 2.9 \\
%                                               |                ||                       |
55265 & 1.6 & 9.0 & 1.4 & 2.8 & 5.3 & 2.6 & 1.5 & 13. & 18. & 15. & 8.4 & 4.1 & 7.0 & 1.3 & 3.8 & 8.3 & 4.6 \\
55266 & 1.9 & 9.0 & 1.7 & 3.4 & 4.8 & 3.4 & 1.7 & 13. & 23. & 15. & 8.7 & 4.1 & 7.0 & 1.4 & 3.8 & 8.8 & 4.6 \\
55267 & 2.1 & 6.5 & 1.3 & 3.8 & 4.8 & 2.8 & 1.4 & 7.9 & 18. & 9.7 & 8.3 & 4.1 & 7.0 & 1.3 & 3.3 & 8.3 & 4.1 \\
55268 & .89 & 12. & 1.1 & 1.6 & 4.8 & 2.2 & 1.1 & 9.5 & 8.8 & 10. & 8.1 & 4.1 & 7.0 & 1.3 & 3.4 & 7.4 & 4.1 \\
55269 & 1.6 & 8.6 & 1.4 & 2.9 & 3.9 & 3.5 & 1.4 & 8.7 & 15. & 10. & 8.4 & 4.1 & 6.9 & 1.3 & 3.4 & 8.0 & 4.1 \\
55270 & 1.3 & 7.6 & 1.0 & 2.4 & 2.7 & 3.7 & 1.1 & 3.4 & 7.3 & 4.2 & 8.0 & 4.1 & 6.8 & 1.3 & 2.8 & 7.2 & 3.5 \\
55271 & 1.6 & 8.4 & 1.3 & 2.9 & 2.7 & 4.8 & 1.4 & 5.0 & 12. & 6.2 & 8.3 & 4.1 & 6.8 & 1.3 & 3.0 & 7.7 & 3.7 \\
55272 & .77 & 9.3 & .71 & 1.4 & 2.7 & 2.6 & .76 & 3.5 & 9.9 & 4.5 & 7.7 & 4.1 & 6.8 & 1.3 & 2.8 & 7.5 & 3.5 \\
55273 & .74 & 6.9 & .50 & 1.3 & 2.7 & 1.9 & .54 & 1.5 & 1.9 & 1.7 & 7.5 & 4.1 & 6.8 & 1.3 & 2.7 & 6.7 & 3.3 \\
55274 & - - & - - & - - & - - & - - & - - & - - & - - & - - & - - & 7.0 & 4.1 & 6.5 & 1.2 & 2.5 & 6.5 & 3.1 \\
55275 & .93 & 6.9 & .63 & 1.7 & 1.7 & 3.6 & .66 & 1.2 & 2.2 & 1.5 & 7.6 & 4.1 & 6.7 & 1.3 & 2.6 & 6.7 & 3.2 \\
55276 & .70 & 8.0 & .56 & 1.3 & 1.7 & 3.2 & .59 & 1.3 & 1.7 & 1.5 & 7.6 & 4.1 & 6.7 & 1.3 & 2.6 & 6.7 & 3.2 \\
55277 & .56 & 7.6 & .42 & 1.0 & 1.7 & 2.4 & .45 & .92 & .95 & 1.0 & 7.4 & 4.1 & 6.7 & 1.2 & 2.6 & 6.6 & 3.2 \\
%                                                 |                 |                       |
%       & -1  &  5  & 43  & 40  & 41  &  1  & 43  & 41  & 40  & 41  & 43  & 43  & 42  & 44  & 42  & 41  & 42
%                                          ||
%  -1     4     43    40    41     1    43   43    43    42    44
%  1.6 & 9.0 & 1.4 & 9.5 & 5.3 & 2.6 & 1.4 & 8.4 & 1.4 & 7.0 & 1.0
%  1.9 & 9.0 & 1.7 & 11. & 4.8 & 3.4 & 1.7 & 8.7 & 1.4 & 7.0 & 1.1
%  2.1 & 6.5 & 1.3 & 13. & 4.8 & 2.8 & 1.4 & 8.3 & 1.4 & 7.0 & 1.0
%  .89 & 12. & 1.1 & 5.4 & 4.8 & 2.2 & 1.1 & 8.1 & 1.4 & 7.0 & 1.0
%  1.6 & 8.6 & 1.4 & 9.7 & 3.9 & 3.5 & 1.4 & 8.4 & 1.4 & 6.9 & 1.0
%  1.3 & 7.6 & 1.0 & 8.1 & 2.7 & 3.7 & 1.0 & 8.0 & 1.4 & 6.8 & 1.0
%  1.6 & 8.4 & 1.3 & 9.5 & 2.7 & 4.8 & 1.3 & 8.3 & 1.4 & 6.8 & 1.0
%  .77 & 9.3 & .71 & 4.7 & 2.7 & 2.6 & .75 & 7.7 & 1.4 & 6.8 & .97
%  .74 & 6.9 & .50 & 4.5 & 2.7 & 1.9 & .54 & 7.5 & 1.4 & 6.8 & .95
%  .93 & 6.9 & .63 & 5.6 & 1.7 & 3.6 & .65 & 7.6 & 1.4 & 6.7 & .97
%  .70 & 8.0 & .56 & 4.3 & 1.7 & 3.2 & .58 & 7.6 & 1.4 & 6.7 & .96
%  .56 & 7.6 & .42 & 3.4 & 1.7 & 2.4 & .44 & 7.4 & 1.4 & 6.7 & .94
\hline
\end{tabular}
\tablefoot{$N_{\rm e}$: total electron number density; $\langle \gamma_{\rm e} \rangle$: mean electron Lorentz factor; $L_{\rm e}$: jet power carried by electrons; $L_{\rm p}$: jet power carried by protons; $L_{\rm B}$: jet power carried by the magnetic field; $U'_{\rm e}/U'_{\rm B}$: ratio of comoving electron and magnetic field energy densities; $L_{\rm jet}$: total jet power; $L_{\rm syn}$: synchrotron luminosity; $L_{\rm IC}$: inverse-Compton luminosity; $L_{\rm ph}$: total photon luminosity from the SSC model. See the calculation explanation in Sect.~\ref{Discussion}. The quantities with the $^{sum}$ superscript report the sums of the quantities from the quiescent and the flaring blob.
}
\end{sidewaystable}

\end{appendix}


\begin{thebibliography}{}
\bibitem[Abdo et al.(2010)]{latsed} Abdo, A.~A., Ackermann, 
M., Agudo, I., et al.\ 2010, \apj, 716, 30 %{2010ApJ...716...30A}
\bibitem[Abdo et al.(2011)]{AbdoMrk421} Abdo, A.~A., Ackermann, 
M., Ajello, M., et al.\ 2011, \apj, 736, 131 %{2011ApJ...736..131A}
\bibitem[Abdo et al.(2014)]{2014ApJ...782..110A} Abdo, A.~A., Abeysekara, 
A.~U., Allen, B.~T., et al.\ 2014, \apj, 782, 110 
%\bibitem[Abdo et al.(2014)]{3FGL} Abdo, A~A., et al. (The Fermi collaboration), 2014, Fermi Large Area Telescope Third Source Catalog, in preparation
\bibitem[Acciari et al.(2009)]{Acciari2009} Acciari, V.~A., Aliu, 
E., Aune, T., et al.\ 2009, \apj, 703, 169 %{2009ApJ...703..169A}
\bibitem[Acciari (2011)]{AcciariPhDThesis} Acciari, V., 2011, PhD thesis, Galway-Mayo Institute of Technology (\url{http://veritas.sao.arizona.edu/documents/Theses/Acciari_}\\\url{Thesis.pdf})
\bibitem[Acciari et al.(2011)]{AcciariMrk4212011} Acciari, V.~A., Aliu, 
E., Arlen, T., et al.\ 2011, \apj, 738, 25 %{2011ApJ...738...25A}
\bibitem[Acciari et al.(2014)]{2014APh....54....1A} Acciari, V.~A., Arlen, 
T., Aune, T., et al.\ 2014, Astroparticle Physics, 54, 1 
\bibitem[Ackermann et al.(2011)]{2011ApJ...741...30A} Ackermann, M., 
Ajello, M., Allafort, A., et al.\ 2011, \apj, 741, 30 
\bibitem[Ackermann et al.(2012)]{Ackermann2012} Ackermann, M., 
Ajello, M., Albert, A., et al.\ 2012, \apjs, 203, 4 
\bibitem[Aharonian et al.(1997)]{Aharonian:1997rm} Aharonian, F.~A., Hofmann, W., Konopelko, A.~K., \& {V{\"o}lk}, H.~J.\ 1997, Astroparticle Physics, 6, 343 %{1997APh.....6..343A}
\bibitem[Aharonian et al.(2002)]{Aharonian2002} Aharonian, F., Akhperjanian, A., Beilicke, M., et al.\ 2002, \aap, 393, 89 %{2002A&A...393...89A}
\bibitem[Aharonian et al.(2003)]{Aharonian2003} Aharonian, F., Akhperjanian, A., Beilicke, M., et al.\ 2003, \aap, 410, 813 %{2003A&A...410..813A}
\bibitem[Aharonian et al.(2005)]{Aharonian2005} Aharonian, F., Akhperjanian, A.~G., Aye, K.-M., et al.\ 2005, \aap, 437, 95 %{2005A&A...437...95A}
\bibitem[Albert et al.(2007)]{Mrk421MAGIC} Albert, J., Aliu, E., Anderhub, H., et al.\ 2007, \apj, 663, 125 %{2007ApJ...663..125A}
\bibitem[Albert et al.(2007)]{Albert2007c} Albert, J., Aliu, E., Anderhub, H., et al.\ 2007, Nuclear Instruments and Methods in Physics Research A, 583, 494 %{2007NIMPA.583..494A}
\bibitem[Aller et al.(1985)]{Aller1985} Aller, H.~D., Aller, M.~F., Latimer, G.~E., \& Hodge, P.~E.\ 1985, \apjs, 59, 513 
\bibitem[Aleksi{\'c} et al.(2012a)]{2012A&A...542A.100A} Aleksi{\'c}, J., Alvarez, E.~A., Antonelli, L.~A., et al.\ 2012, \aap, 542, A100
\bibitem[Aleksi{\'c} et al.(2012b)]{Ale12} Aleksi{\'c}, J., Alvarez, E.~A., Antonelli, L.~A., et al.\ 2012, Astroparticle Physics, 35, 435 %{2012APh....35..435A}
\bibitem[Aleksi{\'c} et al.(2015)]{2015arXiv150202650M} Aleksi{\'c}, J., Ansoldi, S., Antonelli, L.~A., et al.\ 2015, arXiv:1502.02650 
\bibitem[Asano et al.(2014)]{2014ApJ...780...64A} Asano, K., Takahara, F., 
Kusunose, M., Toma, K., \& Kakuwa, J.\ 2014, \apj, 780, 64 
\bibitem[Atwood et al.(2009)]{Atwood2009} Atwood, W.~B., Abdo, A.~A., Ackermann, M., et al.\ 2009, \apj, 697, 1071 %{2009ApJ...697.1071A}
\bibitem[B{\l}a{\.z}ejowski et al.(2005)]{Blazejowski2005} B{\l}a{\.z}ejowski, M., Blaylock, G., Bond, I.~H., et al.\ 2005, \apj, 630, 130 %{2005ApJ...630..130B}
\bibitem[{Bloom \& Marscher}(1996)]{bloom}Bloom, S. D. \& Marscher, A. P. 1996,
\apj, 461, 657 
\bibitem[Boone et al.(2002)]{2002ApJ...579L...5B} Boone, L.~M., Hinton, 
J.~A., Bramel, D., et al.\ 2002, \apjl, 579, L5 
\bibitem[Bradt et al.(1993)]{RXTERef} Bradt, H.~V., Rothschild, R.~E., \& Swank, J.~H.\ 1993, \aaps, 97, 355 
\bibitem[Breiman(2001)]{Breiman2001} Breiman, L. 2011, Machine Learning, 45, 5
\bibitem[Brinkmann et al.(2005)]{Brinkmann05} Brinkmann, W., Papadakis, I.~E.,
Raeth, C., Mimica, P., \& Haberl, F.\ 2005, \aap, 443, 397 
\bibitem[Burrows et al.(2005)]{Burrows2005} Burrows, D.~N., Hill, 
J.~E., Nousek, J.~A., et al.\ 2005, \ssr, 120, 165 %{2005SSRv..120..165B}
\bibitem[Celotti \& Ghisellini(2008)]{2008MNRAS.385..283C} Celotti, A., \& Ghisellini, G.\ 2008, \mnras, 385, 283 
\bibitem[Chen et al.(2011)]{2011MNRAS.416.2368C} Chen, X., Fossati, G.,  Liang, E.~P., \& B{\"o}ttcher, M.\ 2011, \mnras, 416, 2368 
\bibitem[Cogan(2006)]{Cogan:2006}{Cogan}, P. 2006, Ph.D. thesis, School of Physics, University College Dublin
\bibitem[Daniel et~al.(2007)]{Daniel:2007kx} {Daniel}, M.~K., et~al. 2007, Proceedings of the $30^\mathrm{th}$ International Cosmic Ray Conference, ed. R. Caballero, et al. (Mexico City, Mexico: Universidad Nacional Aut\'onoma de M\'exico), Vol.~3, pages~1325
\bibitem[Dermer \& Schlickeiser(1993)]{dermer} Dermer, C. D. \& Schlickeiser, R.
1993, \apj, 416, 458 
\bibitem[Dom{\'{\i}}nguez et al.(2011)]{Dominguez11} Dom{\'{\i}}nguez, A., Primack, J.~R., Rosario, D.~J., et al.\ 2011, \mnras, 410, 2556 %{2011MNRAS.410.2556D}
\bibitem[Edelson et al.(2002)]{Edelson2002} Edelson, R., Turner, 
T.~J., Pounds, K., et al.\ 2002, \apj, 568, 610 %{2002ApJ...568..610E}
\bibitem[Finke et al.(2008)]{2008ApJ...686..181F} Finke, J.~D., Dermer, C.~D., \& B{\"o}ttcher, M.\ 2008, \apj, 686, 181
\bibitem[Finke et al.(2010)]{finke10} Finke, J.~D., Razzaque, S., \& Dermer,
C.~D.\ 2010, \apj, 712, 238 
\bibitem[Fitzpatrick(1999)]{Fitzpatrick99} Fitzpatrick, E.~L.\ 1999, \pasp, 111, 63 
\bibitem[Franceschini et al.(2008)]{franc08} Franceschini, A., Rodighiero, G., \& Vaccari, M.\ 2008, \aap, 487, 837 %{2008A&A...487..837F}
\bibitem[Fomin et al.(1994)]{fomin1994} Fomin, V.~P., Fennell, 
S., Lamb, R.~C., et al.\ 1994, Astroparticle Physics, 2, 151 %{1994APh.....2..151F}
\bibitem[Fossati et al.(2008)]{fossati08} Fossati, G., Buckley, 
J.~H., Bond, I.~H., et al.\ 2008, \apj, 677, 906 %{2008ApJ...677..906F}
\bibitem[Fukugita et al.(1995)]{Fukugita1995} Fukugita, M., Shimasaku, K., \&
Ichikawa, T.\ 1995, \pasp, 107, 945
\bibitem[Gaidos et al.(1996)]{1996Natur.383..319G} Gaidos, J.~A., Akerlof, 
C.~W., Biller, S., et al.\ 1996, \nat, 383, 319 
\bibitem[Georganopoulos \& Kazanas(2003)]{2003ApJ...594L..27G} Georganopoulos, M., \& Kazanas, D.\ 2003, \apjl, 594, L27 
\bibitem[Ghisellini et al.(2005)]{2005A&A...432..401G} Ghisellini, G., Tavecchio, F., \& Chiaberge, M.\ 2005, \aap, 432, 401
\bibitem[Henri \& Saug{\'e}(2006)]{2006ApJ...640..185H} Henri, G., \& Saug{\'e}, L.\ 2006, \apj, 640, 185 
\bibitem[Hillas(1985)]{Hillas:1985ta} {Hillas}, A.~M. 1985, Proceedings of the $19^\mathrm{th}$ International Cosmic Ray Conference, San Diego (CA), USA, Vol.~3, pages~445
\bibitem[Horan et al.(2009)]{Horan2009} Horan, D., Acciari, V.~A., Bradbury, S.~M., et al.\ 2009, \apj, 695, 596 %{2009ApJ...695..596H}
\bibitem[Kalberla et al.(2005)]{Kalberla2005} Kalberla, P.~M.~W., Burton, W.~B., Hartmann, D., et al.\ 2005, \aap, 440, 775 %{2005A&A...440..775K}
\bibitem[Katarzy{\'n}ski et al.(2003)]{Katar2003} Katarzy{\'n}ski, K., Sol, H.,
\& Kus, A.\ 2003, \aap, 410, 101 
\bibitem[Katarzy{\'n}ski et~al.(2005)]{Katarzynski05} Katarzy{\'n}ski, K., Ghisellini, G., Tavecchio, F., Maraschi, L., Fossati, G. \& Mastichiadis, A.  2005, \aap 433, 479-496
\bibitem[Kildea et al.(2007)]{2007APh....28..182K} Kildea, J., Atkins,
R.~W., Badran, H.~M., et al.\ 2007, Astroparticle Physics, 28, 182
\bibitem[Krawczynski et al.(2002)]{2002MNRAS.336..721K} Krawczynski, H.,  Coppi, P.~S., \& Aharonian, F.\ 2002, \mnras, 336, 721 
\bibitem[Krawczynski et~al.(2006)]{Krawczynski:2006ts} {Krawczynski}, H., {Carter-Lewis}, D.~A., {Duke}, C., {Holder}, J., {Maier},  G., {Le Bohec}, S.,  \& {Sembroski}, G. 2006, Astroparticle Physics, 25, 380 
\bibitem[Kneiske et al.(2004)]{kneiske04} {Kneiske, T.~M., Bretz, T.,Mannheim,
K., \& Hartmann, D.~H.\ 2004, \aap, 413, 807} 
\bibitem[Komissarov \& Falle(1997)]{1997MNRAS.288..833K} Komissarov, S.~S., \& Falle, S.~A.~E.~G.\ 1997, \mnras, 288, 833
\bibitem[Krennrich et al.(2002)]{Krennrich2002} Krennrich, F., Bond, 
I.~H., Bradbury, S.~M., et al.\ 2002, \apjl, 575, L9 %{2002ApJ...575L...9K}
\bibitem[Lefa et al.(2011)]{Lefa2011ApJ740p64} Lefa, E., Rieger, F.~M., 
\& Aharonian, F.\ 2011, \apj, 740, 64 %{2011ApJ...740...64L}
\bibitem[Lin et al.(1992)]{lin92} Lin, Y.~C., Bertsch, D.~L., 
Chiang, J., et al.\ 1992, \apjl, 401, L61 %{1992ApJ...401L..61L}
\bibitem[Mannheim(1993)]{1993A&A...269...67M} Mannheim, K.\ 1993, \aap, 269, 67
\bibitem[Maraschi {et~al.}(1992)]{1992ApJ...397L...5M} Maraschi, L., Ghisellini,
G., \& Celotti, A.\ 1992, \apjl, 397, L5
\bibitem[Maraschi et al.(1999)]{1999ApJ...526L..81M} Maraschi, L., Fossati, 
G., Tavecchio, F., et al.\ 1999, \apjl, 526, L81 
\bibitem[Marscher et al.(2008)]{2008Natur.452..966M} Marscher, A.~P., 
Jorstad, S.~G., D'Arcangelo, F.~D., et al.\ 2008, \nat, 452, 966 
\bibitem[Marscher(2014)]{2014ApJ...780...87M} Marscher, A.~P.\ 2014, \apj,  780, 87 
\bibitem[Mattox et al.(1996)]{1996ApJ...461..396M} Mattox, J.~R., Bertsch, 
D.~L., Chiang, J., et al.\ 1996, \apj, 461, 396
\bibitem[Mankuzhiyil et al.(2011)]{2011ApJ...733...14M} Mankuzhiyil, N., 
Ansoldi, S., Persic, M., \& Tavecchio, F.\ 2011, \apj, 733, 14 
\bibitem[Moralejo et al.(2010)]{2010ascl.soft11004M} Moralejo, R.~A., Gaug, 
M., Carmona, E., et al.\ 2010, Astrophysics Source Code Library, 11004 
\bibitem[M{\"u}cke et al.(2003)]{Mucke} M{\"u}cke, A., Protheroe, R.~J., Engel, R., Rachen, J.~P., \& Stanev, T.\ 2003, Astroparticle Physics, 18, 593 %{2003APh....18..593M}
\bibitem[Nandra et al.(1997)]{Nandra1997} Nandra, K., George, I.~M., Mushotzky, R.~F., Turner, T.~J., \& Yaqoob, T.\ 1997, \apj, 476, 70 %{1997ApJ...476...70N}
\bibitem[Nilsson et al.(2007)]{Nilsson2007} Nilsson, K., Pasanen, M., Takalo, L.~O., et al.\ 2007, \aap, 475, 199 %{2007A&A...475..199N}
\bibitem[Nolan et al.(2012)]{Nolan2012} Nolan, P.~L., Abdo, A.~A., Ackermann, M., et al.\ 2012, \apjs, 199, 31
\bibitem[Okumura et al.(2002)]{2002ApJ...579L...9O} Okumura, K., Asahara, 
A., Bicknell, G.~V., et al.\ 2002, \apjl, 579, L9 
\bibitem[Perkins \& Maier(2009)]{Perkins2009} Perkins, J. \& Mayer, G. 2009, eConf Proceedings C091122, astro-ph:0912.3841
\bibitem[Piner et al.(2010)]{Piner10} Piner, B.~G., Pant, N., 
\& Edwards, P.~G.\ 2010, \apj, 723, 1150 %{2010ApJ...723.1150P}
\bibitem[Poutanen et al.(2008)]{Poutanen08} Poutanen, J., Zdziarski, A. A., \& Ibragimov, A. 2008, MNRAS, 389, 1427
\bibitem[Poole et al.(2008)]{Poole2008} Poole, T.~S., Breeveld, 
A.~A., Page, M.~J., et al.\ 2008, \mnras, 383, 627 %{2008MNRAS.383..627P}
\bibitem[Punch et al.(1992)]{1992Natur.358..477P} Punch, M., Akerlof, 
C.~W., Cawley, M.~F., et al.\ 1992, \nat, 358, 477 %{1992Natur.358..477P}
\bibitem[Rebillot et al.(2006)]{Mrk421Whipple2006} Rebillot, P.~F., 
Badran, H.~M., Blaylock, G., et al.\ 2006, \apj, 641, 740 %{2006ApJ...641..740R}
\bibitem[Richards et al.(2011)]{Richards2011} Richards, J.~L., Max-Moerbeck, W., Pavlidou, V., et al.\ 2011, \apjs, 194, 29 
\bibitem[Richards et al.(2013)]{Richards2013} Richards, J.~L.,  Hovatta, T., Lister, M.~L., et al.\ 2013, European Physical Journal Web of  Conferences, 61, 04010
\bibitem[Roming et al.(2005)]{Roming2005} Roming, P.~W.~A., 
Kennedy, T.~E., Mason, K.~O., et al.\ 2005, \ssr, 120, 95 %{2005SSRv..120...95R}
\bibitem[Schlegel et al.(1998)]{schlegel98} Schlegel, D.~J., Finkbeiner, D.~P., \& Davis, M.\ 1998, \apj, 500, 525
\bibitem[Schlickeiser (1985)]{Schlickeiser1985AA143p431} Schlickeiser, R.\ 1985, \aap, 143, 431
\bibitem[Sokolov et al.(2004)]{2004ApJ...613..725S} Sokolov, A., Marscher, 
A.~P., \& McHardy, I.~M.\ 2004, \apj, 613, 725 
\bibitem[Stawarz \& Petrosian (2008)]{Stawarz2013ApJ681p1725} Stawarz, L. \& Petrosian, V.\ 2008, \apj, 681, 1725
\bibitem[Stroh \& Falcone(2013)]{Stroh2013} Stroh, M.~C., \& Falcone, A.~D.\ 2013, \apjs, 207, 28
\bibitem[Tagliaferri et al.(2008)]{2008ApJ...679.1029T} Tagliaferri, G., 
Foschini, L., Ghisellini, G., et al.\ 2008, \apj, 679, 1029 
\bibitem[Takami(2011)]{Takami11} {Takami\ 2011, \mnras, 413,1845} 
\bibitem[Tavecchio et al.(1998)]{1998ApJ...509..608T} Tavecchio, F., 
Maraschi, L., \& Ghisellini, G.\ 1998, \apj, 509, 608
\bibitem[Teraesranta et al.(1998)]{Terasranta1998} Teraesranta, H., Tornikoski, M., Mujunen, A., et al.\ 1998, \aaps, 132, 305 %{1998A&AS..132..305T}
\bibitem[Tramacere et al.(2007)]{Tramacere07} Tramacere, A., Massaro, F., \&
Cavaliere, A.\ 2007, \aap, 466, 521
\bibitem[Tramacere et al.(2009)]{Tramacere09} Tramacere, A., Giommi, P., Perri,
M., Verrecchia, F., \& Tosti, G.\ 2009, \aap, 501, 879
\bibitem[de Vaucoleurs et~al.(1991)]{1991trcb.book.....D} de Vaucouleurs, G., de
Vaucouleurs, A., Corwin, H.~G., Jr., Buta, R.~J., Paturel, G., \& Fouque, P.\
1991, Volume 1-3, XII, 2069 pp.~7 figs..~ Springer-Verlag Berlin Heidelberg New
York
\bibitem[Vaughan et al.(2003)]{Vaughan2003} Vaughan, S., Edelson, 
R., Warwick, R.~S., \& Uttley, P.\ 2003, \mnras, 345, 1271 %{2003MNRAS.345.1271V}
\bibitem[Villata et al.(1998)]{Villata1998} Villata, M., Raiteri, C.~M., Lanteri, L., Sobrito, G., \& Cavallone, M.\ 1998, \aaps, 130, 305
\bibitem[Villata et al.(2008)]{Villata2008} Villata, M., Raiteri, C.~M., Larionov, V.~M., et al.\ 2008, \aap, 481, L79 %{2008A&A...481L..79V}
\bibitem[Villata et al.(2009)]{2009A&A...504L...9V} Villata, M., Raiteri, C.~M., Gurwell, M.~A., et al.\ 2009, \aap, 504, L9
\bibitem[Weekes et al.(2002)]{Weekes:2002pi} Weekes, T.~C., Badran, H., Biller, S.~D., et al.\ 2002, Astroparticle Physics, 17, 221 %{2002APh....17..221W}
\end{thebibliography}
\end{document}